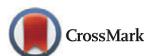

PAPER

# Topological and quantum stability of low-dimensional crystalline lattices with multiple nonequivalent sublattices[*]

Pavel V Avramov[1,**] and Artem V Kuklin[2]

[1] Department of Chemistry, College of Natural Sciences, Kyungpook National University, 80 Daehak-ro, Buk-gu, Daegu 41566, Republic of Korea
[2] Department of Physics and Astronomy, Uppsala University, Box 516, SE-751 20 Uppsala, Sweden
[*] Dedicated to the memory of our close friend, fellow colleague, former student (PA) and teacher (AK) Prof. A A Kuzubov (1974–2016).
[**] Author to whom any correspondence should be addressed.

E-mail: paul.veniaminovich@knu.ac.kr

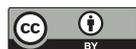

## Abstract

The terms of topological and quantum stabilities of low-dimensional crystalline carbon lattices with multiple non-equivalent sublattices are coined using theoretical analysis, multilevel simulations, and available experimental structural data. It is demonstrated that complex low-dimensional lattices are prone to periodicity breakdown caused by structural deformations generated by linear periodic boundary conditions (PBC). To impose PBC mandatory limitations for complex low-dimensional lattices, the topology conservation theorem (TCT) is introduced, formulated and proved. It is shown that the lack of perfect filling of planar 2D crystalline space by structural units may cause the formation of (i) structure waves of either variable or constant wavelength; (ii) nanotubes or rolls; (iii) saddle structures; (iv) aperiodic ensembles of irregular asymmetric atomic clusters. In some cases the lattice can be stabilized by aromatic resonance, correlation effects, or van-der-Waals interactions. The effect of quantum instability and periodicity breakdown of infinite structural waves is studied using quasiparticle approach. It is found that both perfect finite-sized, or stabilized structural waves can exist and can be synthesized. It is shown that for low-dimensional lattices prone to breakdown of translation invariance (TI), complete active space of normal coordinates cannot be reduced to a subspace of TI normal coordinates. As a result, constrained TI subspace structural minimization may artificially return a regular point at the potential energy surface as either a global/local minimum/maximum. It is proved that for such lattices, phonon dispersion cannot be used as solid and final proof of either stability or metastability. It is shown that *ab initio* molecular dynamics (MD) PBC Nosé–Hoover thermostat algorithm constrains the linear dimensions of the periodic slabs in MD box preventing their thermostated equilibration. Based on rigorous TCT analysis, a flowchart algorithm for structural analysis of low-dimensional crystals is proposed and proved to be a powerful tool for theoretical design of advanced complex nanomaterials.

## Contents

IOP Publishing    *New J. Phys.* **24** (2022) 103015    P V Avramov and A V Kuklin# 1. Introduction

Existence of low-dimensional lattices is limited by fundamental thermodynamic instability of one- (1D) and two-dimensional (2D) crystals at finite temperatures (Landau–Peierls [1] and Mermin–Wagner [2] theorems). Both theorems state that for extended 1D and 2D crystals the crystallographic long-order cannot be maintained because of logarithmic unconvergency of mean square displacements of lattice atoms even at $T = 0$ (1D case) or at the lower limit of integration (2D case). According to the theorems, long-wavelength fluctuations destroy the long-range order of low-dimensional crystals. 2D lattices embedded in a 3D space have a tendency to be crumpled [3] by temperature fluctuations. The anharmonic coupling between bending and stretching modes leads to suppression of temperature fluctuations finally allowing 2D crystals to be exist exhibiting strong height fluctuations [3–5].

Theoretical predictions of the atomic and electronic structure and properties of key nanoclusters have played a leading role in the development of a new revolutionary class of low-dimensional (0D, 1D and 2D) materials. Prediction of fullerenes $C_{60}$ [6, 7] and $C_{20}$ [7] was made back in 1970 and 1973, respectively, using simple structural consideration of aromatic systems [6] and Hückel electronic structure calculations [7]. Experimental discovery of $C_{60}$ buckminsterfullerene in 1985 [8] excellently confirmed all predicted features of atomic and electronic structures of the fullerenes.

The layered crystalline structure of graphite was well known long before experimental exfoliation of graphene [9], so one can consider the first theoretical investigation of graphene electronic structure made by tight-binding model [10, 11], linear combination of atomic orbitals (LCAOs) tight-binding [12] and local density approximation (LDA) [13] within periodic boundary conditions (PBCs) as pioneering studies of band structures of 2D crystalline lattices. Later, theoretical results were confirmed with high accuracy by the ARPES technique [14].

Just after the famous discovery of carbon nanotubes [15], the electronic structure of the tubulenes was successfully independently predicted using tight-binding [16, 17] and DFT [18, 19] approximations within the PBC approach. Later, theoretical predictions were confirmed by scanning tunneling spectroscopy studies [20].

Inspired by discovery of carbon nanotubes, the atomic and electronic structure of *h*-BN nanotubes were predicted and studied theoretically back in February [21] and November [22] of 1994 using tight-binding and LDA calculations within PBC approximation. Experimental discovery of the tubes followed shortly in 1995 [23, 24] was a great experimental achievement based on brilliant theoretical prediction.

Another fascinating prediction of structure and properties of diamanes, the thinnest diamond films, which consist of two ⟨111⟩ cubic or ⟨001⟩ hexagonal diamond planes, respectively, was also made based on density functional (DFT) plane wave (PW) PBC calculations [25]. Just recently, following predicted reaction pathway fluorinated single-layer diamond film was synthesized by fluorination of bilayer graphene [26].

Biphenylene-based novel 2D carbon $\pi$-conjugated material [27] composed of a combination of four-, six- and eight-membered carbon fragments, as well as nanoribbons and nanotubes on its base was proposed and studied using *ab initio* DFT simulations. The low-dimensional atomic lattices were predicted to demonstrate promising hole/electron mobility values typical for ambipolar organic semiconductors with band-gap being significantly smaller than that for the 1D polymer ribbons containing the same number of biphenylene units. The bottom-up approach [28] was used for growth of 2D biphenylene $sp^2$-hybridized carbon network through an interpolymer lateral dehydrofluorination fusion of benzenoid polyphenylene chains on top of Au(111) surface with consequent C–C coupling to form a biphenylene regular lattice. It was experimentally characterized by scanning probe methods, which revealed its metallic nature. The synthesis was accompanied [28] by PBC DFT electronic structure calculations, d$I$/d$V$ differential conductance spectra and theoretical simulation of biphenylene network nc-AFM images using $9 \times 1$ supercell. Comparison of experimental structural data and PBC DFT calculations demonstrated perfect coincidence of experimental and theoretical results. One should mention theoretical prediction [29] and followed experimentally discovery [30] of 2D borophenes as well.

An existence of standing waves in graded 1D InAs/GaAs core–shell nanowires under temperature gradient was predicted using nonequilibrium molecular dynamics simulations [31], and it was found that the standing wave greatly enhance nanowire thermal rectification effect. Using non-equilibrium molecular dynamics structural waves were predicted as well in (0001) oriented wurtzite InAs nanowire and zinc-blend (110)-oriented Si nanowire [32]. In particular, it was shown that the length of the nanowires determines whether or not a standing wave can be formed and the nanowire cross-sections can influence the natural frequencies along the lateral directions. Using nonequilibrium molecular dynamics simulations [33] standing waves also were predicted for non-equilibrium steady-state for graphene and h-BN superlattices. A detailed analysis of the phonon spectra shows that this anomalous thermal conductivity behavior is a result of strong phonon wave interference.

As mentioned above, most theoretical predictions were made using electronic structure calculations based on the PBC approach. The PBC approach is a powerful and widely used tool in solid-state physics and material science to study atomic and electronic structure and spectroscopic characteristics of a vast variety of crystalline lattices, including superlattices. In fact, the PBC approximation can be considered as mandatory symmetry operation, which generates linear 1D, planar 2D and full space 3D crystalline lattices. Since PBC is a kind of symmetry restriction, one can speculate about its improper application for calculations of low-dimensional crystalline lattices with non-zero curvature, which may result in artificial linear crystalline lattices.

Phonon spectra calculation is a well-known, popular and mandatory test of structural stability of the lattices studied within the PBC approach (see, for example, [34]), where the absence of imaginary modes (all force constant matrix elements are non-negative) is considered as a solid and final proof of stability of proposed crystals. It is presumed that in crystals the atoms move around their equilibrium positions following harmonic law. The harmonic approximation works well for 3D crystalline lattices in the vicinity of structural global minima, while obviously it is not the case for 2D/1D materials when one can easily consider a number of elastic modes with linear or even constant dependence of the lattice total energy upon the small bending or twisting deformations perpendicular to the main crystalline plain/axis due to negligibly small corresponding force constants.

In recent years a vast number of low-dimensional nanomaterials were predicted with promising properties based on theoretical PBC calculations (among others see, for example, [35, 36]). Penta-graphene [35] is likely one of the most famous members of the penta-family made of only pentagons. It is characterized by a three-layer thick structure and resembles Cairo pentagonal tilling with central $sp^3$ carbon sublattice and two perpendicular symmetrically nonequivalent $sp^2$ carbon sublattices located in the top and bottom layers. The constituting elements can vary and may include Si, N, and other either metallic or non-metallic ones. In contrast with graphene, which is, in fact, one-atom-thick material, penta-graphene has non-zero thickness (1.2 Å). Penta-graphene was theoretically proposed [35] to be thermodynamically and mechanically stable carbon allotrope, which stability was proved as well by phonon spectrum calculations (no imaginary modes) [35]. However, further theoretical investigations showed that it is mechanically [37], thermodynamically [38], symmetrically and topologically [39, 40], energetically and kinetically [41] unstable. In particular, it was shown [39], that mutually perpendicular top and bottom sublattices of penta-graphene create uncompensated mechanical stress, which causes strong topological instability and bending of the 2D crystalline lattice. Though some unstable materials can be stabilized by a suitable substrate but this is not the case of penta-graphene because of large internal stress [39] which cannot be compensated by van-der-Waals interactions by any kind of supports.

Another perfectly planar 2D carbon allotrope, so-called 'phagraphene' was predicted to be stable in reference [36] with rectangular unit cell parameters $a_1 = 8.0946$ Å and $a_2 = 6.6518$ Å. Its crystalline lattice consists of a regular combination of 5-, 6- and 7-member carbon rings with each single pentagon–heptagon pair surrounded by 8 carbon hexagons. The stability of phagraphene was proved by calculations of phonon spectra [36, 42] (no imaginary modes) using PHONON [43–45] and GULP [46, 47] codes, respectively. Later, using tight-binding and density functional theory it was shown [48] that phagraphene has at least two possible configurations, namely planar and wavy one, which are nearly degenerate in energy.

Different disclinations (see, for example, [49–52]) and regular 2D $sp^2$ carbon lattices composed of a combination of 5-, 6-, and 7-member carbon rings were proposed long before phagraphene [36] (some of the lattices will be considered below). Regular planar 2D Stone–Wales (SW) lattice [49, 53, 54] (pentaheptite or $R_{5,7}$ Haeckelite) composed of only carbon pentagons and heptagons was proposed and the atomic and electronic structure was studied using tight-binding and LDA PW PBC approaches. The $R_{5,7}$ can be generated by the rotation of each single carbon–carbon bond by 90° angle in the center of four fused carbon hexagons of graphene lattice. Significantly more complex planar 2D regular lattices composed of 5-, 6-, and 7-member rings, so-called Haeckelites (namely hexagonal $H_{5,6,7}$, and oblique $O_{5,6,7}$ lattices) were proposed and studied using the tight-binding approach in reference [49]. Among three types of Haeckelites, the hexagonal one, $H_{5,6,7}$, is of special interest because it is composed of linear 5/6/7 cores arranged in regular tilling of hexagonal symmetry. Other regular planar one-atom-thick lattices of intercalated compounds $C_6Ca$ and $C_6Yb$ composed by 5-, 6-, 7-, and 8-member carbon rings under high pressure (>18 GPa) were studied [55–57] using the DFT PW PBC approach and it was found that the most stable *Cmmm* lattice (which was obtained by rotation of two carbon–carbon bonds) is composed of a combination of five- and eight-member rings and the next in energy *Pmma* lattice is composed by lines of 5/7 cores (which is in fact are lines of fused SW defects) separated by lines of carbon hexagons of one hexagon width. Planar and out-of-plain [52, 58–62] $sp^2$ conjugated carbon lattices with SW defects were also considered as well.

One can easily find dozens of predicted unusual/fascinating/fantastic one-, two- and three-atom thick 2D crystalline lattices composed by 3-, 4-, 5-, 6-, 7-, 8-, and even 9-member rings of different light and heavy elements like boron, carbon, nitrogen, silicon, different metals etc somewhere in the literature (in this work the authors are not intended to make the references on disputed publications without any obvious urgent need, nor all of the mentioned hypothetical materials will be critically considered in this study). Mostly the predictions were made using PBC calculations at different levels of theory with calculated phonon spectra as solid and final proof of the stability of the lattices.

Using atomistic simulations [63] it was found that in pristine narrow graphene ribbons, disruption of the aromatic bond network results in depopulation of covalent orbitals and tends to elongate the edge, with an effective force of $f_e \sim 2$ eV Å$^{-1}$. In the case of narrow ribbons, this effect favors structural spontaneous twisting, resulting in the parallel edges forming a double helix with a pitch of about 15–20 lattice parameters. For wide graphene ribbons the twist abruptly vanishes and instead the corrugation localizes near the edges. Using elastic plate theory [64], it was found, that the edge stresses at both zigzag and armchair edges introduce intrinsic ripples in freestanding graphene sheets even in the absence of any thermal effects. It causes out-of-plane warping to attain several degenerate mode shapes. The scaling laws for the amplitude and penetration depth of edge ripples as a function of wavelength were identified. It was found that the edge stresses can lead to twisting and rolling of the nanoribbons.

In this work, a theoretical model to consider mechanical tension in low-dimensional lattices was proposed and it was found that multiple nonequivalent sublattices may generate internal structural stress caused by uncompensated torques. Based on symmetry and structural analysis, a topology conservation theorem (TCT) was formulated and proved for 1D and 2D crystalline lattices. According to TCT, to avoid the uncompensated mechanical stress perpendicular to the main plane or axis of a low-dimensional lattice, the constituting structural fragments must perfectly fit the low-dimensional space without implying any external forces. Due to the leading contribution of the stretching force constants to total energy, any small structural mismatch leads to bending of low-dimensional crystalline lattices in the perpendicular direction to the main plain or axis to eliminate the resulting mechanical stress. It was shown that PBC used to introduce linear translation invariance (TI) symmetry restrictions may cause artificial structural stabilization of linear 1D and planar 2D lattices which are prone for structural deformations. This may lead to multiplication of the translation periods with formation of superperiodic structural waves or even breakdown of the lattice TI in the forms of screwing, bending, or rolling. In the worst case, the PBC may artificially stabilize low-dimensional lattices which are otherwise structurally unstable with breakdown of fundamental short-order structural parameters like bond symmetry, coordination numbers and nature of chemical bonding itself like hybridization character. It was found that in many cases the energy of artificial structural stabilization surpasses the energy of van-der-Waals interactions, which makes it impossible to stabilize the low-dimensional nanoclusters on any kind of substrate and consequently synthesize them. Cluster optimizations of different sizes of finite flakes at various levels of theory result either in bent lattices with multiplication of translation vectors, breakdown of periodicity, structure waves of variable or constant wave lengths, 1D nanotubes or rolls, or even barrier-free transformations to separate graphene-like flakes or aperiodic irregular carbon clusters. Hessian calculations of finite clusters constituted by 5-, 6- and 7-member rings revealed imaginary vibrational frequencies of all considered planar conformers, which unequivocally prove their transition state nature. PBC optimizations of 1D nanoribbons of one-atom-thick wavy carbon allotropes confirm significant energy gain of wave-like structures in comparison with planar conformers. It was shown that structure of standing waves of finite-length flakes can be described by homogeneous D'Alamber second-order differential equations with predetermined boundary conditions. The structure of infinite free-standing lattice waves can be described by inhomogeneous D'Alamber second-order differential equations, for which the solutions can be written as infinite decomposition over planar waves with non-zero external forces acting on the lattice atoms. In contrast to infinite free-standing wave lattices, the finite lattice waves were considered as quantum-stable structures due to perfectly localized positions of the atoms. The solutions of inhomogeneous equations for free-standing infinite lattices were interpreted in terms of quantum instability due to complete delocalization of atomic positions of lattice constituting atoms. It was shown that artificial stabilization and lack of imaginary modes in phonon dispersions of planar conformers prone for breakdown of periodicity may be caused by constrained structural minimization in a subspace of TI normal coordinates (TI SS NC) of true Hilbert-space potential energy surfaces far away from any kind of extremum points. It consequently leads to improper localization of a regular point in complete active space of normal coordinates (CAS NC) as a potential energy surface extremum with consequent erroneous phonon dispersion law calculated at constrained extremmum. It was concluded that the absence of imaginary modes in phonon dispersion laws cannot be considered as a solid and final proof of either structural stability or metastability of low-dimensional crystalline lattices with multiple nonequivalent sublattices. Almost perfect coincidence of the cluster optimizations at various levels of theory combined with PBC PW DFT consideration completely excludes the electronic factor in artificial structural stabilization and directly indicates the symmetry nature of the effect. It was found that in some special cases the structural stress caused by departure of the lattice from mandatory TCT requirements can be almost completely compensated by aromatic resonance and correlation effects. It was shown that ab initio molecular dynamics (MD) PBC Nosé–Hoover thermostat algorithm constrains the linear dimensions of the periodic slabs in MD box preventing their thermostated equilibration. Based on TCT analysis, a flowchart algorithm to study structure and properties of low-dimensional crystals was proposed and proved to be a powerful tool for theoretical design of advanced complex nanomaterials. Proposed theoretical results perfectly correspond to available structural experimental data.

## 2. Structural models and computational methods

To study artificial structural stabilization of low-dimensional lattices caused by translation symmetry restrictions imposed by PBC, one-, two- and three-atom thick low-dimensional carbon $sp^2$ and $sp^3$ allotropes with multiple nonequivalent sublattices were designed by regular combination of rectangular, five-, six-, seven-, and eight-member structural fragments. Along with a set of complex low-dimensional carbon crystals with multiple sublattices, a number of finite graphene flakes (figure 1(a)) were considered as a reference to check the accuracy of atomic and electronic structure calculations. It is necessary to note that in this study a set of low-dimensional carbon $sp^2/sp^3$ lattices were chosen for simplicity since it is well known that symmetry and atomic structure of the lattices are mostly determined by hybridization nature of constituting carbon atoms which allows one to easily separate symmetrical and electronic structural contributions.

The SW lattice [49, 53, 54] which can be obtained from graphene lattice through C–C bonds rotation caused by SW transformation [53] is presented in figure 1(b). The SW transformation converts all hexagons of graphene lattice into pentagon/heptagon pairs, effectively eliminating all graphene hexagons. In $R_{5,7}$ Haeckelite all pentagons and heptagons (and 5/7 pairs as well) share common C–C edges.

The 'phagraphene' lattice [36] is presented in figure 1(c). It was designed by a combination of planar five-, six- and seven-member carbon rings. The main structural feature of phagraphene is pairs of carbon penta- and heptagons which share one common carbon–carbon bond (5/7 fragment), surrounded by 6 hexagons. The neighboring 5/7 pairs contact each other by the vertices of pentagons and heptagons through one C–C bond.

Another phagraphene-like lattice can be designed by insertion of one hexagon between pentagons and heptagons effectively separating the 5 and 7 fragments (figure 1(d)). Let us introduce $p$(pentagon) $h$(heptagon)$C$(carbon), $phC(n, m)$, notations to denote 2D lattices composed of 5-, 6-, and 7 carbon fragments, where $n$ index denotes the number of hexagons, which separate 5 and 7 rings, and $m$ index indicates the number of hexagons which separate the 5/7 pairs. In these notations, the phagraphene lattice can be denoted as $phC(0,1)$, the SW lattice, $R_{5,7}$ (no hexagons in the lattice), must be indexed as $phC(0,0)$, the next one, with one hexagon separating pentagons and heptagons and without C–C bonds separating 5/6/7 cores can be denoted as *ph*C(1,0), and the last considered lattice with completely separated pentagons and heptagons by the hexagons (each single pentagon surrounded by 5 hexagons and each single heptagon surrounded by 7 hexagons) is denoted as *ph*C(1,1) (figure 1(e)). All four *ph*C(0,1), *ph*C(0,0), *ph*C(1,0), and *ph*C(1,1) lattices are investigated in details in this study. One can simply imagine an infinite number of regular *ph*C(*n*,*m*) lattices with arbitrary large *n* and *m* indexes (see, for example [59] for some of such structures).

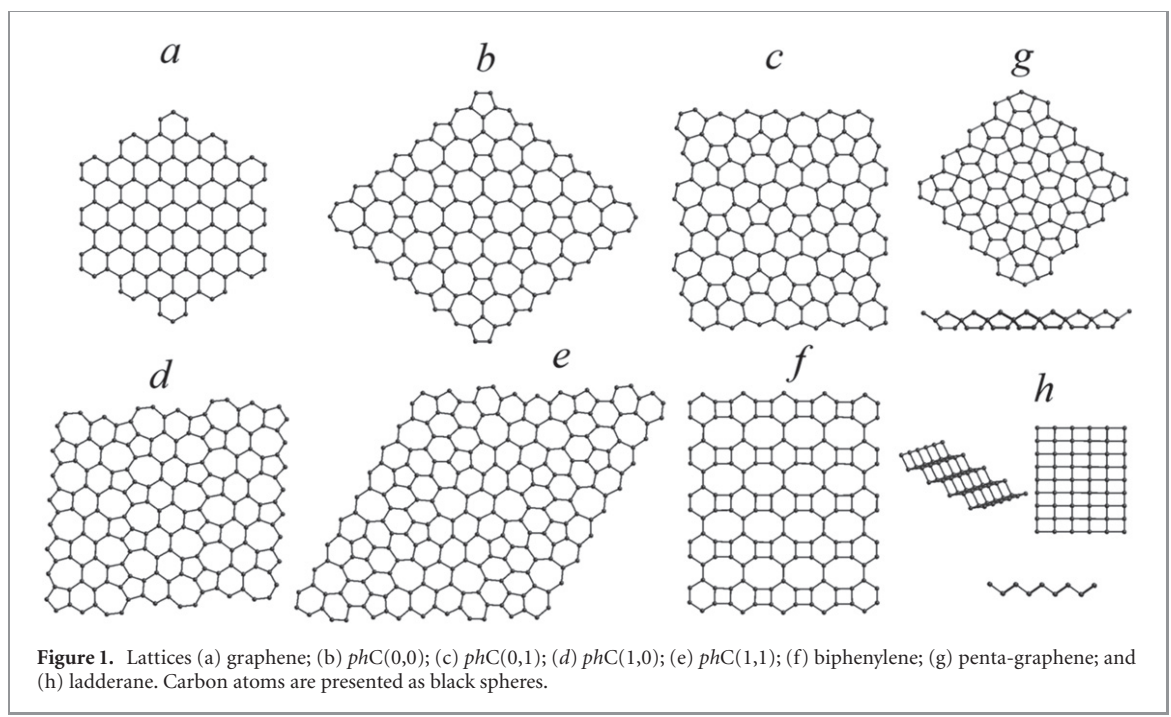

**Figure 1.** Lattices (a) graphene; (b) *ph*C(0,0); (c) *ph*C(0,1); (d) *ph*C(1,0); (e) *ph*C(1,1); (f) biphenylene; (g) penta-graphene; and (h) ladderane. Carbon atoms are presented as black spheres.

The one-atom-thick 2D biphenylene lattice [27, 28] is composed by fused biphenylene fragments and is a combination of four-, six- and eight-membered carbon rings (figure 1(f)). Each single elongated hexagon in 2D biphenylene lattice is surrounded by 4 elongated planar octagons and 2 four-member carbon rings rather close to regular squares. Each single carbon square is surrounded by 2 hexagons and 2 octagons and the octagons are surrounded by 2 other octagons, 2 squares and 4 hexagons. It is necessary to note that despite of very specific electronic effects which will be described below, the lattice can be considered as a planar low-dimensional crystal.

Three-atom-thick mixed $sp^2$/$sp^3$ penta-graphene (figure 1(g), [35]) which consists of a combination of only pentagonal fragments was tested as well for artificial structural stabilization caused by PBC translation symmetry restrictions. Penta-graphene was the first among many others three-atom-thick (one can simply find dozens of publications devoted to theoretical predictions of penta-like 2D materials) theoretically 'discovered' 2D crystalline lattices constituted only by pentagonal fragments, which stability was 'confirmed' by phonon spectra calculations in the framework of translation symmetry restrictions imposed by linear PBC approximation.

The last regular hypothetical 2D carbon allotrope, ladderane [65], consists of $sp^3$ atoms arranged in a two-atom thick buckled rectangular lattice (figure 1(h)). Ladderane can be characterized by two different types of four-coordinated C–C bonds. The first type consists of two distorted single carbon–carbon bonds with perfect 180° angle between them which follow linear pattern forming perfect atomic lines which belong to a single atomic plane (two atomic planes in total). Two other single C–C bonds keep close to perfect $sp^3$ bond angle between them bonding atomic lines from neighboring atomic planes. In total, each single carbon atom of ladderane lattice has two single C–C bonds with the partners in the atomic line and two bonds which connect the atomic planes.

The finite flakes and nanoribbons cut out from parent 2D crystals with multiple sublattices were considered using the cluster approach. First, to elucidate the role of symmetry restrictions and exclude the electronic factors in determination of the main features of the lattice structure, the model MM+ potential (a development of the MM2 method [66]) was used to simulate the atomic structure of finite extended clusters. Semiempirical PM3 [67] and hybrid DFT B3LYP [68, 69] coupled with Pople split-valence 6-31G* basis set [70, 71] lattice as implemented in the GAMESS package [72] were used for structural optimization of finite carbon clusters. For some finite clusters of *ph*C(1,0) *ph*C(0,0) and *ph*C(1,1) lattices, combination of 3-21G and 6-31G* basis sets was used as well. Model potential and semiempirical PM3 approaches were used to calculate the atomic structures of extended (up to 1000 carbon atoms) nanoclusters and to reveal the role of electronic factors and accuracy of structure determination as references to *ab initio* DFT B3LYP approach. All cluster electronic structure calculations were performed without taking into account Grimme corrections [73] for van-der-Waals interaction since disperse interactions is an extence characteristic which makes impossible to comare the results of electronic structure calculations for the clusters of different sizes.

The atomic and electronic structure and vibration spectra of a number of graphene flakes ($C_{42}H_{18}$, $C_{114}H_{30}$, $C_{222}H_{42}$, $C_{366}H_{54}$, $C_{546}H_{66}$, and $C_{762}H_{78}$) of hexagonal symmetry with armchair edges (figure 1(a)) were calculated to test reliability of Hessian calculations of extended low-dimensional carbon clusters at PM3, *ab initio* B3LYP/3-21G and B3LYP/6-31G* levels of theory. It was shown that Hessian calculations of all small flakes (up to PM3 $C_{222}H_{42}$) returned only real vibration frequencies starting from 0.01 cm$^{-1}$ as the smallest ones. Depending on the accuracy of structural optimization (which should be $10^{-5}$ kcal/mol/Å or even better for Hessians making it difficult to achieve), for extended flakes ($C_{366}H_{54}$, $C_{546}H_{66}$, and $C_{762}H_{78}$), Hessian PM3 calculations may return one or two imaginary modes of $\sim$2i cm$^{-1}$, that may be considered as practical accuracy of vibrational mode calculations.

The PBC calculations were performed using Vienna *ab initio* simulation package (VASP) [74, 75] coupled with projector augmented wave (PAW) [76] basis set and GGA PBE functional [77] with wavefunction cutoff energy equal to 400 eV. The first Brillouin-zone for periodic calculations was sampled depending on structural features according to the Monkhorst–Pack scheme [78]. The convergence tolerances of forces and electronic minimizations were $10^{-3}$ eV Å$^{-1}$ and $10^{-5}$ eV, respectively. All periodic images were separated by a vacuum distance at least of 15 Å to avoid artificial interactions of the lattice images along *z*-direction. Several finite clusters were calculated at Γ-point of the Brillouin zone. The visualization for electronic and structural analysis software (VESTA) [79] was used for visualization of the results. For selected 2D and 1D crystalline lattices, the PHONOPY code [34] was used to calculate phonon spectra. For all PBC calculations weak disperse interactions were taken into account using Grimme empirical corrections [73].

The *ab initio* molecular dynamics (AIMD) simulations were performed at 300 K temperature controlled by a Nosé–Hoover thermostat [80, 81] (NVT) during 5 ps with a step interval of 1 fs. To reduce the boundary effects, extended 4 × 4 × 1 and 10 × 1 × 1 supercells for 2D and 1D phagraphene lattices, respectively, were used for AIMD simulations of 2D planar phagraphene (*phC*(0,1)) and both planar and wave conformers of 1D phagraphene *phC*(0,1) 10*A* × 1*B* nanoribbons. The wave conformer of *phC*(0,1) 5*A* × 1*B* $C_{144}H_{40}$ and planar conformer of 0D *phC*(0,1) 10*A* × 1*B* $C_{248}H_{56}$ clusters, respectively, were used for AIMD simulations of 0D flakes. During PBC calculations, the slab images were separated by at least 10 Å in all non-periodic directions to avoid artificial interactions between the partners.

In contrast to PBC optimization procedure, during which both atomic coordinates and translation vectors are minimized, all thermostat algorithms conserve topology of simulated ensembles. In particular, NVT algorithm keeps the volume of MD box constant, keeping constant translation vectors, and recalculating atomic coordinates at each step of MD simulation.Effectively, it freezes the MD box dimensions preventing the dimension changes of 2D and 1D objects confined in the box. MD simulations of 0D objects like molecules and clusters (it could be nanocrystals and nanoflakes, for example), do not have such disadvantage since finite atomic clusters are not strictly confined by fixed MD box dimensions, which allow them to change freely their dimensions and orientation.

## 3. Theoretical model of internal mechanical stress of 1D and 2D lattices with multiple nonequivalent sublattices

### 3.1. 1D *h*-BN zig-zag narrow nanoribbon case

A hypothetical 1D *h*-BN zig-zag narrow nanoribbon (*h*-BN ZNR) of one $B_3N_3$ hexagonal fragment width (figure 2) may be considered as the simplest case of a low-dimensional lattice with multiple non-equivalent sublattices. Without loss of generality one can introduce $a_1$, $a_2$, and $b_1$, $b_2$ nonequivalent sublattices for N and B basis atoms, respectively, associated with symmetrically nonequivalent $\mathbf{a}_{a1}$, $\mathbf{a}_{a2}$, $\mathbf{a}_{b1}$, $\mathbf{a}_{b2}$ translation vectors oriented along *X* direction. In particular, both external atomic rows have coordination numbers equal to 2 with symmetrically different environment, and different neighborhood, namely $a_1$ N atoms have 2 $b_2$ B neighbors while $b_1$ B atoms have 2 $a_2$ N neighbors. The coordination numbers of $a_2$ N and $b_2$ B atoms are equal to 3 with 2 $b_1$ and 1 $b_2$ boron atoms and 2 $a_1$ and 1 $a_2$ nitrogen neighbors, respectively. For simplicity all interatomic bond lengths can be considered equal to $R_{NB}$. Since the basis atoms, the boundary conditions, coordination numbers and types of environment of $a_1/a_2$ and $b_1/b_2$ N and B sublattices are nonequivalent, one can write:

$$\mathbf{a}_{a1} \neq \mathbf{a}_{a2} \neq \mathbf{a}_{b1} \neq \mathbf{a}_{b2}.$$

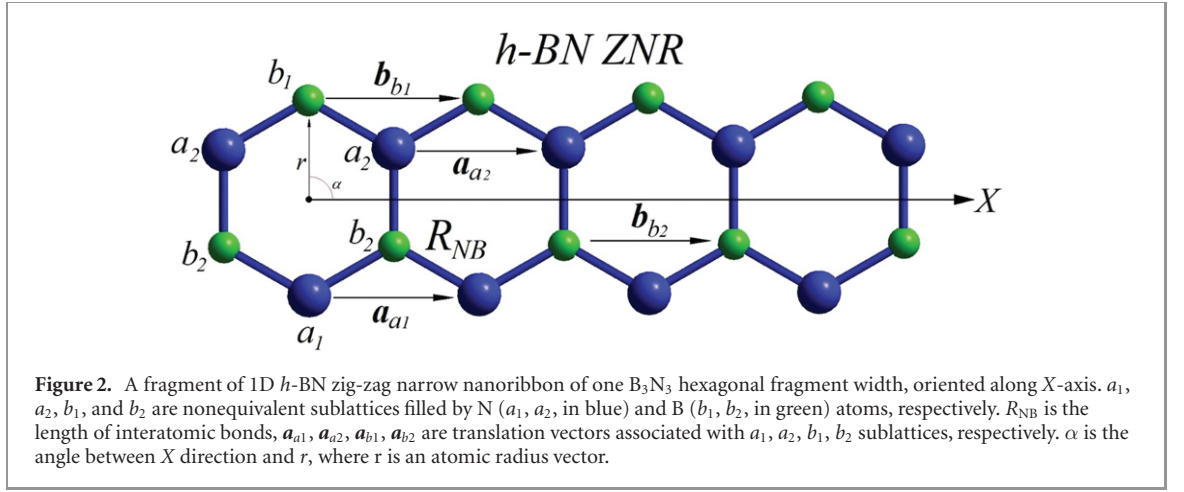

**Figure 2.** A fragment of 1D $h$-BN zig-zag narrow nanoribbon of one $B_3N_3$ hexagonal fragment width, oriented along $X$-axis. $a_1$, $a_2$, $b_1$, and $b_2$ are nonequivalent sublattices filled by N ($a_1$, $a_2$, in blue) and B ($b_1$, $b_2$, in green) atoms, respectively. $R_{NB}$ is the length of interatomic bonds, $\boldsymbol{a}_{a1}$, $\boldsymbol{a}_{a2}$, $\boldsymbol{a}_{b1}$, $\boldsymbol{a}_{b2}$ are translation vectors associated with $a_1$, $a_2$, $b_1$, $b_2$ sublattices, respectively. $\alpha$ is the angle between $X$ direction and $r$, where r is an atomic radius vector.

The N $a_1$ and B $b_1$ sublattices with coordination numbers equal to 2 possess different force constants $Q_1$, $Q_2$ of vibrations along $X$ direction because of different symmetry and environment of the edge N and B atoms even $a_2$–$b_1$–$a_2$ and $b_2$–$a_1$–$b_2$ angles and $R_{NB}$ length of all N–B bonds keep initial parameters of perfect 2D $h$-BN. The force constant $Q$ along the $X$ direction and $R_{NB}$ length of B–N bonds for $a_2$ and $b_2$ sublattices with coordination numbers equal to 3 keep the initial $Q_i$ value of perfect 2D $h$-BN lattice. For perfect hexagonal lattice $|\boldsymbol{a}_{a1}| = |\boldsymbol{a}_{b1}| = \sqrt{3}R_{NB}$. So, for $\boldsymbol{a}_{a2}$ and $\boldsymbol{a}_{b2}$ translation vectors and the force constants the following relationships can be written:

$$Q_1 \neq Q_2 \neq Q$$

$$\boldsymbol{a}_{a2} = \boldsymbol{a}_{b2}.$$

In harmonic approximation and taking into account symmetry restrictions, for the stress energies $E_i$ and forces acting on $a_i$ and $b_i$ sublattices one can write:

$$E_{a2} = E_{b2}$$

$$E_{a1} = E_{b1}$$

$$F_{a2} = F_{b2}$$

$$F_{a1} \neq F_{b1}.$$

Symmetrical non-equivalency of the forces acting on the boundary B and N atoms ($F_{a1} \neq F_{b1}$) creates mechanical stress and structural curvature of ultranarrow $h$-BN ZNR and other similar zig-zag heteroatomic nanoribbons [82].

For perfect hexagonal lattice the torques $\tau_i$ caused by uncompensated forces acting on nonequivalent sublattices of $h$-BN ZNR in respect to the motionless center of the mass can be written as

$$\tau_{a2} = R_{NB} \cdot |F_{a2}| \cdot \sin(1/6\pi) = R_{NB}|F_{a2}|/2$$

$$\tau_{b2} = R_{NB} \cdot |F_{b2}| \cdot \sin(-1/6\pi) = -R_{NB}|F_{a2}|/2$$

$$\tau_{b2} = -\tau_{a2}$$

and $a_2$, $b_2$ associated torques mutually compensate each other. For $a_1$, $b_1$ sublattices of the opposite N and B atoms the $\alpha$ angles are equal to $-\pi/2$ and $\pi/2$, respectively, (figure 2), and

$$|r_{a1}| = |r_{b1}| = R_{NB},$$

so,

$$\tau_{a1} = R_{NB} \cdot |F_{a1}| \cdot \sin(-\pi/2) = -R_{NB}|F_{a1}|$$

$$\tau_{b1} = R_{NB} \cdot |F_{b1}| \cdot \sin(\pi/2) = R_{NB}|F_{b1}|.$$

The $\tau_{b1}$ and $\tau_{a1}$ are oriented in opposite directions. The total torque acting on the nanostructure is the sum of the torques associated with the sublattices. Since $F_{a1} \neq F_{b1}$

$$\tau_{a1} + \tau_{b1} = R_{NB}|F_{b1}| - R_{NB}|F_{a1}| = R_{NB}(|F_{b1}| - |F_{a1}|) \neq 0$$

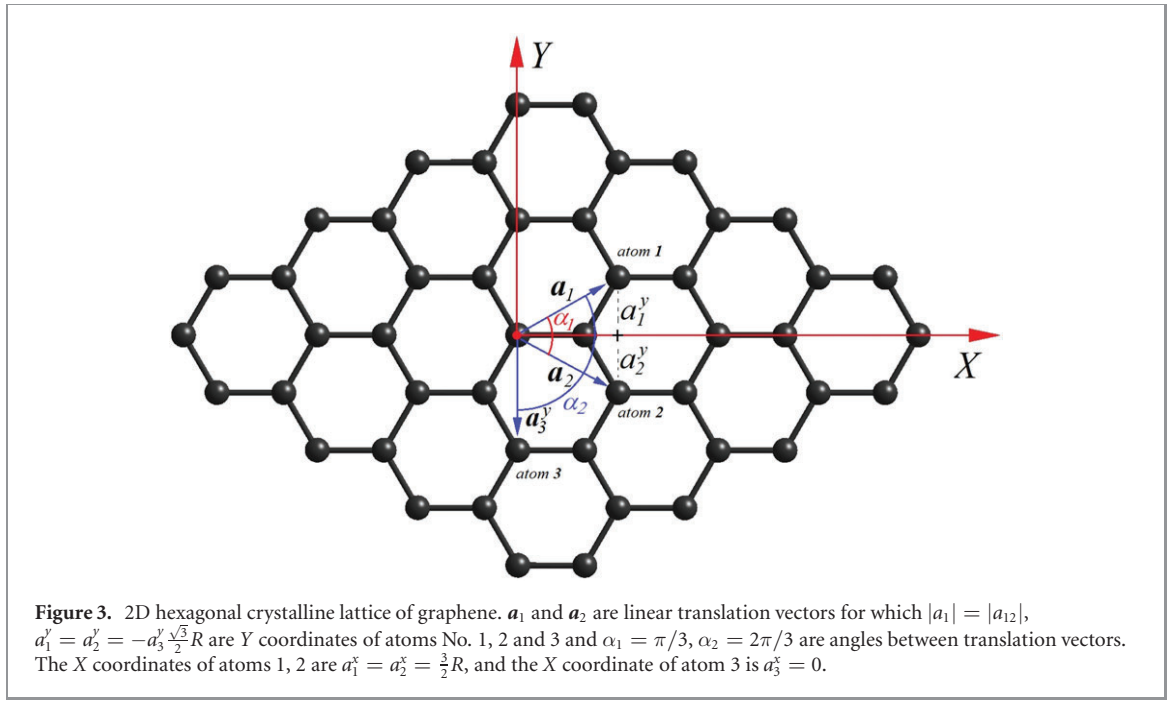

**Figure 3.** 2D hexagonal crystalline lattice of graphene. $\boldsymbol{a}_1$ and $\boldsymbol{a}_2$ are linear translation vectors for which $|a_1| = |a_{12}|$, $a_1^y = a_2^y = -a_3^y \frac{\sqrt{3}}{2} R$ are $Y$ coordinates of atoms No. 1, 2 and 3 and $\alpha_1 = \pi/3$, $\alpha_2 = 2\pi/3$ are angles between translation vectors. The $X$ coordinates of atoms 1, 2 are $a_1^x = a_2^x = \frac{3}{2} R$, and the $X$ coordinate of atom 3 is $a_3^x = 0$.

and the torques do not compensate each other. Since both nonequivalent torques associated with $a_1$ and $b_1$ sublattices are perpendicular to the plane of $h$-BN ZNR lattice and oriented in opposite directions, they lead to displacements of the atoms with the formation of a cone fragment of one $B_3N_3$ width and breaking down the linear translation symmetry [39, 82]. Detailed derivation of the formulas are presented in Supplemetary Data SI2 section.

### 3.2. 1D graphene zig-zag narrow nanoribbon case
The model can be easily extended for the case of 1D zig-zag graphene narrow nanoribbon, which has equivalent edges with equivalent force constants $Q_1$, $Q_2$, so the associated torques perfectly compensate each other and the nanoribbon keeps its linear translation symmetry.

### 3.3. 2D graphene and *h*-BN cases
Both 2D graphene and *h*-BN crystalline lattices have two basis atoms in hexagonal unit cells associated with 2 translation vectors $\boldsymbol{a}_1$ and $\boldsymbol{a}_2$ with absolute values $|a_1| = |a_{12}| = \sqrt{3}R$, where $R$ is the length of either C–C or B–N chemical bonds. The angle between translation vectors, $\alpha = \pi/3$ or $\alpha = 2\pi/3$ can be chosen in two different ways depending on the choice of the unit cell. For the sake of simplicity let us consider the case of the first graphene sublattice (figure 3) with $\alpha_1 = \pi/3$ and the force constant $Q$.

Because of the symmetry of the lattices, the partial energy components acting on atoms 1, 2, 3 are equal to each other:

$$E_{a1}^x = E_{a2}^x = E_{a3}^x$$

$$E_{a1}^y = E_{a2}^y = E_{a3}^{xy}$$

which can be satisfied only if all components are equal to 0, so, for pristine hexagonal lattices the $X$ and $Y$ components of the forces mutually compensate each other and the total stress of the lattice is equal to 0. For the torques acting on atoms 1 and 2 in respect to the origin of the coordinate system one can write:

$$\tau_1 = -\tau_2$$

and the torques mutually compensate each other as well, keeping TI and planar 2D structure. Detailed derivation of the formulas are presented in SI2 section.

## 4. Structural effects of 5, 7, 5/7, and 5/6/7 carbon cores embedded into graphene lattice

### 4.1. Inclusion of pentagons and heptagons into hexagonal carbon lattice
According to *Theorema egregium*, geometrically, an introduction of one pentagonal fragment in perfect hexagonal lattice leads to formation of positive 30° Gaussian curvature [83, 84], whereas introduction of heptagon leads to negative 30° Gaussian curvature [85–92]. It is necessary to note that both planar and curved conformers of all low-dimensional carbon clusters considered in this paragraph were optimized by MM+, PM3 and *ab initio* B3LYP/6-31G* methods without applying of any symmetry restrictions. The $C_{20}H_{10}$ corannulene or [5]circulene (figure 4(a)) [93] and $C_{28}H_{14}$ [7]circulene [94, 95] (figure 4(b)) are the well-known examples of the molecules with distinctively pronounced convex and saddle shapes of $C_{5v}$ and $C_s$ symmetries, respectively, in their ground states. For both of these molecules, two equivalent curved global minima are separated by planar $D_{5h}$ and $D_{7h}$ transition states, respectively.

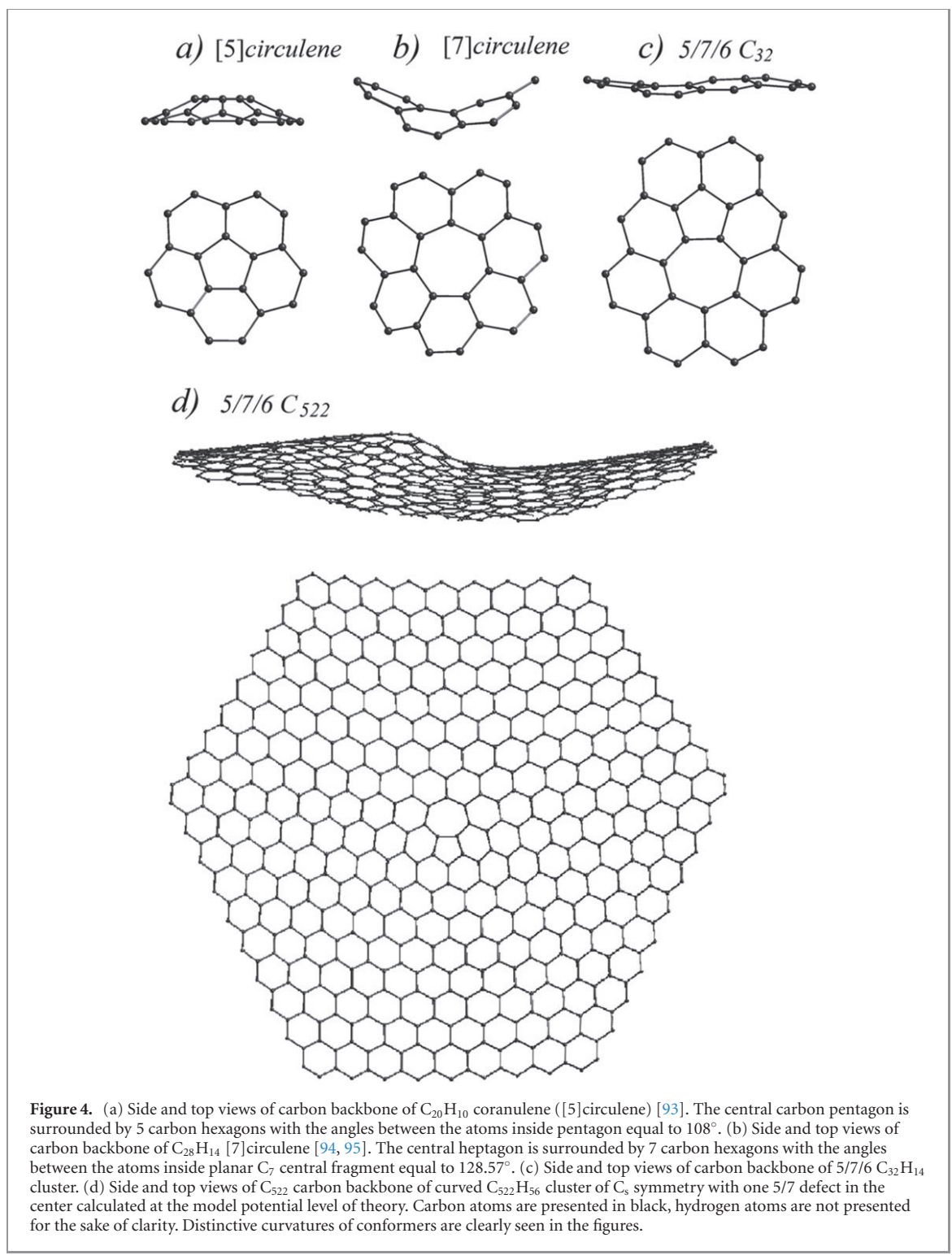

**Figure 4.** (a) Side and top views of carbon backbone of $C_{20}H_{10}$ coranulene ([5]circulene) [93]. The central carbon pentagon is surrounded by 5 carbon hexagons with the angles between the atoms inside pentagon equal to 108°. (b) Side and top views of carbon backbone of $C_{28}H_{14}$ [7]circulene [94, 95]. The central heptagon is surrounded by 7 carbon hexagons with the angles between the atoms inside planar $C_7$ central fragment equal to 128.57°. (c) Side and top views of carbon backbone of 5/7/6 $C_{32}H_{14}$ cluster. (d) Side and top views of $C_{522}$ carbon backbone of curved $C_{522}H_{56}$ cluster of $C_s$ symmetry with one 5/7 defect in the center calculated at the model potential level of theory. Carbon atoms are presented in black, hydrogen atoms are not presented for the sake of clarity. Distinctive curvatures of conformers are clearly seen in the figures.

Fusion of one pentagon results in formation of a perfect planar $C_{10}$ cluster (5/7 core) of $C_{2v}$ symmetry. Consequent adding of 8 $C_6$ rings around 5/7 core leads to formation of $C_{32}$ cluster (5/7 core surrounded by 8 hexagonal rings, 5/7/6 flake, figure 4(c)) with the $C_{2v}$ symmetry of planar conformer. One could speculate that fusion of +30° and −30° curved clusters should lead to cancellation of the curvature of the resulting flake with restoration of perfectly planar $sp^2$ carbon lattice. Initial $C_{10}$ and $C_{32}$ flakes were designed keeping the structure and symmetry of 5/7 core and the lengths of all C–C bonds equal to 1.4 Å. In particular, the pentagonal fragment of the 5/7 core is characterized by internal 108.00° angles between the lattice nodes, as well as the heptagonal fragment keeps original 128.57°. The external angles between the nodes at 5–7 boundary are equal to 123°. One can easily discover significant departures of the angles in the hexagons which surround the 5/7 core (figure 4(c)) from perfect 120° to 104°–132°. The angle between the lines connected external 3 nodes of both sides of the $C_{32}$ cluster is equal to 21°. Significant departure of hexagon internal angles in the 5/7/6 flake from the parent hexagonal lattice angle of 120° makes impossible perfect filling of the hexagonal lattice by introducing the 5/7 defect. Optimization of $C_{32}H_{14}$ 5/7/6 flake at model potential and B3LYP/6-31G* levels of theory keeps its perfect planar structure of C2v symmetry.

Semiempirical PM3 optimization of 5/7/6 $C_{32}H_{14}$ flake revealed planar $C_{2v}$ as well as two equivalent curved $C_s$ conformers (figure 4(c)), which display distinctive uncompensated curvature of complex nature caused by mechanical stresses introduced by 5/7 core. At the PM3 level of theory, curved conformers are symmetrically equivalent global minima without imaginary frequencies in vibrational spectrum and the smallest real frequency 28.43 of cm$^{-1}$). Planar conformer is a transition state with the barrier 0.418 kcal mol, which is much smaller than the accuracy of the PM3 method [96]) between the minima with one imaginary vibration frequency of 39.78i cm$^{-1}$ and first real frequency of 22.00 cm$^{-1}$. Comparing PM3 relative energies, mechanical stress of 5/7/6 $C_{32}H_{14}$ cluster caused by interaction of 5- and 7-fragments competes in energy with aromatic resonance effect, making both conformers almost equal in energy. Here and after the PM3 method is used to reveal the role of electronic factors taking into account at different levels of rigidity in determination of fine details of atomic structure of low-dimensional lattices prone to deformations caused by internal mechanical stress. At restricted R-B3LYP/6-31G* level of theory the 5/7/6 $C_{32}H_{14}$ cluster retains perfectly planar $C_{2v}$ symmetry which might indicate that the $\pi$-system resonance energy exceeds the deformation energy conserving perfectly planar cluster geometry. Comparison of closed-shell R-B3LYP/6-31G* and opened shell U-B3LYP/6-31G* singlets and triplet states returned closed-shell singlet as the ground state with 35.317 kcal/mol singlet–triplet splitting.

Introduction of 5/7 core in extended graphene flake like $C_{522}H_{56}$ cluster with 8 $C_6$ rows around the central 5/7 $C_{10}$ fragment (figure 4(d)) leads to departure the symmetry from parent perfect $D_{6h}$ of graphene flake to $C_s$ point group caused by insertion of one additional carbon hexagon to both left and right bottom edges of the flake with formation of regions with uncompensated positive and negative curvatures (the number of $C_6$ layers can be either increased or decreased without loss of generality). At the model potential level of theory, the perfectly planar transition state has relative energy $E_{\text{planar}}^{\text{Rel}} = 51.343$ kcal mol in respect to two equivalent curved $C_{522}H_{56}$ global minima with relative energy $E_{\text{curved}}^{\text{Rel}} = 0.0$ kcal mol.

### 4.2. Structural units and fragments of *phC*(0,1) lattice with pair of antiparallel 5/7 cores

The unit cell of phagraphene [36] (or in terms introduced above, the *phC*(0,1) lattice) consists of two antiparallel 5/7 fragments separated by 3 carbon hexagons (the smallest *phC*(0,1) cluster constituted by two 5/7 pairs separated by 3 hexagonal rings with additional 3 + 3 hexagons from left and right, $C_{40}$, is presented in figure 5(a)). The unrelaxed $C_{50}$ planar fragment of phagraphene (figure 5(b)) with additional surrounding 10 atoms was designed keeping all C–C bonds equal to 1.4 Å and bond angles of the 5/7 core (figure 4) of 108° and 128° for pentagons and heptagons, respectively. The hexagon between 5/7 partners (just in the center of figure 5(b)) which separates 5/7 partners keeps the same angles as the most distorted hexagon in $C_{32}$ cluster of single 5/7 core (figure 4(c)), namely 105°, 123° and 132° which strongly depart from 120° angles of perfect hexagons. In spite of strong distortions of hexagons, all external zigzag edges of the unit cell are perfectly parallel to each other due to squeezing of the central hexagon between 5/7 cores and stretching of the external sides of its two adjusting hexagons. Comparison of unrelaxed phagraphene $C_{50}$ fragment with the perfect hexagonal lattice (figure 5(b)) reveals significant multiple structural departures up to 0.919 Å of the 5/7 cores along both zigzag and armchair directions of graphene lattice in addition to the distortions of the hexagons which constitute the phagraphene fragment. In particular, at the equator, the unrelaxed phagraphene unit cell is 1.428 Å wider than the parent hexagonal lattice.

Optimization of $C_{40}H_{16}$ phagraphene fragment (figure 5(a)) at model MM+ potential level of theory returned planar cluster of $C_{2h}$ symmetry. At the PM3 level of theory two symmetrically equivalent curved global minima conformers of $C_2$ symmetry with relative energy $E_{\text{curved}}^{\text{Rel}} = 0.0$ kcal mol and planar transition state of $C_{2h}$ symmetry (two imaginary frequencies of 71.31i and 69.97i cm$^{-1}$ and the first real frequency of 51.07 cm$^{-1}$) with relative energy $E_{\text{planar}}^{\text{Rel}} = 12.758$ kcal mol were located. The *ab initio* R-B3LYP/6-31G* also returned two curved $C_2$ equivalent global minima ($E_{\text{curved}}^{\text{Rel}} = 0.0$ kcal mol) and one planar transition state of $C_{2h}$ symmetry with relative energy $E_{\text{planar}}^{\text{Rel}} = 6.233$ kcal mol. Improper torsion angles along and perpendicular to 5/7 cores for PM3 curved global $C_{40}H_{16}$ conformers are equal to 48.67° and 40.83°, respectively with B3LYP/6-31G* corresponding values 50.96° and 46.56°. The U-B3LYP/6-31G* calculations of both open-shell singlet and triplet states confirmed closed-shell singlet as the ground state with singlet–triplet splitting of 8.813 kcal/mol.

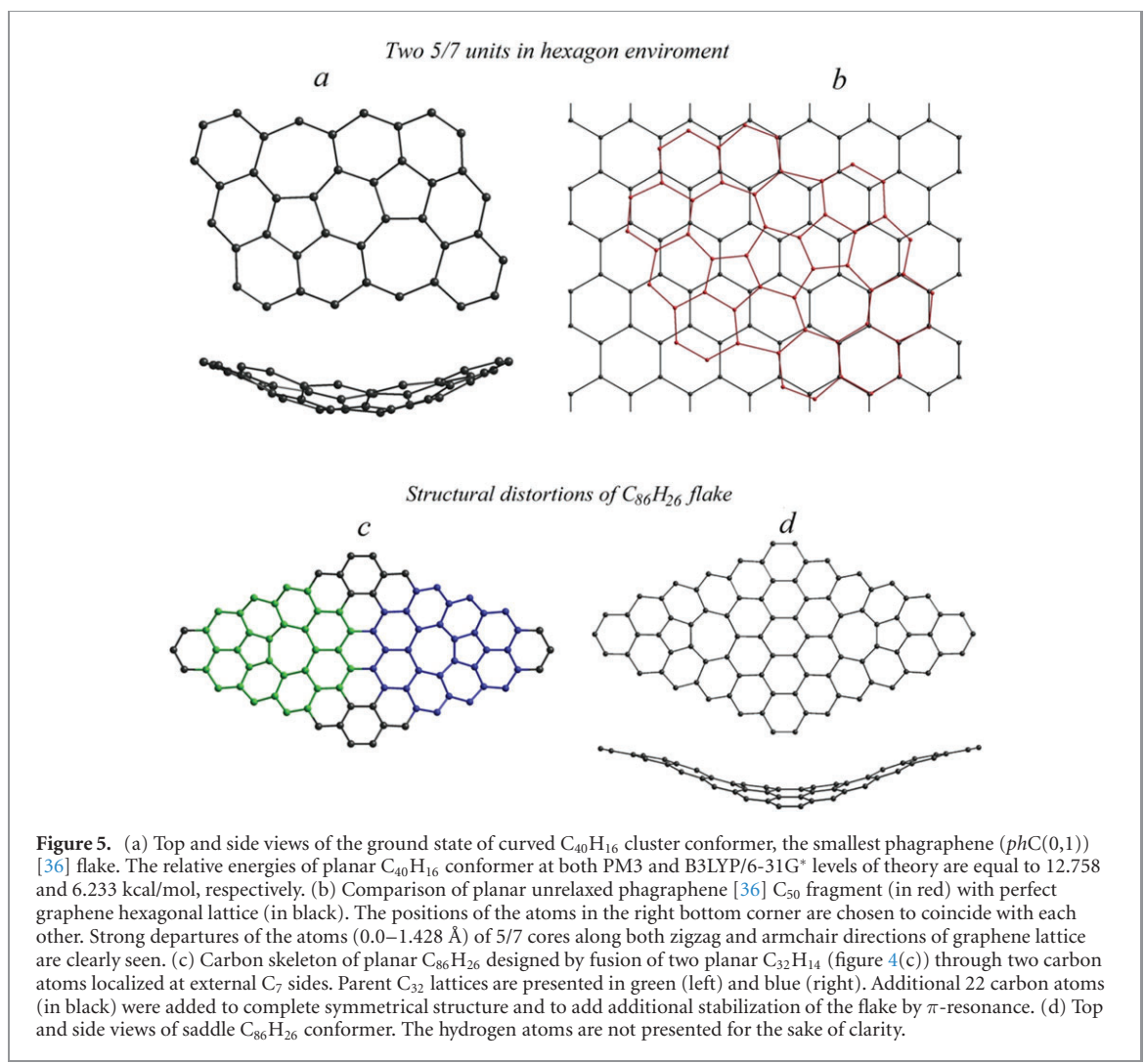

**Figure 5.** (a) Top and side views of the ground state of curved $C_{40}H_{16}$ cluster conformer, the smallest phagraphene ($phC(0,1)$) [36] flake. The relative energies of planar $C_{40}H_{16}$ conformer at both PM3 and B3LYP/6-31G* levels of theory are equal to 12.758 and 6.233 kcal/mol, respectively. (b) Comparison of planar unrelaxed phagraphene [36] $C_{50}$ fragment (in red) with perfect graphene hexagonal lattice (in black). The positions of the atoms in the right bottom corner are chosen to coincide with each other. Strong departures of the atoms (0.0–1.428 Å) of 5/7 cores along both zigzag and armchair directions of graphene lattice are clearly seen. (c) Carbon skeleton of planar $C_{86}H_{26}$ designed by fusion of two planar $C_{32}H_{14}$ (figure 4(c)) through two carbon atoms localized at external $C_7$ sides. Parent $C_{32}$ lattices are presented in green (left) and blue (right). Additional 22 carbon atoms (in black) were added to complete symmetrical structure and to add additional stabilization of the flake by $\pi$-resonance. (d) Top and side views of saddle $C_{86}H_{26}$ conformer. The hydrogen atoms are not presented for the sake of clarity.

To examine the perfect environment of 5/7 cores of phagraphene, the $C_{50}H_{18}$ cluster was considered (figure SI1-1), in which two central 5/7 cores are separated by three $C_6$ rings and surrounded by additional 10 $C_6$ fragments. At the model potential level of theory, optimization returned the planar $C_{50}H_{18}$ cluster. In contrast, PM3 returned two symmetrically equivalent curved global minima of $C_2$ symmetry, which correspond to global minima ($E_{\text{curved}}^{\text{Rel}} = 0.0$ kcal mol, no imaginary frequencies, with the first real frequency 31.97 cm$^{-1}$) and planar transition state of $C_{2h}$ symmetry with relative energy $E_{\text{planar}}^{\text{Rel}} = 5.641$ kcal mol, two imaginary frequences of 67.88i and 53.64i cm$^{-1}$, and the first real mode of 31.56 cm$^{-1}$. A small increase of the relative energy of planar $C_{50}H_{18}$ cluster in comparison with $C_{40}H_{16}$ one is caused by enhanced $\pi$-resonance due to increase of the number of surrounding $\pi$-conjugated $C_6$ rings from 8 ($C_{40}H_{16}$) to 10 ($C_{50}H_{18}$). The R-B3LYP/6-31G* relative energy of planar transition state of $C_{2h}$ symmetry in respect to equivalent curved global minima conformers of $C_2$ symmetry is very small and equal to $E_{\text{planar}}^{\text{Rel}} = 0.504$ kcal mol. Comparison of the relative energies of planar $C_{2h}$ transition states of $C_{40}H_{16}$ and $C_{50}H_{18}$ clusters clearly demonstrates increasing of $\pi$-resonance energy with increasing of the number of $C_6$ fragments in cluster backbone, which competes with deformation energy caused by introduction of 5/7 cores in the carbon lattices of graphene flakes. PM3 improper torsion angles along and perpendicular to 5/7 cores (figure SI1-1) for curved global $C_{50}H_{18}$ conformers are equal to 111.25° and 95.15°, respectively with corresponding R-B3LYP/6-31G* values equal to 128.01° and 129.01°. U-B3LYP/6-31G* calculations confirmed the closed-shell singlet as the ground state with singlet–triplet splitting equal to 30.952 kcal/mol.

An extended $C_{96}H_{24}$ cluster was considered to study structural relaxation of phagraphene core with two 5/7 cores embedded into hexagon lattice environment (figure SI1-2). Two central 5/7 cores are separated by 3 $C_6$ rings and surrounded by additional external 30 $C_6$ fragments. Model potential optimization returns two equivalent global minima of $C_i$ or $C_s$ symmetry with relative energy $E_{curved}^{Rel} = 0.0$ kcal mol (no imaginary modes, the minimal frequency 37.23 cm$^{-1}$), and planar $C_{2h}$ transition state (relative energy $E_{flat}^{Rel} = 2.10$ kcal mol, one imaginary mode of 27.66i cm$^{-1}$, the first real frequency 31.37 cm$^{-1}$) which is much smaller than the accuracy of the force field approach itself. Extension of $C_{50}$-based cluster up to $C_{96}$ leads to prevalence of mechanical stress over $\pi$-resonance on the formation of the distinctive wave-like cluster even at the force field level of theory. The PM3 method returns a significantly larger potential barrier of $E_{flat}^{Rel} = 44.33$ kcal mol between the ground state minima (no imaginary modes, the first real mode is 17.33 cm$^{-1}$, $E_{curved}^{Rel} = 0.0$ kcal mol) and planar transition state, which has 3 imaginary modes of 126.78i, 77.22i, and 66.03i cm$^{-1}$ with the first real one of 27.87i cm$^{-1}$.

Two planar $C_{32}H_{14}$ clusters of $C_{2v}$ symmetry (figure 4(c)) can be fused through the carbon atoms localized at $C_7$ external sides along 5–7 direction (figure 5(c)). One can expect that the fusion and completion of the lattice by additional 22 carbon may lead to formation of planar extended rhombus $C_{86}H_{26}$ flake. The atomic structure of the flake was optimized by model force field, semiempirical PM3 and *ab initio* DFT B3LYP/6-31G* with both restricted and unrestricted $M = 1$ and 3 multiplicities. At all levels of theory, two symmetrically equivalent saddle shape conformers which correspond to equivalent global minima (figure 5(d)) of $C_{2v}$ symmetry with relative energy $E_{curved}^{Rel} = 0.0$ kcal mol were revealed. The planar transition state of $D_{2h}$ symmetry which separates the minima has MM+, PM3 and U-B3LYP/6-31G* relative energies $E_{planar}^{Rel} = 15.85, 0.557$ and 0.329 kcal mol/atom. The U-B3LYP/6-31G* open-shell triplet is the ground state of both planar and saddle conformers with singlet–triplet splitting equal to 13.708 kcal/mol for saddle conformer. The MM+ and PM3 Hessian calculations return groups of imaginary frequencies from 64.94i to 43.52i cm$^{-1}$ and from 78.51i to 9.46i cm$^{-1}$, respectively, for planar transition state with the first real modes of 18.37 cm$^{-1}$ and 13.76 cm$^{-1}$. Structural stress of the $C_{86}H_{26}$ cluster is caused by the symmetrical mismatch of 5/7 cores with a perfect hexagonal lattice of graphene. Distortion of initial planar structure and high relative energies of planar transition state $C_{86}H_{26}$ conformer at all 3 levels of theory in the range of 15.9–0.3 kcal/mol/atom unequivocally demonstrate strong non-linear response of the atomic structure for extension of the carbon lattice due to the local character of the $\pi$-resonance stabilization in comparison with completely delocalized mechanical stress over the whole flake.

### 4.3. Structure of carbon lattices with 5/6/7 cores

Two *phC*(1,1) $sp^2$ carbon flakes, namely $C_{40}H_{16}$ and $C_{62}H_{20}$, formed by embedding of one or two 5/6/7 cores into hexagonal environment are presented in figures 6(a) and (b), respectively. In figure 6, the central 5/6/7/cores, formed by fusion of penta- and hepta-rings through carbon hexagon are marked by blue, (pentagon fragment), green (hexagon fragment) and red (heptagon) for the sake of clarity. The $C_{2v}$ symmetry $C_{14}H_{14}$ cluster which represents the 5/6/7 core keeps its perfect planar geometry at MM+, PM3 and *ab initio* DFT B3LYP/6-31G*. The electronic ground state is closed-shell singlet with B3LYP/6-31G* singlet–triplet splitting of 12.35 kcal/mol. As well as the $sp^2$ $C_{10}H_8$ 5/7 core, the $\pi$-conjugated electron subsystem keeps its perfectly planar topology.

Let us consider both planar and curved conformers of $C_{2v}$ symmetry $C_{40}H_{14}$ flake composed of 5/6/7 core (figure 6(a)) surrounded by 10 carbon hexagons. The PM3 planar conformer is open-shell singlet with relative energies of closed-shell singlet and triplet states equal to 24.078 and 2.196 kcal mol, respectively. The B3LYP/6-31G* revealed that the planar $C_{40}H_{16}$ is a closed-shell singlet with relative energy of triplet state equal to 12.349 kcal/mol. At the PM3 level of theory, both singlets and triplet of the planar conformer have 4 imaginary vibrational frequencies (table SI1-1), making the conformer being a transition state between two symmetrically equivalent degenerated curved global minima (figure 6(a)).

The departure from $C_{40}H_{16}$ planar $C_{2v}$ transition state, to distinctive non-planar $C_s$ conformer (figure 6(a)) leads to visible energy gain of 37.254 kcal/mol for PM3 ground open-shell singlet and 17.408 kcal/mol for B3LYP/6-31G* ground closed-shell singlet. The relative energies of closed-shell singlet and triplet states at the PM3 level of theory are equal to 27.398 and 0.401 kcal mol, respectively, making open-shell singlet and triplet states almost degenerate with very small singlet–triplet splitting, which is smaller than the accuracy of the PM3 method. The B3LYP/6-31G* relative energy of the triplet state is equal to 6.667 kcal/mol. At the PM3 level of theory, no imaginary frequencies in the vibration spectrum of the non-planar $C_{40}H_{16}$ cluster of $C_s$ symmetry for closed-shell singlet and triplet states are observed, making them the global and local minima at potential energy profile of $C_s$ $C_{40}H_{16}$ conformer.

Two antiparallel 5/6/7 cores can be fused through 4 carbon hexagons with formation of extended $sp^2$ $C_{62}H_{20}$ cluster (figure 6(b)), in which the double 5/6/7 core is surrounded by 12 hexagons. The PM3 open-shell singlet is energetically favourable than the closed-shell singlet and triplet states with relative energies of 40.034 and 3.164 kcal/mol, respectively. The PM3 vibration spectra of closed- and open-shell singlets revealed five and seven imaginary frequencies, respectively (table SI1-2), which indicate that the planar conformer is a transition state in both spin states. At the B3LYP/6-31G* level of theory, only a closed-shell singlet planar conformer was located.

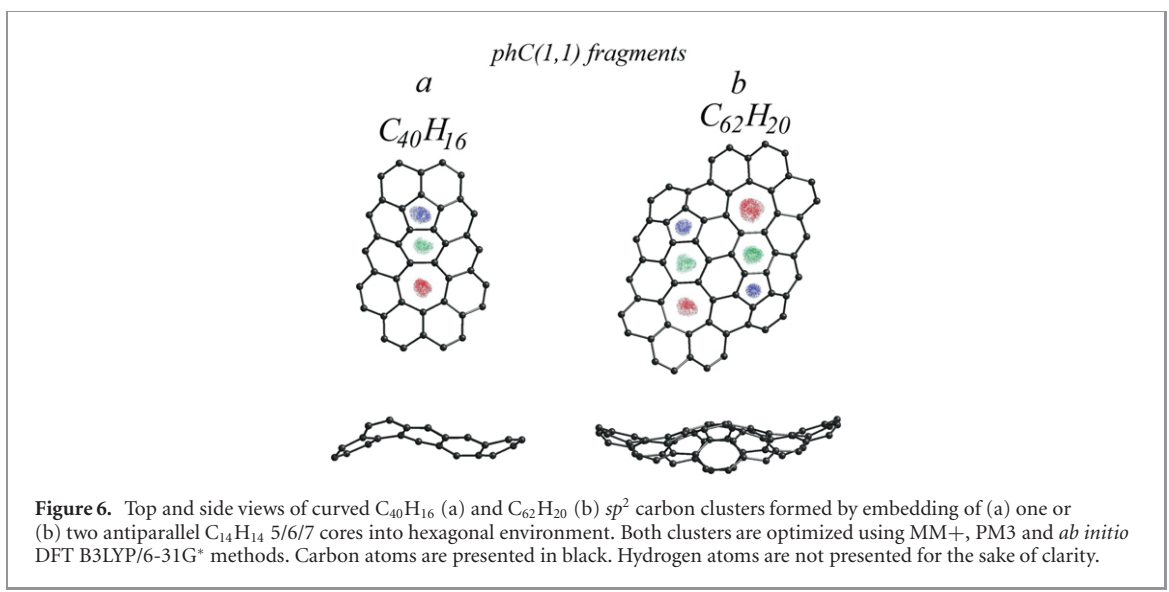

**Figure 6.** Top and side views of curved $C_{40}H_{16}$ (a) and $C_{62}H_{20}$ (b) $sp^2$ carbon clusters formed by embedding of (a) one or (b) two antiparallel $C_{14}H_{14}$ 5/6/7 cores into hexagonal environment. Both clusters are optimized using MM+, PM3 and *ab initio* DFT B3LYP/6-31G* methods. Carbon atoms are presented in black. Hydrogen atoms are not presented for the sake of clarity.

Departure from $sp^2$ planar $C_{2h}$ symmetry leads to localization (force field, PM3 and B3LYP/3-21G) of curved $C_2$ $C_{62}H_{20}$ conformer of complex shape (figure 6(b)). Reduction of symmetry from PM3 planar $C_{2h}$ transition state to two equivalent opened-shell singlet curved $C_2$ global PM3 minima leads to visible energy gain of 65.056 kcal/mol with relative energies of non-planar closed-shell singlet and open-shell triplet states of 49.245 and 13.157 kcal/mol, respectively. No negative frequencies in vibrational spectra for all 3 spin states of non-planar conformers were observed.

For all considered finite clusters, it was found that embedding of the fragments which contain pentagon- and heptagon carbon rings in hexagonal environment in all cases leads to breakdown of perfect planar hexagonal symmetry of $\pi$-conjugated lattices with formation of curved flakes. Comparison of model potential and electronic structure calculations unequivocally demonstrates the leading role of mechanical stress in formation of the fine details of the lattices atomic structure. Following structural and symmetry analysis and using the Hessian calculations, it was found that planar conformers are just transition states with one or several imaginary vibration modes, which connect two symmetrically equivalent curved ground state minima on potential energy surfaces.

## 5. Topological and quantum stability of 1D and 2D *phC(n,m)* lattices

### 5.1. *phC(0,0)* ($R_{5,7}$ Haeckelite) lattice

Perfect planar 2D *phC*(0,0), or regular SW lattice (other names pentaheptite, hexagonal $R_{5,7}$ Haeckelite [49]) belongs to DG47 (*cmm*) diperiodic group [58, 59, 97–99] (figure 7). It constitutes of regularly arranged SW defects formed by rotation of one central carbon–carbon bond of each single pyrene fragment (figure 7(a)) of graphene by 90° [53] which causes its transformation into dicyclopenta[*ef*, *kl*]heptalene (alternative names azupyrene or SW–pyrene) unit. It is necessary to note, that formation of SW fragments cannot be proceeded without distortion of both pentagonal and heptagonal azupyrene fragments and departure of the initial angle between translation vectors of graphene (*P6/mmm* space group) from 60° to 74.02° of *phC*(0,0).

The heptagon external angle of 127.53° of a single freestanding azupyrene (figure 7(b)) is very close to one of freestanding regular planar heptagon (128.57°). The heptagon vertex of 7/5-5 junction is adjusted to two fused regular planar pentagons through the common zig-zag edge with the angle of 144.00°. It is necessary to note, that for 2D *phC*(0,0) lattice this angle is equal to 141.2°. As one can simply estimate, a structural combination of heptagon and fused pentagons does not satisfy the crystallography restriction theorem (CRT) requirements for perfect filling of 2D space since the perfect heptagon angle (128.57°) does not match the angle between two fused perfect pentagons (144.00°) with the difference 144.00°–127.53° = 15.43° (alternative value is 141.2°–127.53° = 13.67°).

It is necessary to note the dominant character of stretching force constant $K_r$ for $sp^2$ carbon atoms which is equal, for example, to 469 kcal/mol/Å$^2$ in respect to $K_\theta$ bending $sp^2$ ∠C–C–C constant (85 kcal/mol/rad$^2$) for AMBER94 force field [100]. The dominant contribution of the stretching mode to the total energy leads to significant distortion of the heptagon angle (128.57° for regular heptagon, see above) in 7/5-5 region of the planar 2D $phC(0,0)$ lattice, making it 141.2°, which is almost equal to the angle adjacent to 5–5 edge (141.3°).

For simplicity, the contribution of aromatic resonance to total energy can be ignored by effectively making the corresponding force constant responsible for keeping planar topology of any aromatic systems equal to 0. The accumulated structural mismatch (15.43°–13.67°) of the units should lead to distortion of planar 2D $phC(0,0)$ lattice with displacement of the atoms from the aromatic plane in the perpendicular direction to compensate the mechanical stress and to conserve the bond lengths minimizing the distortions of bending angles.

So, for low-dimensional cases (1D, 2D) a Topology Conservation Theorem (TCT) can be formulated in the following way: '*To conserve the planar topology of one- or few-atomic layer one-unit-cell-thick low-dimensional crystals with small or zero stabilizing force constants acting in the perpendicular direction to the lattice plain, and to avoid uncompensated mechanical stress, the free-standing constituting fragments (unit cells) must perfectly fit planar low-dimensional space. Due to the leading contribution of the stretching force constants to the total energy, mechanical stress accumulated by small regular structural mismatch of planar structural units should be compensated by out-of-plane structural deformations coupled with either multiplication of translation periods or breakdown of periodicity in one or two dimensions*'.

Experimentally discovered perfect planar hexagonal and triangular lattices of graphene [9], $h$-BN [101], graphane [102], graphdiynes [103, 104], gamma- [105] and holey-graphynes [106], g-$C_3N_4$ [107], and g-$C_4N_3$ [108] perfectly satisfy mandatory requirements of TCT since the structural fragments of low-dimensional crystals perfectly fill in the planar 2D space. It is necessary to note that hypothetic free-standing one-side hydrogenated graphan, so-called graphone [109], does not satisfy TCT mandatory requirements due to strong structural anisotropy. In fact, experimental graphone lattice is stabilized by Ni(111) surface [109] by weak van-der-Waals interactions and, probably, by rather energetic chemisorbtion forces induced by direct interactions of dangling carbon bonds with the substrate in interface region. Reversible dehydrogenation of graphone resulted in complex desorption processes with coverage-dependent changes in the activation energies for the associative desorption of hydrogen as molecular $H_2$ [109].

Two TCT corollaries can be formulated as well: *first*, any kind of external either positive or negative pressure may effectively stabilize planar 2D lattices prone for deformations by annihilation of the internal structural stress; *second*, in stochastic atomistic lattices which are formed by structural units which do not perfectly fill the low-dimensional space, the structural stress and consequent structural regular deformations cannot be accumulated because of mutual compensation of displacements in opposite directions.

The truthfulness of the *first corollary* can be illustrated by experimental fabrication of InGaAs/GaAs micro- and nanotubes with controllable inner diameter (from 4 $\mu$m to 4 nm) by selective etching of stabilizing sacrificial AlAs substrate [110]. The GaAs and InAs pair demonstrate 7.2% lattice structural mismatch, so after freeing of the (2ML)GaAs/(2ML)InAs vertical heterostructure by breakdown of stabilizing substrate, the heterostructure spontaneously rolls into a nanotube with GaAs fragment forming the inner layer of the nanotube. In this method, the AlAs substrate effectively provides stabilizing negative pressure which keeps the external GaAs/InAs vertical heterostructure perfectly planar during initial stages of the synthesys.

The truthfulness of the *second corollary* can be illustrated by the experimental synthesis of an $sp^2$-hybridized one-atom-thick flat carbon membrane with random arrangement of four-, five, six and seven carbon polygons by electron irradiation of graphene [61, 111]. Random distribution of the defects effectively eliminates periodicity of parent $sp^2$ graphene lattice and accumulation of mechanical stress caused by introduction of different types of regular defects.

A set of both screw and planar conformers of finite 1D $n \times 2$ $phC(0,0)$ nanoribbons was studied using PM3 ($n = 1$–18), B3LYP/6-31G$^*$ ($n = 1$–8) and B3LYP/3-21G ($n = 9$–12) methods. The atomic lattices of the lowest in energy $phC(0,0)$ 16 × 2 screw conformer which corresponds to screw half-rotation at all levels of theory, and the highest in energy $phC(0,0)$ 16 × 2 planar conformer are presented in figures 7(d) and (e), respectively.

In figure 7(f) the PM3 and B3LYP/6-31G$^*$ dependencies of stabilization energy per unit cell for formation of screw 1D $n \times 2$ $phC(0,0)$ conformers with respect to the planar ones are presented. At PM3 and B3LYP/6-31G$^*$ levels of theory, only planar 1/2 × 2 and 1 × 2 flakes were located (0 energy gain). The PM3 energy gain can be easily divided into two regions, with the first one with $n = 1$–12 (skipping 11 × 2 flake which belongs to the second region) and the second region with $n = 11$–18 (skipping 12 × 2 flake).

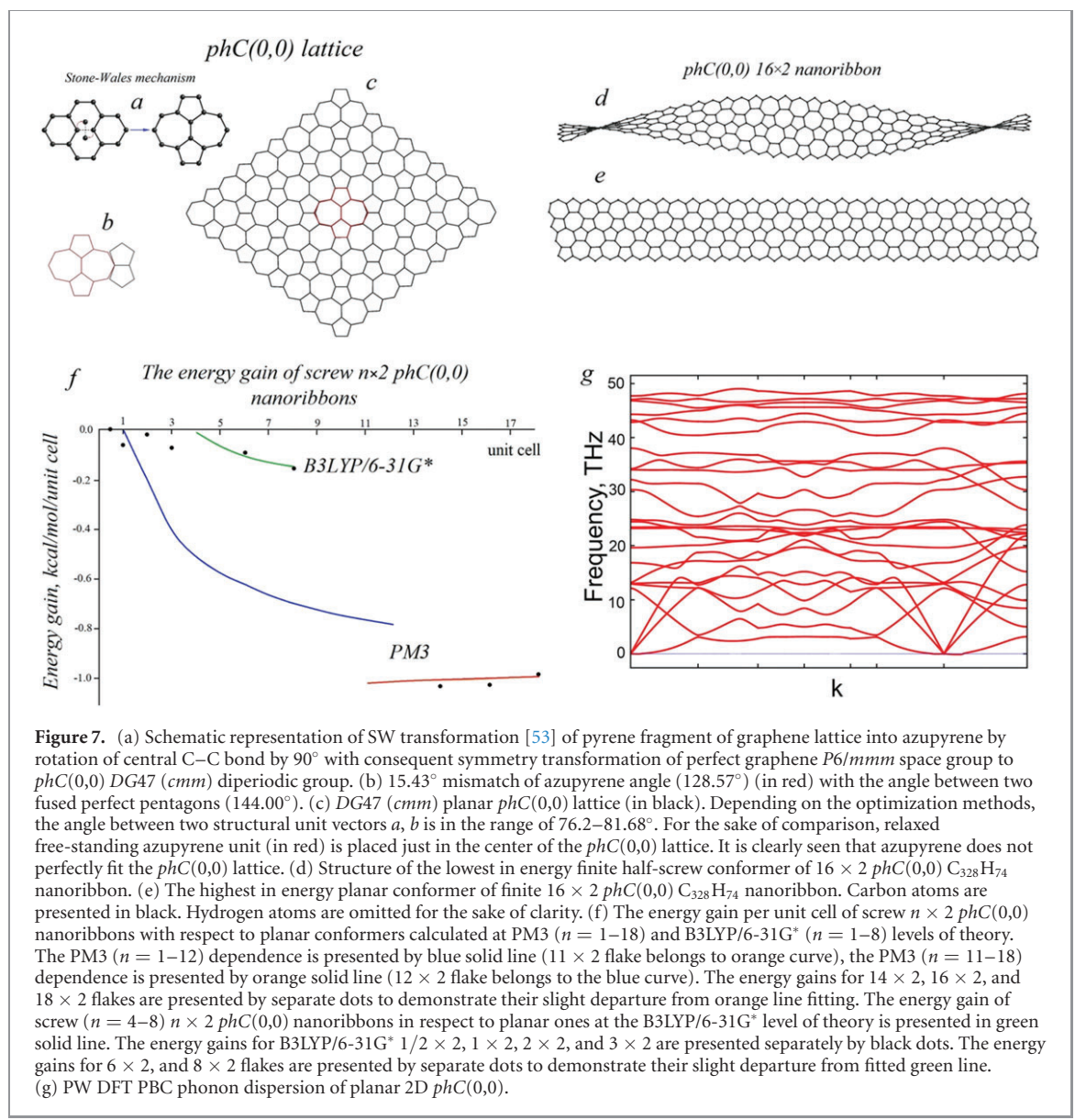

**Figure 7.** (a) Schematic representation of SW transformation [53] of pyrene fragment of graphene lattice into azupyrene by rotation of central C–C bond by 90° with consequent symmetry transformation of perfect graphene $P6/mmm$ space group to $phC(0,0)$ $DG47$ ($cmm$) diperiodic group. (b) 15.43° mismatch of azupyrene angle (128.57°) (in red) with the angle between two fused perfect pentagons (144.00°). (c) $DG47$ ($cmm$) planar $phC(0,0)$ lattice (in black). Depending on the optimization methods, the angle between two structural unit vectors $a$, $b$ is in the range of 76.2–81.68°. For the sake of comparison, relaxed free-standing azupyrene unit (in red) is placed just in the center of the $phC(0,0)$ lattice. It is clearly seen that azupyrene does not perfectly fit the $phC(0,0)$ lattice. (d) Structure of the lowest in energy finite half-screw conformer of $16 \times 2$ $phC(0,0)$ $C_{328}H_{74}$ nanoribbon. (e) The highest in energy planar conformer of finite $16 \times 2$ $phC(0,0)$ $C_{328}H_{74}$ nanoribbon. Carbon atoms are presented in black. Hydrogen atoms are omitted for the sake of clarity. (f) The energy gain per unit cell of screw $n \times 2$ $phC(0,0)$ nanoribbons with respect to planar conformers calculated at PM3 ($n = 1$–18) and B3LYP/6-31G* ($n = 1$–8) levels of theory. The PM3 ($n = 1$–12) dependence is presented by blue solid line ($11 \times 2$ flake belongs to orange curve), the PM3 ($n = 11$–18) dependence is presented by orange solid line ($12 \times 2$ flake belongs to the blue curve). The energy gains for $14 \times 2$, $16 \times 2$, and $18 \times 2$ flakes are presented by separate dots to demonstrate their slight departure from orange line fitting. The energy gain of screw ($n = 4$–8) $n \times 2$ $phC(0,0)$ nanoribbons in respect to planar ones at the B3LYP/6-31G* level of theory is presented in green solid line. The energy gains for B3LYP/6-31G* $1/2 \times 2$, $1 \times 2$, $2 \times 2$, and $3 \times 2$ are presented separately by black dots. The energy gains for $6 \times 2$, and $8 \times 2$ flakes are presented by separate dots to demonstrate their slight departure from fitted green line. (g) PW DFT PBC phonon dispersion of planar 2D $phC(0,0)$.

The first region clearly indicates the extent of aromatic stabilization of π-system up to 12 unit cell length (up to ∼58 Å) with clear $1/x$ energy dependence. From 11 unit cell length (∼52 Å) the energy gain per unit cell demonstrates perfect linear dependence with an almost constant value close to −1 kcal/mol/unit cell. The $14 \times 2$ and $16 \times 2$ nanoribbons demonstrate some secondary aromatic antiresonances which slight decrease of the energy gain (∼0.03 kcal/mol/unit cell). The last $18 \times 2$ nanoribbon demonstrates a small positive deviation from perfect linear dependence due to the lack of accuracy of optimization. For the sake of clarity, the $14 \times 2$, $16 \times 2$ and $18 \times 2$ nanoribbons are presented as separate dots on the PM3 curve.

The $n = 4$–8 $phC(0,0)$ B3LYP/6-31G* energy gain upon length (figure 7(f)) also demonstrates clear $1/x$ energy dependence with slight deviations for $6 \times 2$ and $8 \times 2$ nanoribbons due to the lack of accuracy of optimization and specific aromatic resonances. Small $1 \times 2$, $2 \times 2$ and $3 \times 2$ flakes demonstrate negative stabilization energies with specific energy gains due to abnormal aromatic resonances as well. B3LYP/3-21G calculations of planar and screwed conformers of $n \times 2$ $phC(0,0)$ finite-length nanoribbons with $n = 9$–12 (data are not presented in figure 7(f)) also demonstrate 0.15–0.18 kcal/mol/unit cell energy gain of screwed conformers. The PM3 Hessian calculations of planar $1/2 \times 2$ and $1 \times 2$ flakes demonstrate no imaginary modes, while for planar $n = 2$–5 Hessian calculations returned 1, 3, 5, and 6 imaginary frequencies, respectively. No imaginary frequencies at the PM3 level of theory for all screw conformers were localized. The B3LYP/6-31G* Hessian calculations of planar $1/2 \times 1$, $1 \times 1$ and $1 \times 2$ flakes returned no imaginary frequencies, whereas planar $2 \times 2$ flake demonstrates one imaginary frequency. No imaginary frequencies for $1 \times 1$, $1 \times 2$ and $2 \times 2$ screw flakes were returned. The Hessian calculations combined with symmetry and structural analysis clearly demonstrate that planar conformers of 1D $n \times 2$ $phC(0,0)$ nanoribbons (except $1 \times 2$) are just transition states between two equivalent global minima screw conformers.

PM3 calculations of extended $n \times n$ ($n = 3–8$) $phC(0,0)$ flakes (figure SI1-3) also demonstrate negative stabilization energy of screw conformers from $-0.32$ to $-0.72$ kcal/mol/unit cell without conclusive energy dependence. The absence of clear $1/x$ dependence (figure 7(f)) may be explained by the possible existence of numerous curved isomers among which the true global minimum is difficult to locate.

The phonon dispersion (figure 7(g)) of 2D planar SW lattice calculated at the PW DFT PBC level of theory reveals no imaginary modes. This contradicts to concerted previous results (structural analysis of $phC(0,0)$ lattice, cluster calculations of finite $n \times 2$ and $n \times n$ $phC(0,0)$ nanoribbons and nanoflakes, Hessian calculations, which reveal imaginary modes for all planar conformers without imaginary modes for curved ones, analysis based on newly formulated TCT theorem) for finite $phC(0,0)$ flakes presented above, which unambiguously demonstrate the transition state nature of planar 2D $phC(0,0)$ conformers. This obvious failure of phonon dispersion calculations can be explained by three major reasons: *first*, TI is a symmetry restriction introduced by PBCs, which obviously leads to artificial stabilization of a perfect planar lattice. *Second*, to calculate the phonons with arbitrary (and large) *k*-vectors, proper potential should be used with a supercell 2–3 times larger than the real translation vector of the relaxed lattice. Since the true period of $phC(0,0)$ is at least 32 unit cells (figures 7(d) and (e)), at least $64 \times 64$ supercell should be used to calculate phonon spectrum of the SW lattice. *Third*, and probably the most important reason is that the minimization of planar 2D lattice using the PBC approach was performed in complete active space of normal coordinates, which includes only subspace of normal coordinates which satisfy TI conditions. It is necessary to note that for periodic lattices which satisfy TI conditions, the dimensionality of TI SS NC coincides with CAS NC by definition. The mismatch of structural units of $phC(0,0)$ leads to its departure from planar 2D lattice with formation of aperiodic (2D case) or superperiodic screw (1D case) conformers for which aperiodic coordinates should be included in CAS NC. It effectively makes CAS NC dimensionality much greater than TI SS NC one. From a mathematical point of view, minimization of $phC(0,0)$ in TI SS NC is in fact constrained minimization which can artificially return a regular point on true CAS NC as a local or global minimum or maximum. Consequent phonon calculations using geometry obtained by constrained minimization procedure, as a result, may artificially return no imaginary modes in the dispersion law.

Summarizing the paragraph, it was found that structural units of $phC(0,0)$ lattice do not fit each other and cannot perfectly fill in its planar low-dimensional crystalline space. Assuming the force constant perpendicular to $phC(0,0)$ plane is negligibly small in combination with a dominant character of stretching force constants in comparison with bending ones, a TCT was formulated and proved. Two theorem corollaries were referenced and illustrated by known experimental results as well. It was shown that a mismatch of azupyrene structural units should lead to breakdown of $phC(0,0)$ planar topology. In accordance with newly formulated TCT, the force field and electronic structure calculations of finite $phC(0,0)$ nanoribbons and flakes resulted in bent and screwed structures as global minima with planar conformers between two symmetrically equivalent global minima as transition states. In particular, at the PM3 level of theory, it was found that the screw period of rotation of $n \times 2$ $phC(0,0)$ nanoribbon is equal to $n = 32$. At all levels of theory, Hessian calculations unequivocally confirmed transition state nature of all finite planar $phC(0,0)$ nanoribbons and flakes revealing single or even multiple imaginary modes in their vibrational spectra. Phonon dispersion calculations of planar 2D $phC(0,0)$ revealed no imaginary modes that was interpreted as a result of constrained minimization of $phC(0,0)$ lattice in a subspace of normal coordinates which satisfy TI conditions. The constrained minimization in finite TI SS NC instead of Hilbert CAS NC leads to erroneous localization of a regular point in complete active space as a global/local minimum with the consequent calculation of phonon dispersion law in it. Since in this case the force matrix is calculated at a wrong point (not at a function extremum) the phonon dispersion law cannot be regarded as valid. It was concluded that for low-dimensional complex crystalline lattices with multiple sublattices phonon spectra calculations cannot be used as final and solid proof of structural stability in the case of broken periodicity. To avoid artificial stabilization caused by linear PBC symmetry restrictions, one must perform structural and symmetry analysis of a lattice first, considering symmetry and possible structural mismatches of structural units to satisfy TCT theorem mandatory requirements. Cluster calculations of finite fragments can be used to support TCT analysis as well.

### 5.2. $phC(0,1)$ (phagraphene) lattice

As it was described above, to generate a regular phagraphene rectangular lattice [36], the constituting 5/7 fragments should be oriented in opposite directions, so structurally the neighboring 5/7 pairs cannot share the same vertices. The unrelaxed geometry of 5/7 pair is characterized by the 123.43° external angle at 5/7 seam, which does not fit the 120° of the neighboring hexagons. According to TCT, to avoid accumulation of internal structural stress, 2D and 1D $phC(0,1)$ lattices should depart from perfectly planar topologies with the formation of a bent structures.

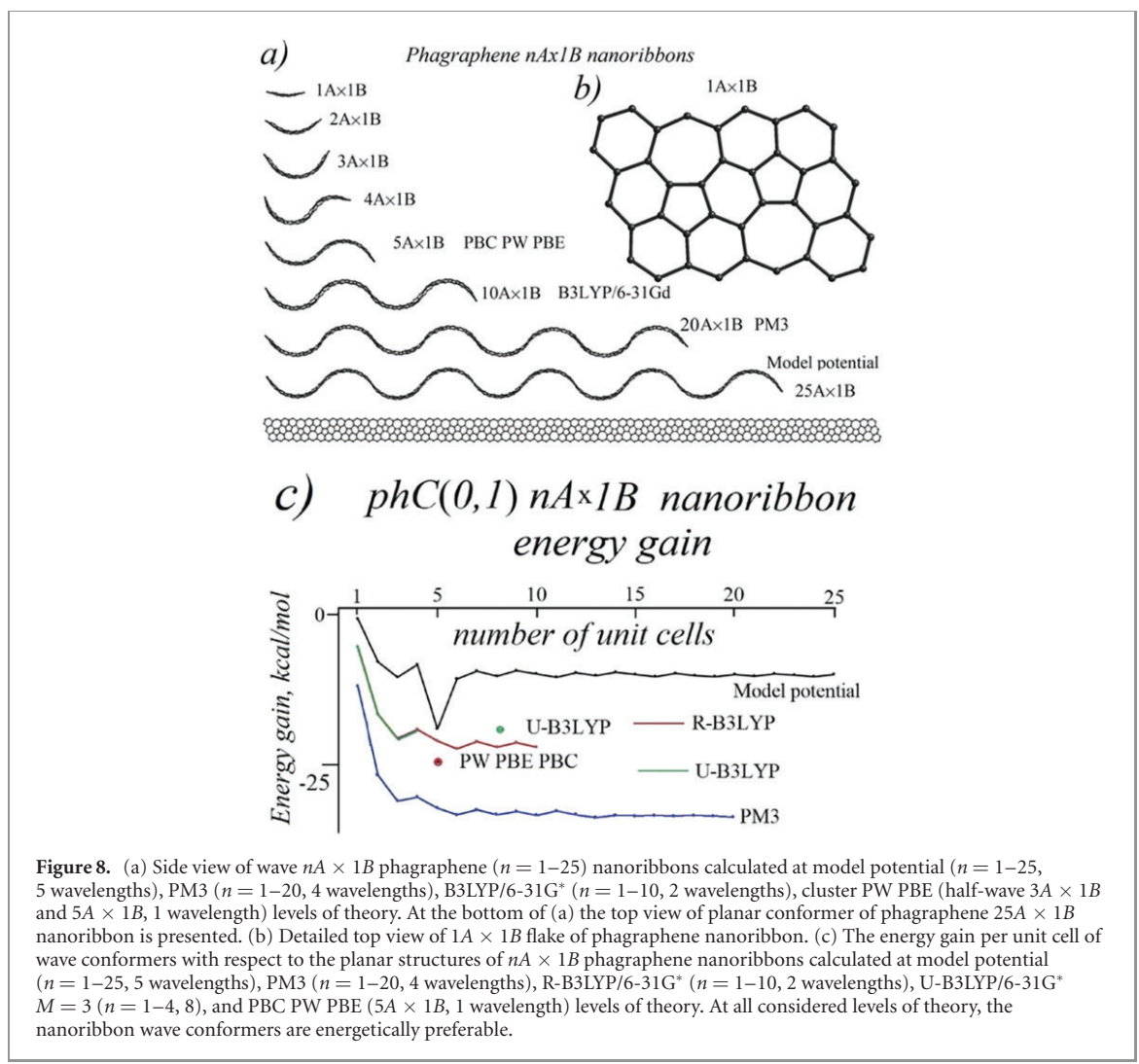

**Figure 8.** (a) Side view of wave $nA \times 1B$ phagraphene ($n = 1-25$) nanoribbons calculated at model potential ($n = 1-25$, 5 wavelengths), PM3 ($n = 1-20$, 4 wavelengths), B3LYP/6-31G* ($n = 1-10$, 2 wavelengths), cluster PW PBE (half-wave $3A \times 1B$ and $5A \times 1B$, 1 wavelength) levels of theory. At the bottom of (a) the top view of planar conformer of phagraphene $25A \times 1B$ nanoribbon is presented. (b) Detailed top view of $1A \times 1B$ flake of phagraphene nanoribbon. (c) The energy gain per unit cell of wave conformers with respect to the planar structures of $nA \times 1B$ phagraphene nanoribbons calculated at model potential ($n = 1-25$, 5 wavelengths), PM3 ($n = 1-20$, 4 wavelengths), R-B3LYP/6-31G* ($n = 1-10$, 2 wavelengths), U-B3LYP/6-31G* $M = 3$ ($n = 1-4, 8$), and PBC PW PBE ($5A \times 1B$, 1 wavelength) levels of theory. At all considered levels of theory, the nanoribbon wave conformers are energetically preferable.

To elucidate the symmetry effects and mechanisms of structural stabilization, planar and curved conformers of *phC*(0,1) finite narrow $nA \times mB$ nanoribbons (*A* and *B* are translation vectors along *X* and *Y* directions, respectively) with $n = 1-25$ and $m = 1-18$ unit cell length and width were studied using model potential MM+, semiempirical PM3, *ab initio* B3LYP/6-31G*, and plane wave PBE methods in cluster approximation. For the sake of comparison, the PBC PW PBE calculations of 1D and 2D *phC*(0,1) $1A \times 1B$ and 1D $5A \times 1B$ periodic supercells of wave (one wavelength) and planar phagraphene nanoribbons were used to reveal the effects of symmetry restrictions introduced by TI conditions of PBC. First, both one-unit cell 1D and 2D calculations resulted in perfectly planar lattices, whereas $5A \times 1B$ periodic supercells revealed two conformers, namely (i) the ground state wave of 5 unit cell wavelength and (ii) the planar high-energy phagraphene nanoribbon.

Optimizations of $nA \times 1B$ nanoribbons for $n = 1-25$ at various levels of theory revealed two distinctly different types of conformers: the lowest in energy ground state wave conformer with the wavelength of 5 unit cells (or 33.844 Å at PM3 level of theory). Each single wave supercell contains 137 carbon atoms of total weight of 1644 Dalton) and planar conformers that are approximately from 20 (model potential) to 34 (PM3) kcal/mol/unit cell higher in energy of corresponding wave structures (figures 8(a) and (b)). Cluster PW PBE optimizations of $1A \times 1B$, $3A \times 1B$ and $5A \times 1B$ flakes returned curved ($n = 1, 3$) or wave conformers as the ground states as well. Both *ab initio* B3LYP/6-31G* and PBC PW PBE approaches reveal close relative energies of 23–25 kcal/mol/unit cell for planar structures as compared to wave-like ones. Using *ab initio* R-B3LYP/6-31G* and U-B3LYP/6-31G* calculations it was found that for *phC*(0,1) ($n = 1-4$) nanoribbons the triplet and closed-shell singlet states are almost degenerated in energy. For *phC*(0,1) $8A \times 1B$ wave conformer, the open-shell singlet is the ground state with the energy of singlet–triplet splitting equal to 8.4 kcal/mol/unit cell. The energy gains per atom ($1.3 \times 10^0$–$7.7 \times 10^{-1}$ kcal/mol/atom) caused by structure wave formation of *phC*(0,1) is comparable or greater than the energy of van der Waals interactions ($9.6 \times 10^{-1}$–$9.6 \times 10^{-2}$ kcal/mol/atom, [112]).

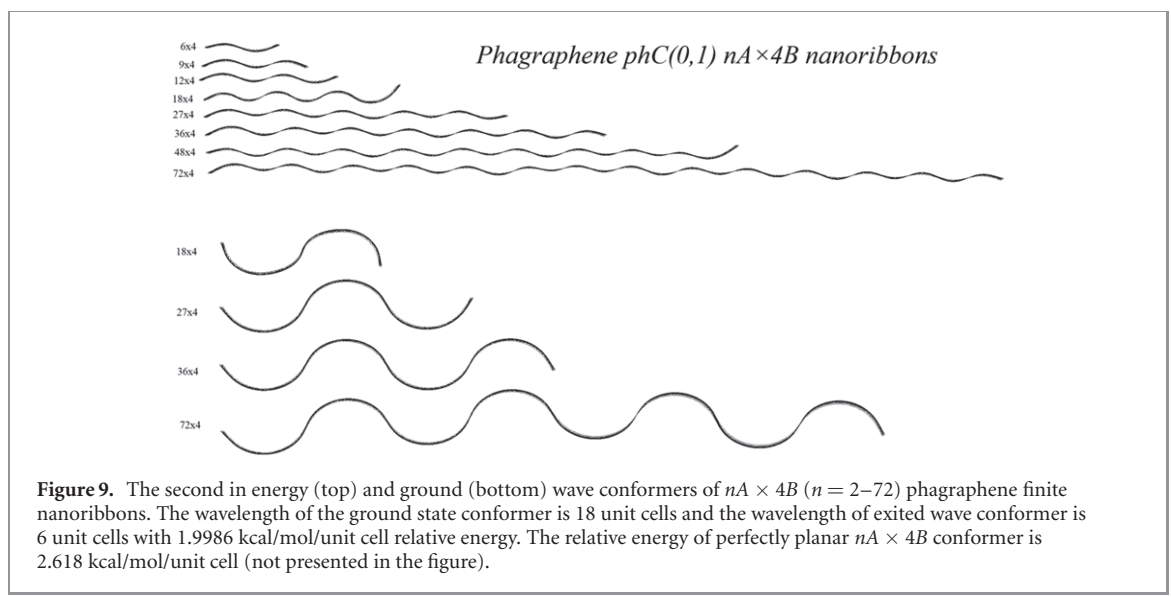

**Figure 9.** The second in energy (top) and ground (bottom) wave conformers of $nA \times 4B$ ($n = 2$–$72$) phagraphene finite nanoribbons. The wavelength of the ground state conformer is 18 unit cells and the wavelength of exited wave conformer is 6 unit cells with 1.9986 kcal/mol/unit cell relative energy. The relative energy of perfectly planar $nA \times 4B$ conformer is 2.618 kcal/mol/unit cell (not presented in the figure).

Perfect correspondence of model potential and semiempirical structural parameters and relative energies of all $nA \times 1B$ flakes with the results of cluster and PBC DFT calculations unequivocally indicates the structural origin of the waving effect. The smallest energy gain is revealed for the shortest $1A \times 1B$ cluster which is equal from 2 (model potential) to 13 (PM3) kcal/mol/unit cell while B3LYP/6-31G* exhibits 6 kcal/mol (figure 8(c)). One can see the sharp drops in relative energy for half- ($n = 3$) and one- ($n = 5$) wavelength structures. Consistent elongation of the length of finite nanoribbons for $n = 10$ (maximum length for B3LYP/6-31G* approach), $n = 20$ (semiempirical PM3) and $n = 25$ unit cells (model potential) leads to concerted oscillations of the energy gain per unit cell with local minima correspondent to integer numbers of half-waves at $n = 8, 10$ (B3LYP/6-31G*, PM3, model potential), 13, (PM3, model potential), 18, 20, 23 and 25 (model potential). For PM3, the elongation of the nanoribbon beyond $n = 14$ leads to almost constant (33.4–33.6 kcal/mol/unit cell) energy differences between wave and planar conformers.

In table SI1-3 the model potential and PM3 vibration frequencies of finite $nA \times 1B$ phagraphene nanoribbons are presented. It was found that the accuracy of model potential structure optimization mostly determines the results of Hessian calculations and should be $10^{-5}$ kcal/mol/Å or even better. The MM+ wave clusters except for $n = 13, 20$ and 25 have no imaginary frequencies in vibration spectra. Small imaginary frequencies (12i–5i cm$^{-1}$, table SI1-3) of MM+ $n = 13, 20$ and 25 finite nanoribbons are likely caused by either poor convergency during structure optimization or probably even some Hessian algorithm flaws. In contrast with the wave conformers, all $n = 1$–$25$ finite planar nanoribbons have several distinctive imaginary frequencies (76i–2i cm$^{-1}$, table SI1-3) in vibration spectra.

At PM3 level of theory, the $n = 1$–$17$ finite $nA \times 1B$ wave nanoribbons conformers have no imaginary frequencies in vibration spectra. Planar conformers with $n = 1$–$18$ have one or more imaginary frequencies (135i–71i, table SI1-3), which directly indicates that they are transition states between two equivalent ground state waves. It is necessary to note that the GAMESS code erroneously returns no imaginary modes for $n = 12$–$15$ and skips many of them for $n = 10$–$18$.

The $phC(0,1)$ $25A \times 2B$ and $15A \times 2B$ nanoribbons calculated by both model potential and PM3 levels of theory keep the same, $n = 5$, wavelengths of structure waves. At the model potential level of theory, the $nA \times 4B$ ($n = 2$–$72$) phagraphene finite nanoribbons (figure 9, table 1) also reveal wave behavior with two localized waves of 18 (ground state, 0.0 kcal/mol/unit cell relative energy) and 6 (1.999 kcal/mol/unit cell) unit cell length, which can be considered as the second in energy excited state. Nanoribbons of 72 unit cell length correspond to 4 wavelengths of the ground wave cluster and 14 wavelengths of the short-wave conformer. The planar conformer with the highest relative energy of 2.618 kcal/mol/unit cell corresponds to the transition state between equivalent ground state waves of either 18 or 6 unit cell wavelengths.

At the model potential level of theory rectangular $nA \times nB$ ($n = 2$–$18$) phagraphene flakes (figure 10) also reveal the same wave behavior with the wavelength of 18 unit cells. For $n = 8$ almost planar structure is located, which is reflected in the half-wave structural antiresonance.

The dependence of the wavelength of phagraphene structure waves upon the width and length of finite nanoribbons, which is in fact the wave boundary conditions, localization of different conformers with different wavelengths and convergence of the ground state waves for different widths to the same wavelength

**Table 1.** Model potential energies (in *kcal/mol/unit cell*) of planar, with one complete standing wave and 3 standing waves $C_{1596}H_{136}$ phagraphene 18 × 4 clusters.

| Conformers of $C_{1596}H_{136}$ phagraphene $72A \times 4B$ nanoribbons | Relative energy, kcal/mol/unit cell |
|---|---|
| Planar | 2.618 |
| 14 waves | 1.999 |
| 4 waves | 0.000 |

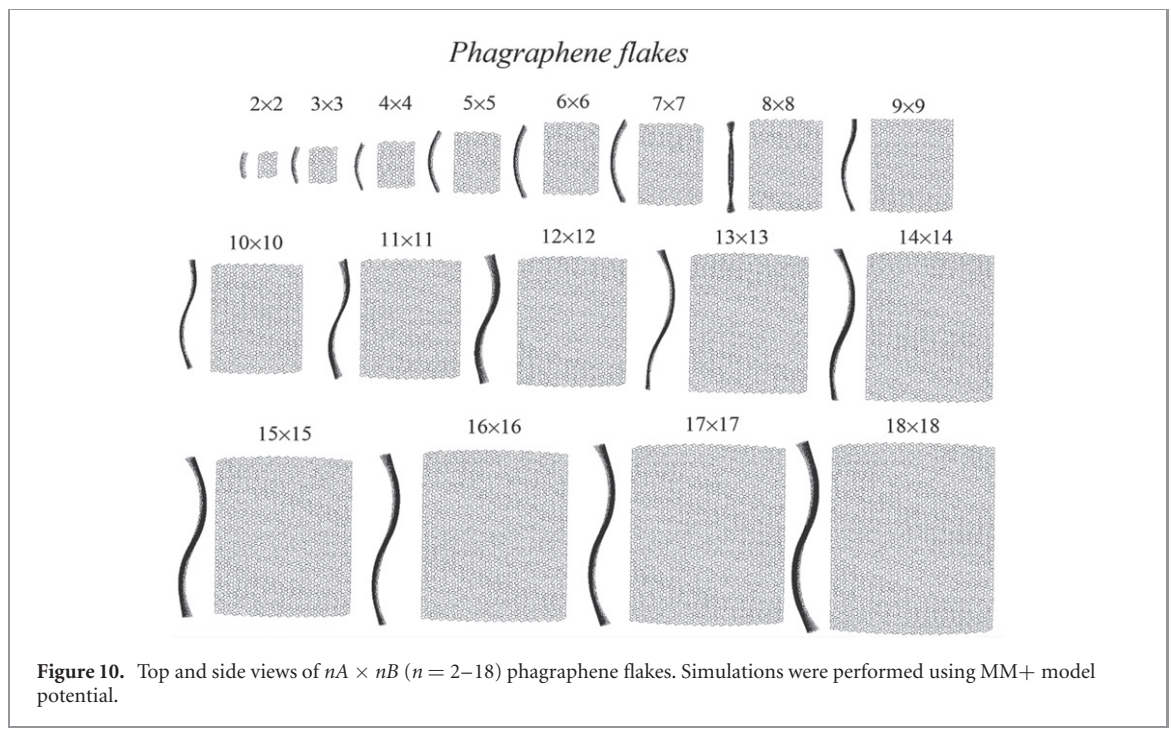

**Figure 10.** Top and side views of $nA \times nB$ ($n = 2$–18) phagraphene flakes. Simulations were performed using MM+ model potential.

(18 unit cells), indicate that this phenomenon can be interpreted in terms of homogeneous D'Alembert wave equation:

$$\Delta u = \frac{1}{v^2} \frac{\partial^2 u}{\partial t^2},$$

where $\Delta$ is Laplace operator, $v$ is phase velocity, and solution $u = \{\varphi_n\}$ is a set of standing waves each of which satisfies preset boundary conditions. In terms of structure waves, the finite phagraphene flakes could exist since the solutions of the homogeneous D'Alembert wave equation can be found. Recent experimental synthesis of visibly curved dibenzo[*a*, *m*]dicyclohepta[*bcde*, *nopq*]rubicene atop of Au(111) [113] confirms this conclusion.

In contrast to homogeneous D'Alembert wave equation for the standing waves with preset boundary conditions, for infinite unconstrained waves the inhomogeneous equation

$$\frac{\partial^2 u}{\partial t^2} = v^2 \Delta u + f(x, t)$$

should be considered, where $f(x, t)$ is a function of non-zero external acting force of one or two independent spatial coordinates. The solution of the inhomogeneous D'Alembert wave equation is an infinite superposition of plane waves each of which is a solution of D'Alembert equation. It is necessary to note that in the particular case of *phC*(0,1) lattice, the structural waves could be regarded as the analogs of acoustic phonons since the lattice atoms demonstrate coherent displacements from the plane without polarization effects (figures 8(a)–10). Infinite unconstrained averaged in time low-dimensional wave-shaped lattices correspond to completely undetermined lattice parameters and atomic coordinates because the plane-wave functions with different wave numbers and phases span the whole momentum and spatial spaces.

The quantum interference of large organic molecules like $C_{60}$, (60 atoms, weght 720 Dalton, size 7 Å), $C_{60}[C_{12}F_{25}]_8$ (356 atoms), PFNS10 $C_{60}[C_{12}F_{25}]_{10}$ (430 atoms, weight 6910 Dalton, size 40 Å), $C_{44}H_{30}N_4$ (78 atoms) $C_{84}H_{26}F_{84}N_4S_4$ (202 atoms), TPPF152 $C_{168}H_{94}F_{152}O_8N_4S_4$ (430 atoms, weight 5310 Dalton, size >60 Å) [114] is a well known effect of quantum nature with effective masses and dimensions of the objects up to

430 atoms, 6910 Dalton
and >60 Å. Each single wave of *phC*(0,1) $nA \times 1B$ nanoribbon (5 unit cells or 34 Å, 137 atoms, 1644 Dalton, see above) is significantly smaller the macromolecules of record mass (PFNS10, 6910 Dalton) and dimensions (TPPF152, >60 Å) which demonstrate quantum interference [114].

Each single structural wave of *phC*(0,1) $nA \times 1B$ is a solution of the wave equation with corresponding de Broglie wave length $\lambda$, eigenstate, and eigenvalue. The wave can be considered as a quantum quasi-particle with defined non-zero absolute value of the impulse $|p| = h/\lambda$ (where $h$ is the Planck constant) with two symmetrically equivalent opposite wave propagation directions which makes the average impulse $p = 0$. Disregarding the effect of spontaneous departure of free-standing graphene from perfectly planar lattice with formation of random intrinsic ripples in graphene perpendicular to the crystalline plane (see, for example, [115–117]), for pristine graphene lattice, which demonstrates the structural wave length $\lambda = \infty$, the absolute value of structural impulse $|p| = h/\infty = 0$, which makes 1D and 2D graphenes dynamically stable lattices. The quasiparticle estimation of structural wave energy of *phC*(0,1) lattice (see SI2 section) returnes negligibly small energy per carbon atom ($6.97 \times 10^{-11}$ kcal $\cdot$ mol$^{-1}$ $\cdot$ C atom$^{-1}$), which makes the lattice dynamically stable.

For the infinite unconstrained averaged in time low-dimensional wave-shaped lattices the infinite superposition of the structural waves of different wavenumbers and phases directly leads to completely undetermined lattice parameters and atomic coordinates. Any kind of structural constraints and/or physical perturbations may lead to wave function of structure wave collapse with instantaneous localization of structural parameters and atomic positions of the atomistic lattice.

Disregarding thermal fluctuations, the energy of a perfect structural wave infinite periodic lattice is completely and exactly determined by its structural parameters like wave length and amplitude. Following the quantum uncertainty principle, the coordinate uncertainty $\Delta x = \hbar/\Delta E = \infty$, where $\Delta E = 0$ is Heisenberg energy uncertainty, which makes the atomic coordinates of the wave completely undetermined before any kind of measurement. During structural spectroscopic experiment the wave function of undisturbed free-standing infinite structural wave quasiparticle should collapse with localization of atomic positions at random coordinates with breakdown of lattice periodicity and formation of an aperiodic atomic lattice. Such type of structural instability can be regarded as *a lattice quantum instability* which can be defined as a lattice instability of wave-shaped low-dimensional infinite free-standing crystalline lattices *caused by complete quantum uncertainty of atomic positions before any kind of structural experiment or another external impact which should lead to the collapse of the wave function of unconstrained and undisturbed structural wave quasi-particle at random atomic positions and lattice parameters.*

Using quasiparticle approach (see SI2 section) and ideal gas laws, the internal pressures/temperatures of two localized infinite *phC*(0,1) structural waves of 5 unit cell wavelength (figure SI1-4) and 18 unit cell wavelength (figure SI1-5) were estimated to be 2466.7 atm/2870 K and 569.2 atm/2668 K, respectively. Based on the above estimations, one may conclude that the unconstrained and undisturbed infinite *phC*(0,1) lattice before the collapse of wavefunction of structural wave quasiparticle during structural experiment is characterized by completely undetermined structural parameters and random distribution of atomic positions with averaged carbon–carbon distances in the range of 2.5–4.0 Å, 2312.5–569.2 atm internal pressure and 2818–2668 K effective temperature. Such conditions seem to be enough [61, 111] for fundamental structural transformations with breakdown of perfect wave structure and periodicity lift-off with the formation of random aperiodic carbon atomic lattices. It makes impossible the existence of unconstrained and undisturbed infinite *phC*(0,1) lattice due to lattice quantum instability. This conclusion can be extended for the case of screw-shaped 1D *phC*(0,0) lattice since regular screws demonstrate periodic behavior as well and can be considered as structure waves which prone to lattice quantum instability effects.

Planar 2D graphene lattice can be considered as a wave with infinite wavelength and 0 amplitude for which the volume of structure wave box is equal to 0. It makes the lattice parameters and atomic positions determined, the effective temperature and pressure of graphene lattice equal to 0 K and 0 atm, respectively, and averaged carbon–carbon distances equal to its solid-state limit of 1.5 Å.

The PW PBE PBC calculations of planar and wave conformers of narrow 1D phagraphene nanoribbon of one unit cell width were carried out for the supercell consisted of 5 unit cells (figure 8(a)), which corresponds to one wavelength of the lowest in energy wave conformer. The crystallographic group of phagraphene is *Pmg* with the rectangular unit cell having *a*, *b* unit vectors equal to 8.095, 6.652 Å, respectively. It was found that the planar conformer is significantly higher in energy with relative energy equal to 24.762 kcal/mol/unit cell. For the sake of comparison, the calculation of the narrow 1D phagraphene nanoribbon of one unit cell width using just one unit cell as a slab was performed, which generated a planar 1D crystalline lattice as it was expected.

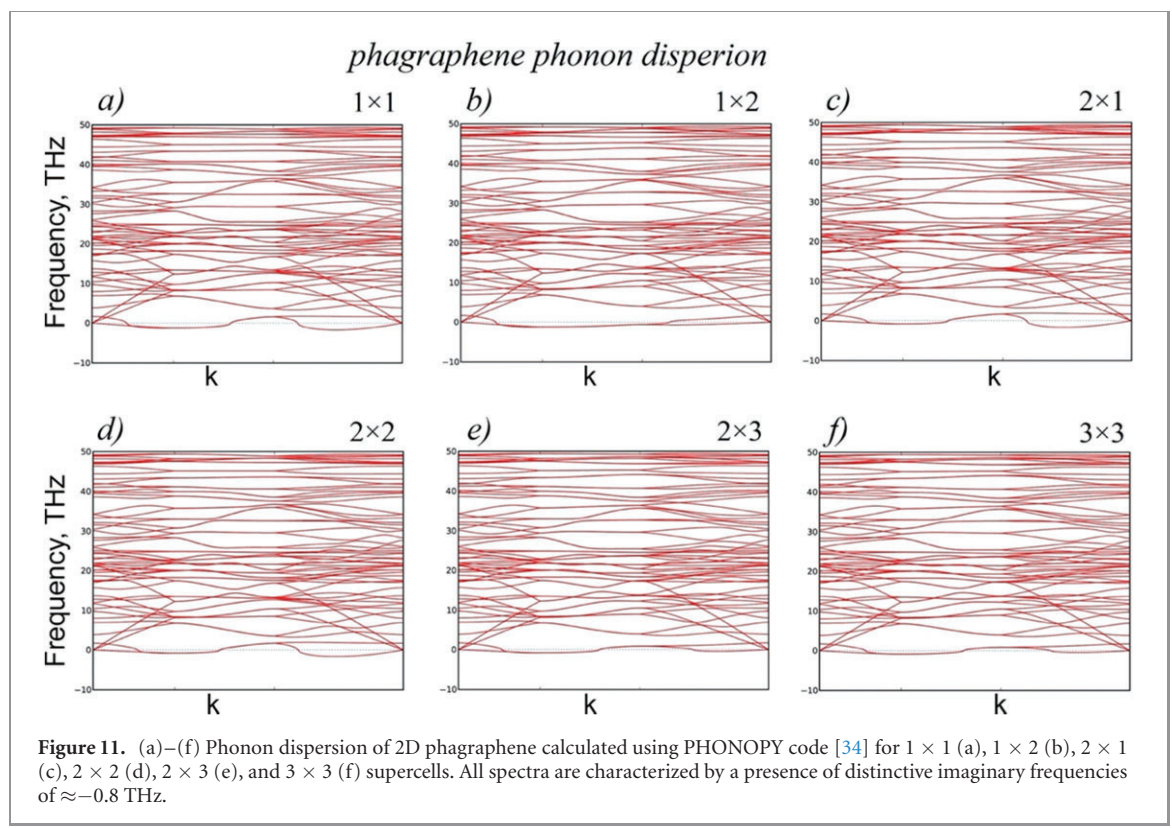

**Figure 11.** (a)–(f) Phonon dispersion of 2D phagraphene calculated using PHONOPY code [34] for $1 \times 1$ (a), $1 \times 2$ (b), $2 \times 1$ (c), $2 \times 2$ (d), $2 \times 3$ (e), and $3 \times 3$ (f) supercells. All spectra are characterized by a presence of distinctive imaginary frequencies of $\approx -0.8$ THz.

The crystalline lattice of planar 2D phagraphene was calculated using the PW PBE PBC approach following the parameters of the original publication [36]. Previously [36, 42] the phonon dispersion of phagraphene was calculated using both PHONON and GULP [46, 47] codes for $2 \times 2$ and $3 \times 3$ [36, 45] supercells and no imaginary frequencies were detected (figures SI1-6(a) and (b)). The phonon dispersions of planar 2D phagraphene were calculated for $1 \times 1$, $1 \times 2$, $2 \times 1$, $2 \times 2$, $2 \times 3$, $3 \times 3$, and $4 \times 1$ supercells (figures 11(a)–(f)). In contrast to previous publications [36, 42] the phonon dispersion for all considered supercells is characterized by distinctive imaginary frequencies. Visualization of the imaginary frequencies demonstrates that they correspond to wave-like vibrational excitations of phagraphene crystalline lattice in the out-of-plane direction. The phonon dispersion of narrow planar 1D phagraphene $4 \times 1$ nanoribbon is characterized by imaginary frequencies as well (not presented in figure 11). One can speculate that obvious failure to determine imaginary frequencies by both codes [34, 44, 46] could be caused by 2 reasons. First, since the structure wavelength of 2D phagraphene (18 unit cells) is much larger than $2 \times 2$ and $3 \times 3$ supercells used for phonon calculations, the atomic vibration potentials used to calculate phonons were not appropriate to determine the imaginary modes. The second reason for the failure may be caused by possible $\pi$-resonance stabilization of extended 2D aromatic lattices with possible transformation of the planar transition state of 2D phagraphene into metastable intermediate conformer. In both cases, routine PBC DFT calculations of 2D $phC(0,1)$ based on optimization of the structure using just one unit cell were failed to locate its true atomic structure determined by wave behavior of the lattice.

The results of *ab initio* NVT MD PW PBE PBC and cluster simulations of various 2D, 1D, and 0D $phC(0,1)$ phagraphene lattices at 300 K are presented in figure 12. 5 ps AIMD NVT PBE PW PBC simulations of 2D phagraphene (figure 12(a)) lead some departure of carbon atoms from the lattice plane with formation of small amplitude irregular wave-like structure. Due to NVT algorithm restrictions which froze the MD box dimensions to optimized PBC sepercell size, no one could expect the lattice can be transformed from the planar to a wave conformer since the formation of structural waves should lead to significant contraction of the lattice. For example, at *ab initio* DFT B3LYP/6-31G* level of theory, the length of typical $phC(0,1)$ $10A \times 1B$ flake contracts from 73.24 Å for planar conformer to 63.60 Å (almost 10 Å absolute difference) for the wave one, which corresponds to relative 13% contraction. During the whole course of 5 ps NVT PBC MD simulations, the total energy of 2D phagraphene lattice fluctuates ($\pm 1.25$ eV) around thermostated equilibrium level of $-2890$ eV without any traces of phase transition.

The AIMD PBC simulations at 300 K of planar 1D phagraphene $phC(0,1)$ $nA \times 1B$ nanoribbon (figure 12(b)) return very close result with some thermal expansion of the lattice with small departures of the atoms from nanoribbon plane and formation of wave-like irregular conformer. Like in the case of AIMD PBC simulations of 2D phagraphene, the MD simulation box imposes strict limitations on the topology, volume of the nanoribbon slab. The length of the ribbon is equal to 73 Å, which is significantly longer the length of the wave conformer (64 Å) and, in fact, is equivalent to strong stretching force. The total energy of 1D *phC*(0,1) $nA \times 1B$ nanoribbon does not reveal any sign of phase transition fluctuating around $\pm 1.0$ eV of 300 K thermostated equilibrium level of $-1269$ eV.

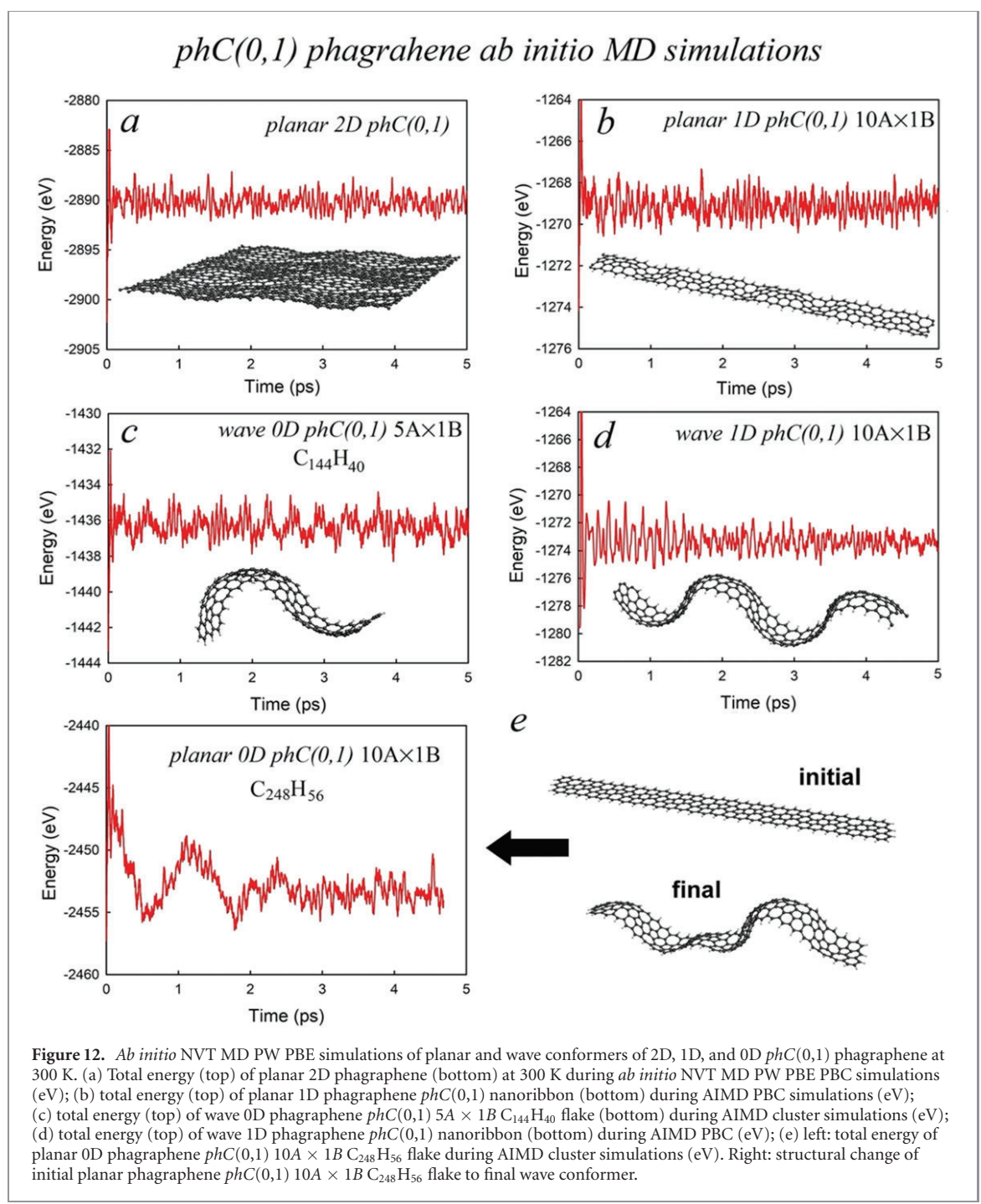

**Figure 12.** *Ab initio* NVT MD PW PBE simulations of planar and wave conformers of 2D, 1D, and 0D *phC*(0,1) phagraphene at 300 K. (a) Total energy (top) of planar 2D phagraphene (bottom) at 300 K during *ab initio* NVT MD PW PBE PBC simulations (eV); (b) total energy (top) of planar 1D phagraphene *phC*(0,1) nanoribbon (bottom) during AIMD PBC simulations (eV); (c) total energy (top) of wave 0D phagraphene *phC*(0,1) $5A \times 1B$ $C_{144}H_{40}$ flake (bottom) during AIMD cluster simulations (eV); (d) total energy (top) of wave 1D phagraphene *phC*(0,1) nanoribbon (bottom) during AIMD PBC (eV); (e) left: total energy of planar 0D phagraphene *phC*(0,1) $10A \times 1B$ $C_{248}H_{56}$ flake during AIMD cluster simulations (eV). Right: structural change of initial planar phagraphene *phC*(0,1) $10A \times 1B$ $C_{248}H_{56}$ flake to final wave conformer.

The AIMD simulations at 300 K of 0D wave conformer of phagraphene *phC*(0,1) $5A \times 1B$ $C_{144}H_{40}$ flake (figure 12(c)) conserve its wave shape. The total energy fluctuates around $\pm 1.0$ eV of thermostated equilibrium level of $-1436$ eV.

The AIMD PBC simulations at 300 K of wave conformer of *phC*(0,1) $nA \times 1B$ nanoribbon (figure 12(d)) conserve its wave shape. Its thermostated equilibrium total energy approximately equal to $-1273.5$ eV which is significantly lower the thermostated equilibrium total energy of corresponding planar *phC*(0,1) $nA \times 1B$ nanoribbon conformer (figure 12(b)) equal to $-1269$ eV. The fluctuations of the total energy of the wave conformer is much smaller and lie in the region of $\pm 0.5$ eV without any traces of phase transitions.

The AIMD simulations of 0D planar phagraphene *phC*(0,1) $10A \times 1B$ $C_{248}H_{56}$ flake (figure 12(e)) demonstrate completely different behavior of both total energy and the lattice structure of the flake upon simulation time. Average total energy of the flake at the beginning of simulation was equal to $\sim -2448$ eV. During the course of simulation the total energy plummeted up to $-2455$ eV (7 eV drop, $\sim 0.5$ ps), rised to $-2450$ eV (5 eV rise, $\sim 1.2$ ps), another drop to $\sim -2456$ eV (6 eV drop, $\sim 1.9$ ps), and finally equilibrated at $\sim -2453$ eV (another 3 eV rise, $\sim 3$ ps). Dramatic energy change during the course of simulations is coupled with planar topology breakdown and structural transformation of initial perfectly planar flake to irregular wave conformer, which can be interpreted in terms of typical phase transition with fundamental change of atomic lattice.

Overall, it was found that the *phC*(0,1) lattice forms structure waves with wavelengths varying from 5 to 18 unit cells depending on the boundary conditions with planar 1D and 2D conformers as transition states between two equivalent ground state waves. At least one second in energy structure wave of 6 unit cell wavelength was localized as well for $nA \times 4B$ nanoribbons. In contrast with previous publications [36, 42], all Hessian and phonon dispersion calculations of *phC*(0,1) revealed imaginary phonon modes for planar conformers. It was shown that NVT AIMD PBC algorithm artificially confines periodic lattices inside MD simulation box of constant dimensions which prevents their thermal relaxation and equilibration. In fact, NVT AIMD PBC approach effectively imposes artificial stretching or compressing forces keeping the dimensions of periodic slabs constant during the whole course of MD simulations. It was found that infinite 1D and 2D *phC*(0,1) lattices suffer quantum instability, whereas finite cluster or confined lattice of *phC*(0,1) may exist and could be experimentally synthesized.

### 5.3. *phC*(1,0) lattice

The next simplest lattice can be introduced for the heptagons and pentagons in one unit cell to be separated by one hexagon and without hexagons separated heptagons and pentagons of neighboring 5/6/7 fragments. The same lattice was proposed earlier [58] among others as conformer 'New5'. A planar $4 \times 4$ and $3 \times 2$ *phC*(1,0) flakes are presented in figures S7 and 13(a). For the sake of simplicity, one can compare, for example, the *phC*(1,0) and *phC*(0,1) lattices depicted in figures 1(c) and (d), respectively. It is clearly seen, that *phC*(1,0) is very different from *phC*(0,1), namely, it contains 5/5 and 7/7 pairs which are connected with each other forming infinite (7/7)/(5/5) lines, separated by armchair ribbons of one hexagon width.

In addition to 5/7/6 fused fragments, the structural mismatch for which is equal to 3.43°, the *phC*(1,0) lattice is characterized by 5/5/7 and 7/7/5 fragments (figures SI1-7 and 13(a)) as well. The unrelaxed geometry of 5/5 pair is characterized by 144° external angle at 5/5 seam, which is far beyond the 128.57° of the perfect heptagon. The 7/7 pair has 102.86° which does not match 108° of the perfect pentagon. So, following TCT, the *phC*(1,0) should suffer internal structural stress, which should lead the lattice to depart from the perfect plane with formation of a bent structure.

The atomic structure and mechanical stress of finite *phC*(1,0) nanoribbons, flakes and nanotubes were studied by model potential, semiempirical PM3, *ab initio* B3LYP/3-21G and PW PBE methods. For the sake of comparison, PBC PW PBE calculations of 2D *phC*(1,0) lattice were performed as well. The *phC*(1,0) bending energies per unit cell of $nA \times 1B$ *phC*(1,0) nanoribbons (figure 13(a)) calculated as $E_b = (E_c - E_p)/n$, where $E_b$ is the binding energy of finite $nA \times 1B$ *phC*(1,0) nanoribbon, $E_c$ and $E_p$ are the total energies of curved and planar conformers, respectively, and $n$ is the length of the nanoribbons in unit cells. At all levels of theory optimizations of one unit cell $1A \times 1B$ *phC*(1,0) fragment returned planar atomic lattice (top-left of figure 13(a)).

The model potential optimization of finite fragments of $nA \times 1B$ *phC*(1,0) nanoribbons ($n = 1 - 14$) returned curved atomic lattices for ($n = 3 - 14$) which may be finally rolled to a multiwalled nanotube (an example of $12A \times 1B$ *phC*(1,0) fragment is presented in figure 13(a)) or fused into a short single-wall *phC*(1,0) (14,0) zig-zag nanotube of 1 unit cell length. At the model potential level of theory, the bending energy varies from 0 kcal/mol/unit cell for $n = 1, 2$ to $-12.53$ kcal/mol/unit cell for $n = 13$ with distinctive $1/x$ behavior for ($n = 3 - 11$). A relatively sharp drop of $E_b$ for ($n = 12, 13$) is caused by the attractive interaction of both ends of the nanoribbon caused by its rolling. Following its universal curvature, the edges of *phC*(1,0) nanoribbon can be fused to form the (14,0) zig-zag nanotube of 21.368 Å diameter. The nanotube bending energy (blue dot in figure 13(a)) belongs to the same $1/x$ ($n = 3 - 11$) pattern due to the absence of vdW interactions between free edges and rather large nanotube diameter (and the number of constituting unit cells) to eliminate structural differences between complete and uncomplete nanotubes.

At PM3 ($n = 1 - 21$) and B3LYP/3-21G ($n = 1 - 7$) levels of theory, elongation of $nA \times 1B$ *phC*(1,0) leads to formation of helix-shaped nanoribbons (see one, two and three helix period length $13A \times 1B$, $25A \times 1B$, and $37A \times 1B$ *phC*(1,0) inserts in figure 13(a)) with typical $1/x$ behavior of $E_b$ stabilization bending energy. A small bump in PM3 $E_b$ at ($n = 12, 13$) is probably caused by repulsive interactions of the edges of helix-shaped $phC(1,0)$ nanoribbon. The PM3 bending energy gain of $nA \times 1B$ $phC(1,0)$ asymptotically approaches to $-26$ kcal/mol/unit cell with B3LYP/3-21G $E_b$ as twice as small ($-12.6$ kcal/mol/unit cell). Both PM3 and B3LYP/3-21G methods revealed the same geometrical characteristics of $nA \times 1B$ $phC(1,0)$ helices with the diameter equal to 13.273 Å, 32.7 Å pitch, and the length of one step of the helix equal to 12 unit cells.

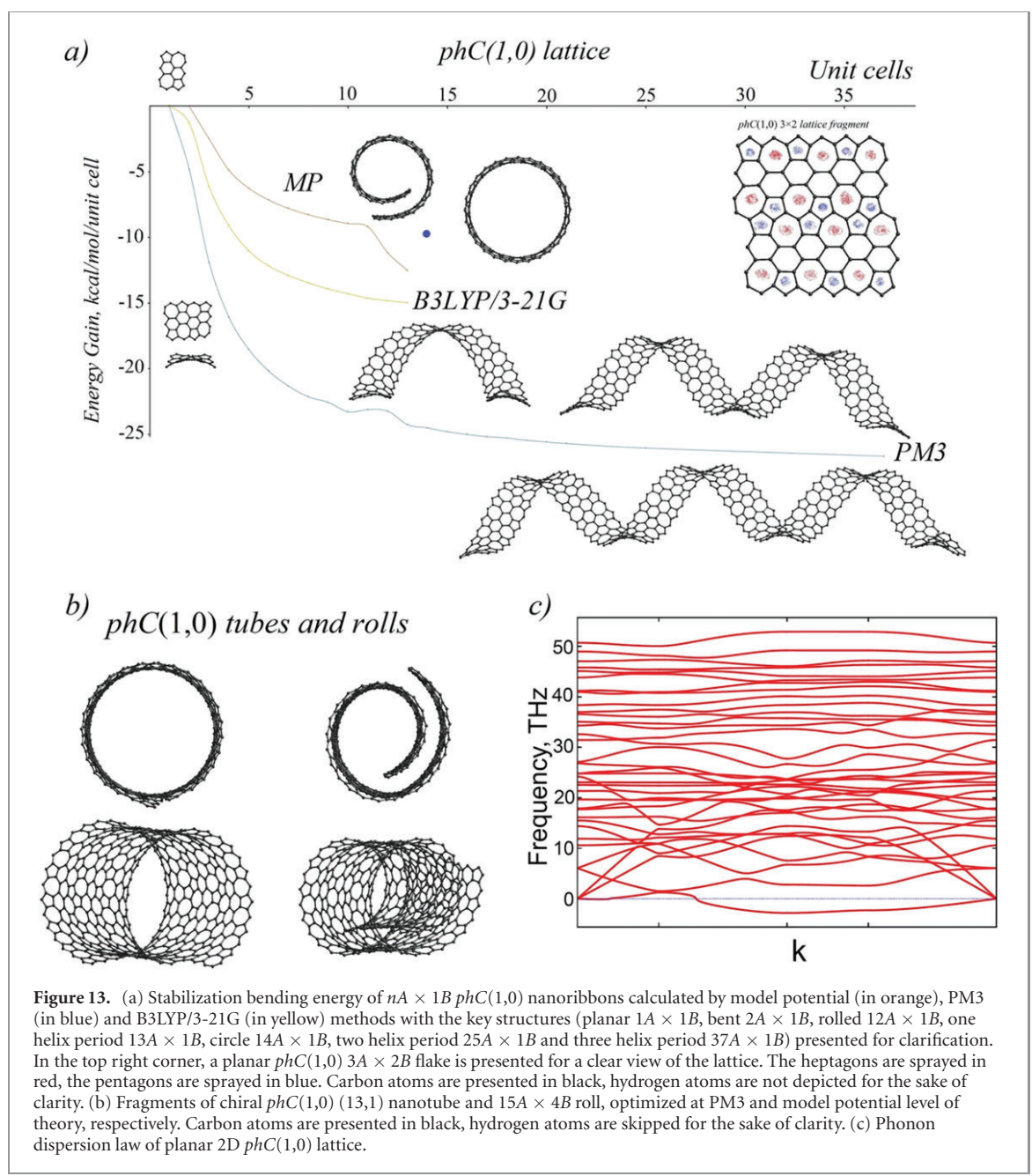

**Figure 13.** (a) Stabilization bending energy of $nA \times 1B$ $phC(1,0)$ nanoribbons calculated by model potential (in orange), PM3 (in blue) and B3LYP/3-21G (in yellow) methods with the key structures (planar $1A \times 1B$, bent $2A \times 1B$, rolled $12A \times 1B$, one helix period $13A \times 1B$, circle $14A \times 1B$, two helix period $25A \times 1B$ and three helix period $37A \times 1B$) presented for clarification. In the top right corner, a planar $phC(1,0)$ $3A \times 2B$ flake is presented for a clear view of the lattice. The heptagons are sprayed in red, the pentagons are sprayed in blue. Carbon atoms are presented in black, hydrogen atoms are not depicted for the sake of clarity. (b) Fragments of chiral $phC(1,0)$ $(13,1)$ nanotube and $15A \times 4B$ roll, optimized at PM3 and model potential level of theory, respectively. Carbon atoms are presented in black, hydrogen atoms are skipped for the sake of clarity. (c) Phonon dispersion law of planar 2D $phC(1,0)$ lattice.

Let us compare the stabilization bending energy for $nA \times 1B$ $phC(1,0)$ nanoribbons with the energy of van der Waals interactions ($9.6 \times 10^{-1}$–$9.6 \times 10^{-2}$ kcal/mol/atom, [112]). Since the unit cell of $phC(1,0)$ contains 22 carbon atoms, vdW energy of interactions of $phC(1,0)$ lattice with a substrate could be from 22 to 2.2 kcal/mol/unit cell. This value is comparable to the absolute value of B3LYP bending ($-12.6$ kcal/mol/unit cell), making the synthesis and stabilization of $phC(1,0)$ lattice on top of a vdW substrate impossible.

Vibration spectra of both planar and curved conformers of $nA \times 1B$ $phC(1,0)$ nanoribbons were calculated at model potential ($n = 1-14$), PM3 (($n = 1-21$) for planar and ($n = 1-14$) for curved conformers) and B3LYP/3-21G ($n = 1-2$) levels of theory. It was found that all methods returned no imaginary vibrational frequencies for all considered curved conformers (except PM3 $n = 12, 13, 14$), which unequivocally indicates that they are structural minima. PM3 vibration spectra of $n = 12, 13, 14$ curved conformers have 2 (1.75i and 1.44i cm$^{-1}$), 1 (1.54i) and 2 (1.64i and 1.24i cm$^{-1}$) small imaginary frequencies which correspond to bending modes caused by interactions of the edges of the nanoribbons and insufficient accuracy of structural optimization of extended narrow nanoribbons.

In contrast to curved conformers, all planar finite-length nanoribbons, except $n = 1, 2$ ones for model potential and $n = 1$ for PM3 and *ab initio* B3LYP/3-21G, demonstrate multiple (for large $n$ even more than 50) imaginary modes, which makes planar conformers the transition states between two symmetrically equivalent minima.

It is necessary to note two very important features of vibration spectra calculations for one-atom-thick extended nanoclusters: (i) to make solid conclusions about stability of one-atom-thick nanoclusters, the quality of optimization must be very high with the gradient $<1 \times 10^{-5}$ kcal/mol/Å; (ii) for one atom thick extended clusters with the number of atoms more than 54, existing algorithms of calculations of vibration spectra are not stable. To avoid artificial absence of imaginary modes in vibration spectra of the clusters, one has to use at least 2 independent codes for calculations of molecular vibrations (in this work GAMESS and HyperChem codes were used to cross control the results), combined with TCT structural analysis.

Model potential and PM3 relaxations of $4A \times 4B$ and $6A \times 6B$ *phC*(1,0) nanoflakes returned high-energy planar and curved ground state conformers (figure SI1-8) with bending stabilization energies equal to $-0.008/-0.238$ and $-0.025/-0.242$ kcal/mol/C atom, respectively. Based on calculations of $nA \times 1B$ *phC*(1,0) nanoribbons (see above) one can speculate that the planar conformers are the transition states between equivalent curved structures.

Elongation and fusion of the edges of PM3-calculated $nA \times 1B, (n \geqslant 14)$ *phC*(1,0) nanoribbons leads to self-assembly of slightly stressed lattice into (13,1) chiral nanotube (NT) (figure 13(b)). The relative energies per carbon atom of *phC*(1,0) (13,1) chiral nanotubes constituted by $n = 14$–42 *phC*(1,0) unit cells in respect to the energy of the shortest *phC*(1,0) (13,1) chiral nanotube constituted by just one $14A \times 1B$ *phC*(1,0) circle (figure 13(a)) is presented in figure SI1-9. The extension of the nanotube leads to fast drop of the energies from 0 kcal/mol/C atom for the shortest NT to almost $-16$ kcal/mol/C atom for NT constituted by 42 unit cells. Let us compare the relative energies of *phC*(1,0) (13,1) NTs constituted by 23 and 24 unit cells (336 and 348 carbon atoms with relative energies $-8.62$ and $-9.25$ kcal/mol/C atom, respectively) in respect to NT(13,1) with 14 unit cell with curved $21A \times 1B$ *phC*(1,0) nanoribbon (342 carbon atoms, $-1.58$ kcal/mol/C atom). It is clearly seen, that formation of (13,1) chiral nanotube is 5.7 times energetically preferable than even formation of *phC*(1,0) helix.

The PM3 vibration spectra calculations of *phC*(1,0) (13,1) chiral nanotubes constituted by 16–23 unit cells (figure 13(b)) revealed no imaginary frequencies, which unequivocally demonstrate their structural stability. The shortest nanotubes of 14 and 15 unit cells length revealed one small imaginary mode each of 4.68i and 2.17i cm$^{-1}$, respectively, which is caused by soft vibrational modes of the nanotube fragments in the vicinity of fusion regions, for which even enhanced accuracy of optimization maybe insufficient to calculate vibration spectra.

One can speculate that *phC*(1,0) can spontaneously form roll nanotubes rather than planar 2D lattice or (13,1) seamless chiral nanotube. For the sake of comparison the *phC*(1,0) (13,1) seamless chiral nanotube is presented in figure 13(b) as well. The *phC*(1,0) $15A \times 4B$ roll was optimized at the model potential level of theory and it is clearly seen that weak vdW interactions form roll structure with curvature even greater than the nanotube.

PW PBE PBC calculations of finite $4A \times 4B$ planar and curved *phC*(1,0) nanoflakes returned $-0.288$ kcal/mol/C atom energy gain for curved conformer, which is very close to PM3 result of $-0.238$ kcal/mol/C atom for $4A \times 4B$ flakes (see above). PW PBE PBC calculations of phonon frequencies of 2D *phC*(1,0) crystalline lattice returned 1 imaginary mode (figure 13(c)) which corresponds to a bending mode perpendicular to the lattice plane that is in perfect agreement with vibration spectra calculations of all planar finite fragments. Therefore, one can conclude that planar 2D *phC*(1,0) is structurally unstable.

Overall, the TCT analysis and computer simulations of 1D and 2D *phC*(1,0) lattices revealed that: (i) structural mismatch of *phC*(1,0) units leads to mechanical stress with consequent formation of 1D helices, (13,1) chiral, or rolled nanotubes; (ii) at all levels of theory curved structures are symmetrically equivalent global minima on potential energy surface separated by planar conformers as transition states; (iii) stabilization bending energy of *phC*(1,0) lattice is comparable or greater the vdW interactions with any kind of support; (iv) calculations of vibrational spectra either for finite clusters or 2D lattices require incredibly high (up to $1 \times 10^{-5}$ kcal/mol/Å gradient) convergency criteria for global minimum optimization. Curved global minima have no imaginary vibrational frequencies at all levels of theory. (v) Phonon spectra calculation of 2D *phC*(1,0) lattice returns one imaginary mode; (vi) calculations of vibrational spectra of finite planar clusters are prone to obvious mistakes which contradicts either each other or TCT analysis.

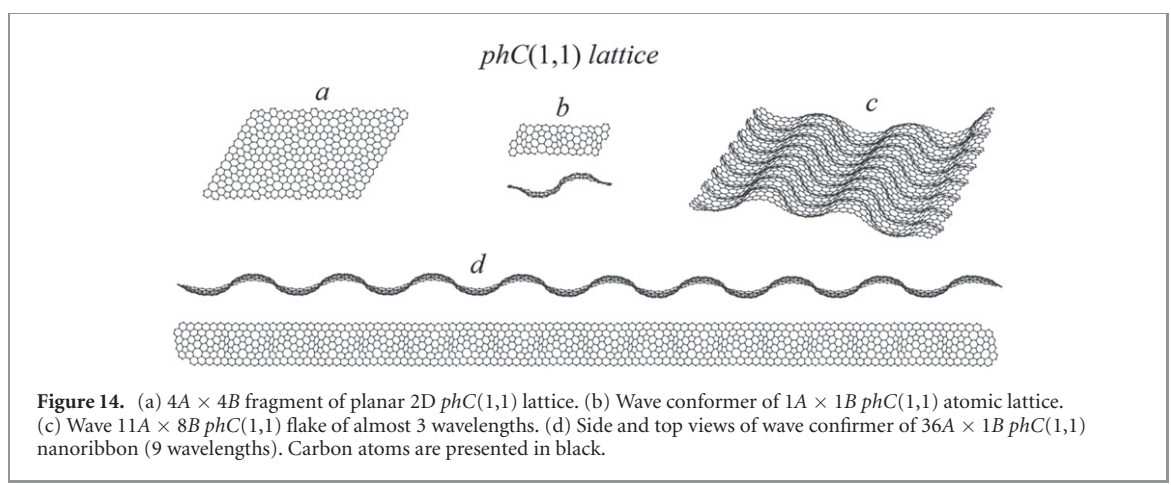

**Figure 14.** (a) $4A \times 4B$ fragment of planar 2D $phC(1,1)$ lattice. (b) Wave conformer of $1A \times 1B$ $phC(1,1)$ atomic lattice. (c) Wave $11A \times 8B$ $phC(1,1)$ flake of almost 3 wavelengths. (d) Side and top views of wave confirmer of $36A \times 1B$ $phC(1,1)$ nanoribbon (9 wavelengths). Carbon atoms are presented in black.

### 5.4. $phC(1,1)$ lattice

In figures 14(a) and SI1-10 a $4A \times 4B$ fragment of a planar $phC(1,1)$ flake is presented. Each single unit cell of $phC(1,1)$ consists of two antiparallel 5/6/7 units separated by 4 fused $C_6$ fragments arranged in stripes and separated from each other by infinite lines of fused carbon hexagons.

The lattice is formed by fused 5/6 and 7/6 fragments both of which confine hexagons at the seam angles. Like in the cases of $phC(0,0)$, $phC(0,1)$, and $phC(0,1)$, the $phC(1,1)$ lattice does not satisfy the TCT requirements. In particular, the angles formed by fusion of pentagons and heptagons with hexagons are equal to 132° and 111.43°, respectively, which is far beyond 120° of perfect hexagons. The $phC(1,1)$ lattice cannot be perfectly filled in by undeformed pentagonal, hexagonal and heptagonal fragments and should be relaxed with the formation of a bent structure even its structural parameters (for example, the model potential $a_1 = 12.935$ Å, $a_2 = 8.013$ Å, $\alpha = 119.317°$ and $\beta = 60.683°$) are rather close to a perfect hexagonal lattice.

Atomic structure, energetic characteristics and vibrational frequencies of finite planar and wave $nA \times mB (n = 0.5 - x, m = 1 - y)$ $phC(1,1)$ finite flakes and nanoribbons were calculated using model potential, PM3 and B3LYP/3-21G methods. It was found that at all levels of theory planar 2D conformers (figures 14(a) and S10) are transition states between two equivalent wave structures (figures 14(b)–(d)) with structure waves propagating along fused hexagonal lines. In particular, the model potential $11A \times 8B$ $phC(1,1)$ flake ($C_{3270}H_{150}$, figure 14(c)) is characterized by bending stabilization energy equal to $-32.574$ kcal/mol/unit cell or $-0.877$ kcal/mol/C atom. The wavelength of structure $nA \times 1B$ $phC(1,1)$ wave (figure 14(b)) is equal to 4 unit cells that is even shorter than the wavelength of $phC(0,1)$ $nA \times 1B$ lattices (5 unit cells) and does not depend upon the width of the ribbon, which can be attributed to confinement effect of lines of hexagons.

Both PM3-calculated planar and bent $1A \times 1B$ conformers converge to open-shell singlets with either closed-shell singlet or triplet states for planar and bent $2A \times 1B$. For B3LYP/6-31G*, $1A \times 1B$ cluster converges to planar closed-shell singlet with either closed-shell singlet or triplet states for planar and bent $2A \times 1B$ one. Both planar and bent conformers of PM3 and B3LYP/6-31G* $3A \times 1B$ and $4A \times 1B$ clusters revealed closed-shell singlet ground states.

All PM3-calculated planar conformers of $nA \times 1B$ ($n = 1–4$) revealed 2, 4, 7 and 11 imaginary vibrational frequencies, which directly indicate their transition state nature. No imaginary modes were detected for any wave conformers. B3LYP/6-31G* minimization of the smallest $1A \times 1B$ cluster returned only planar conformer without imaginary modes in vibration spectrum (for larger clusters $nA \times 1B$ ($n = 2, 3$) vibration spectra were not calculated). At PM3 and B3LYP/6-31G* levels of theory stabilization energy of formation of $3A \times 1B$ wave conformer in respect to the planar transition state is equal to $-1.447$ and $-1.170$ kcal/mol/C atom, respectively, which is close enough to $-0.877$ kcal/mol/C atom model potential stabilization energy for $11A \times 8B$ $phC(1,1)$ flake. The $4A \times 1B$ cluster (figure 14(b)) was studied only at the PM3 level of theory, for which stabilization bending energy is equal to $-1.716$ kcal/mol/C atom.

Overall, it was found that the planar 2D $phC(1,1)$ lattice does not satisfy TCT requirements which makes it topologically unstable suffering structural deformations perpendicular to the main lattice plane. Using model potential, PM3 semiempirical, and *ab initio* B3LYP/3-21G* simulations it was found that $phC(1,1)$ lattice forms structure waves which propagate along narrow zig-zag ribbons formed by fused carbon hexagons. The ribbons confine the wave behavior of the lattice keeping the wavelength of $phC(1,1)$ structure waves constant and equal to 4 unit cells. Stabilization bending energy of *phC*(1,1) lattice is estimated to be 1–2 kcal/mol/C atom, which is greater than the vdW energy, so planar 2D *phC*(1,1) lattice cannot be stabilized by any kind of supports. Because of structure wave behavior, an infinite free-standing 2D *phC*(1,1) sheet should be prone to quantum instability effects.

## 6. Penta-graphene topological instability

The topological and structural instability of penta-graphene [35] was recently studied [37–39] and it was shown that at the one-electron level of theory its planar conformer is just a regular point on the potential energy curve [37] between two equivalent non-periodic saddle structures. The penta-graphene lattice instability is entirely caused by structural reasons and it should be thoroughly investigated to reveal the true nature of its atomic lattice.

The $sp^3$ central sublattice of penta-graphene is characterized by strong departure of bonding angles (99.262–100.800°) from a perfect undisturbed angle of 109.471° formed between $sp^3$ atoms. Top and bottom carbon dimer sublattices of penta-graphene also demonstrate strong departure of bonding (110.974–112.601°) and torsion (127.168°) angles from perfect $sp^2$ environment (120° and 180°, respectively), which means the penta-graphene 2D planar lattice does not satisfy mandatory TCT requirements and should suffer strong structural distortion due to internal mechanical stress.

Both planar and curved conformers of $C_{72}H_{28}$ (3 × 3) and $C_{120}H_{36}$ (4 × 4) (figure 15(a)) penta-graphene clusters were calculated at PM3 and *ab initio* B3LYP/6-31G* levels of theory. $C_{72}H_{28}$ and $C_{120}H_{36}$ planar conformers were minimized using symmetry restrictions keeping constant *z*-coordinates of central $sp^3$ carbon sublattice and relaxing *x* and *y* coordinates of all C atoms at particular *z*-coordinates of non-equivalent top and bottom $sp^2$ carbon sublattices. Global minima for planar clusters were minimized at $z = \pm 0.60555$ and $z = \pm 0.60731$ Å for carbon atoms which belong to top and bottom dimer sublattices for (3 × 3) $C_{72}H_{28}$ and (4 × 4) $C_{120}H_{36}$ clusters, respectively. It is necessary to note, that the thickness of both clusters (1.211 and 1.215 Å, respectively) just coincide with high accuracy with PBC results [35].

At B3LYP/6-31G* level of theory both curved 3 × 3 $C_{72}H_{28}$ and 4 × 4 $C_{120}H_{36}$ clusters demonstrate −3.814 and −3.366 kcal/mol/C atom stabilization energy, respectively, which is way beyond vdW and even most covalent bonding. At the PM3 level of theory stabilization energy of 4 × 4 $C_{120}H_{36}$ cluster was determined to be equal to −2.766 kcal/mol/C atom. Taking into account boundary effects (see SI2 section), the B3LYP/6-31G* stabilization energy was estimated to be equal to 1.879 kcal/mol/C atom, which is way beyond vdW binding energy of $9.6 \times 10^{-1}$–$9.6 \times 10^{-2}$ kcal/mol/atom, [112].

To evaluate the curvature of saddle conformers of penta-graphene narrow nanoribbons cut out in *a* and *b* directions (figures SI1-11(g) and (l), respectively) of two unit cell width were calculated using PM3 method. Both type of nanoribbons spontaneously form short fragments of nanotubes (which can be also considered as rings) of either 5.98/7.14 Å (the ring circumference of 8 unit cells), or 9.94/11.19 Å (the ring circumference of 17 unit cells) of inner/outer radii $r_i$. It is necessary to note that the main axis of a nanotube formed by *a*-oriented nanoribbon is parallel to *b*-dimension and the main axis of a nanotube formed by *b*-oriented nanoribbon is parallel to *a*-direction, forming *b*-, and *a*-nanotubes, respectively. Corresponding curvatures $\kappa_i = \frac{1}{r_i}$ and the total curvature of the saddle point $K = \kappa_{bi}\kappa_{ao} = -0.015$ Å$^{-2}$, where $b_i$ and $a_o$ indexes correspond to the inner radius of *b*-oriented ring and outer radius of *a*-oriented ring, respectively. To prove the stability of the rings, unlocked *b* and *a* rings (figures SI1-11(f) and (k), respectively) were calculated as well and it was shown that both rings keep their perfect round shapes and radius values.

The ratios of inner and outer radii of *b*- and *a*-rings are pretty close to each other and are equal to 0.602 and 0.638, respectively. Different radii of *b*- and *a*-oriented nanotubes indicate significantly different bending force constants associated with either *b*-, or *a*-directions of one-unit cell thick penta-graphene, making the lattice significantly anisotropic. Following the saddle shape of the penta-graphene flakes (figures 15(a), SI1-11) and taking into account significantly different curvature radii of *b*- and *a*-nanotubes, one can conclude that *b*- and *a*-associated bending force constants have opposite directions and different absolute values. It is necessary to note that since the inner and outer radii of *a*-oriented ring are significantly larger the *b* one, corresponding bending force constant should be visibly smaller.

Elongation of *b*-oriented ring (figure SI1-11(g)) up to 5 unit cells generates perfect nanotube with the same inner/outer radii $r_i$ of 5.98/7.14 Å. Unlocking the *b*-nanotube (figure SI1-11(h)) resulted in some increase of the effective radius of the tube due to competition and partial mutual compensation of the *b*- and *a*-bending force constants.

One can speculate that during a hypothetical growth of one unit cell-thick penta graphene lattice, fusion of opposite lattice edges is hardly possible due to kinetic reasons and misorientation of opposite edges. Instead of the nanotubes and rings, formation of corresponding *b*- and *a*-oriented 1D rolls (figures SI1-11(j) and (m)) is very plausible. Since the bending energy of one-unit cell thin penta-graphene lattice is approximately as twice as large as vdW binding energy, one could speculate that after achievement of certain radius at which the resulting curvature of external penta-graphene roll would be rather close to planar structure, strong bending stress should form secondary roll structure with structural passivation of the growing edge, preventing formation of large-diameter penta-graphene rolls.

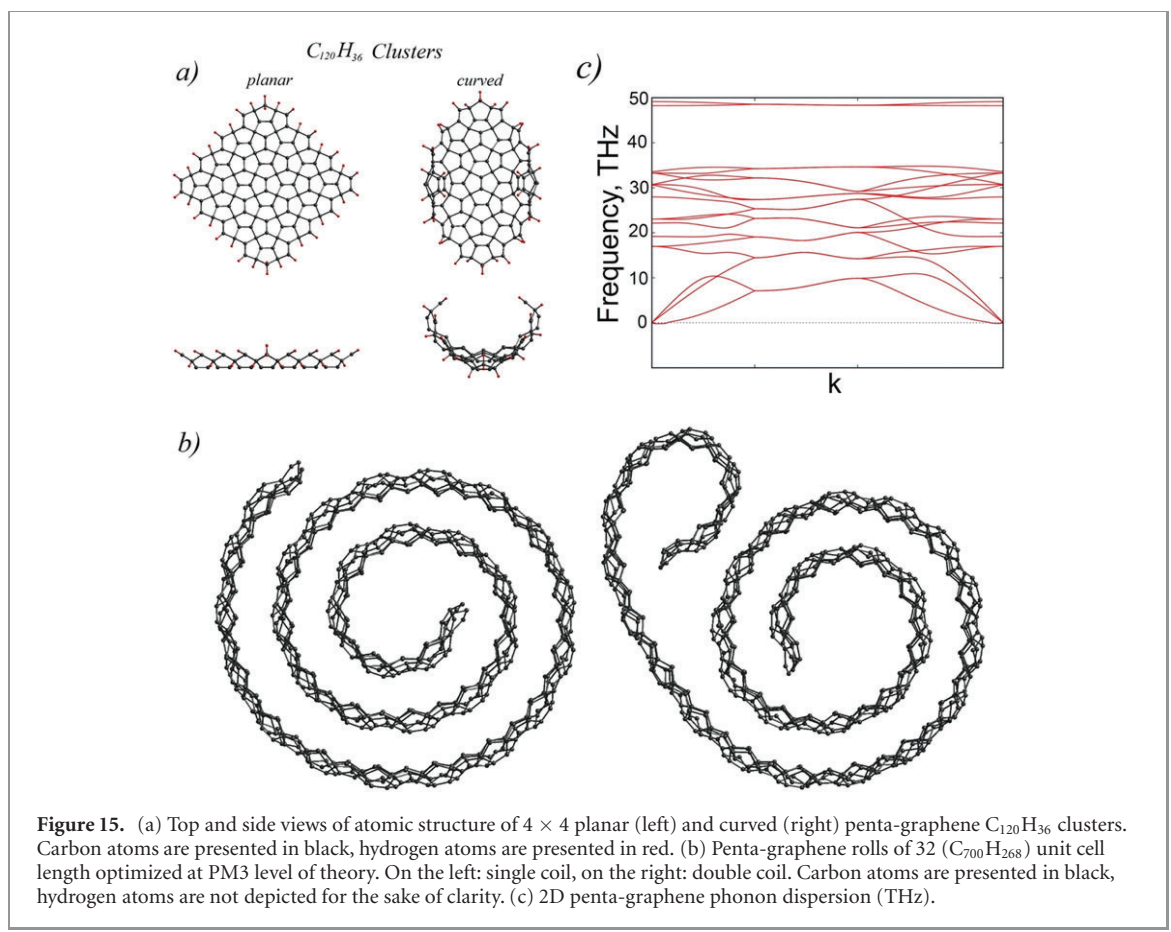

**Figure 15.** (a) Top and side views of atomic structure of 4 × 4 planar (left) and curved (right) penta-graphene $C_{120}H_{36}$ clusters. Carbon atoms are presented in black, hydrogen atoms are presented in red. (b) Penta-graphene rolls of 32 ($C_{700}H_{268}$) unit cell length optimized at PM3 level of theory. On the left: single coil, on the right: double coil. Carbon atoms are presented in black, hydrogen atoms are not depicted for the sake of clarity. (c) 2D penta-graphene phonon dispersion (THz).

Considering $sp^2$ one-atom thick lattices, the negatively curved surfaces may form several regular 0D, 1D or 3D solids of regular structure. 0D species contain quasi-spherical round giant and star-shaped closed-shell fullerenes and tori (see, for example, [118, 119]), which have both negatively and positively curved regions. 1D regular species [118], namely corkscrew-shaped nanotubes also demonstrate both negatively and positively curved fragments. From mathematical point of view, the penta-graphene negatively curved lattice cannot form neither 0D or 1D closed-shell solids due to two reasons: *first*, small penta-graphene flakes (figure 15(a)) do not have positively-curved fragments; *second*, both 0D tori and 1D corkscrew-shaped nanotubes have non-zero thick minor radii, so the length of extra torus/corkscrew fragments of negative curvature should be longer the inner positively curved ones which is not possible without introduction/cutout of extra three-atom thick unit cells from external/internal positively/negatively curved fragments of the lattices.

3D triply periodic minimal surfaces (TPMS) were mathematically introduced and studied by Schwarz [85] and Schoen [86, 120] long before discovery of any complex forms of nanocarbons back in 1890 and 1970. A number of regular 3D $sp^2$-carbon TPMS schwarzites or buckygyms consisted of just negatively curved lattices were introduced (see, for example, [88–90]). Considering the TPMSes and carbon foams on the base of them, Hyde *et al* [121] mentioned that 'we find we have entered an enormously complex, and virtually unexplored, area of geometry'.

All schwarzites which were developed on the base of TPMSs have 3D periodic structures (see, for examples, figure SI1-12) with 0 thickness of negatively curved continuous non-crossing surfaces constituted by hexagons and larger polygons (heptagons, octagons and nonagons) without any residual mechanical stress. The last property implies that the force constants normal to any TPMS surface can be ignored. Zero thickness of the surface also means that the surface integrals of both sides of the surface should be symmetrically equivalent. The penta-graphene lattice do not mathematically satisfy both TPMS properties since it has non-zero thickness with strongly anisotropic force constant field.

Extended penta-graphene $na \times 8b$ ($n = 5$–$8$) flakes (figures SI1-11(a)–(e)) demonstrate complex structural behavior upon the flake composition. Initially, the $5a \times 8b$ flake (figure SI1-11(b)) demonstrate distinguished $a$-related curvature. An extension of the lattice in $a$ direction (figures SI1-11(c)–(e)) leads to graduate mutual compensations of the force constants (figure SI1-11(a)) oriented in opposite direction with cancellation of overall negative curvature of the $na \times mb$ flakes with formation of $b$-oriented linear 1D penta-graphene nanotube (figure SI1-11(i)) considered above. It is necessary to note that since both types of relaxed minimal energy penta-graphene nanotubes are highly symmetrical, they cannot demonstrate any kind of curvature in any direction.

Model potential and PM3 optimization of extended $a$-oriented nanoribbons of 32 ($C_{700}H_{268}$) (figures 15(b)) and 162 ($C_{3560}H_{1308}$) (figure SI1-13) unit cell length returned penta-graphene rolls oriented along $b$-axis with 30, 31 and 67 Å external diameters. For 32 unit cell roll both PM3 and MM+ approaches returned close structural parameters with 3 rotations of $a$-nanoribbon around the main $b$-oriented axis. The 162 unit cell roll demonstrate 7 rotations around $b$-oriented main axis. For both model potential and PM3 32 nanoribbons the double coil rolls were also considered (figures 15(b) and SI1-13). It necessary to note, that elongation of 162 roll by just one unit cell, up to 163 roll spontaneously forms the secondary coil at the end of the nanoribbon developed by the competition of decreased curvature of the lattice at the external roll layer and strong bending force constant. The structural parameters of double coil 32 and 163 conformers are $26 \times 36$ Å (32, both model potential and PM3), and $65 \times 73$ Å (163, model potential) demonstrate that both model potential and semiempirical PM3 methods return very close structural parameters of the rolls. In fact, formation of the secondary coil should prevent the lattice growth due to passivation of the nanoribbon edge which in turn should limit the maximum diameter of penta-graphene rolls from 26 to 65 Å. Since the secondary coil introduces strong lattice anisotropy, it should generate non-zero bending force perpendicular the main $b$-oriented roll axis, which should lead to complete breakdown of the lattice periodicity in any direction.

Phonon spectrum of planar 2D penta-graphene (figure 15(c)) calculated using PBC PBE PAW approach reveals a small (less than $1i$ cm$^{-1}$) imaginary mode around $\Gamma$-point, that is in contrast to [35] where no imaginary frequencies were revealed. The imaginary part has a gentle slopping and is associated with out-of-plane vibrations that are normally presented in various 2D materials. Therefore such negative part around $\Gamma$-point may be considered as numerical error and ignored with considering a low-dimensional lattice as stable according to the PBC approach.

Infinite multidimensional complete active space of normal coordinates of penta-graphene includes both finite TI and infinite *non*-TI subspaces. From a mathematical point of view, PBC constrains energy minimization of the penta-graphene lattice to finite TI subspace of normal coordinates, which leads to artificial localization of a regular point in true CAS NC as the structural global minimum. Consequent phonon dispersion calculations using the geometry obtained by constrained minimization procedure may artificially return no imaginary modes witnessing false stability of the lattice.

Within symmetrical TI constraints, 2D planar penta-graphene [35] consists of 3 non-equivalent sublattices with one central $sp^3$ atom and two (top and bottom) $sp^2$ carbon dimer sublattices oriented perpendicular to each other. Since non-equivalent $sp^2$ sublattices are separated from each other by 1.2 Å, they generate uncompensated torque (see section 3), which as a consequence leads to dramatic structural distortion of the lattice. In fact, the linear PBC approximation is a symmetry restriction that forbids (if applied) folding of any kind of low-dimensional lattices especially for non-periodic bend structures or structure waves with a wavelength greater than one unit cell or, in general, applied supercells. One should take into account that the linear PBC approach leads to artificial stabilization of low-dimensional crystalline lattices which may suffer a complete breakdown of periodicity due to, for example, formation of curved structures, with consequent artificial luck of imaginary modes in phonon spectrum.

Overall, it was shown that: (i) mathematically one unit cell thin penta-graphene lattice can not exist in either perfectly planar 2D lattice, or 0D closed-shell cages like spherical fullerenes, carbon tori or 1D corkscrew coils; (ii) penta-graphene lattice can form either saddle-shaped 0D flakes of negative curvature of $K = 0.015$ Å$^{-2}$, or 1D perfect linear minimal energy 1D nanotubes of either 5.98/7.14 Å or 9.94/11.19 Å inner/outer diameters or rolls of finite diameters and complex nature; (iii) due to strong bending force constants perpendicular to the lattice, the rolls have very asymmetric structure due to formation of secondary coil which should passivate the lattice growth, limits external roll diameter up to 65–70 Å, introduce strong lattice anisotropy and generate non-zero bending along the main axis, which should, in turn, completely eliminate periodicity of the rolls in any direction; (iv) PBC approximation causes artificial topological stabilization of 2D penta-graphene due to restrictions imposed by linear translation symmetry; (v) for small flakes, two perpendicular $sp^2$ carbon dimer sublattices of penta-graphene separated by 1.2 Å cause uncompensated torque which breaks down topology and periodicity with the formation of aperiodic saddle structures; (vi) the stress energy of carbon atoms in planar flakes with preserved coordination is twice as great the vdW energy, making impossible to stabilize planar penta-graphene clusters at any kind of supports; (vii) constrained TI SS NC energy minimization of penta-graphene artificially returns a regular point in CAS NC surface as the global minimum; (viii) phonon dispersion law calculations of penta-graphene at constrained minimum artificially return no imaginary modes, which compromise the phonon spectrum calculations as a solid and final proof of stability of low-dimensional crystalline lattices; (ix) small bent $n \times m$ ($n, m \leqslant 5$) penta-graphene clusters or nanorolls could exist and in principle may be synthesized.

## 7. Topological instability of ladderane lattice

Originally, the ladderane lattice [65] was designed following cyclobutane motif (figure 16(a)). The *Pmma* orthorhombic unit cell of ladderane consists of two non-equivalent carbon atoms. The 2D crystalline structure was calculated using the PW PBE PBC approach and it was found that C–C bond lengths are equal to 1.640 and 1.511 Å, respectively, with $\alpha = \beta = \gamma = 90$, which is rather close to the original data of [65] (1.650 and 1.513 Å with $\alpha = \beta = \gamma = 90$). Four-fold coordination of each carbon atom neither belongs to $sp^3$- nor $sp^2$-hybridization, so the lattice cannot be perfectly fitted either by tetrahedral $sp^3$ or triangular $sp^2$ fragments due to symmetrical reasons. One can conclude, that 2D planar ladderane does not satisfy the mandatory TCP requirements and could be artificially stabilized by the PBC approach. Phonon dispersion of ladderane (figure 16(b)) reveals no imaginary modes in the whole momentum space, which perfectly coincides with original results [65].

To check topological stability of the lattice, minimization of two finite unsaturated $C_{54}$ and saturated $C_{54}H_{30}$ ladderane $4 \times 5$ clusters was performed using the PW PBE approach (figure 16(c)). Minimization of unsaturated $C_{54}$ cluster revealed dramatic restructuration of ladderane lattice converting initial $4 \times 5$ buckled ladderane $C_{54}$ cluster of $C_{2v}$ symmetry to perfectly planar $D_{2h}$ hexagonal lattice (figure 16(c) left). Complete restructuration followed by symmetry breakdown of the initial 2D ladderane lattice may lead either to formation of 2D graphene or separate 0D graphene flakes, that may lift the long-range crystalline order of regular covalent bonds between the atoms spanned all over the crystal. Like in the case of penta-graphene, PBC artificially stabilizes rectangular buckled ladderane lattice keeping its shape and translational symmetry invariance. PW PBE energy minimization of saturated $C_{54}H_{30}$ cluster (figure 16(c) right) leads to a complete breakdown of its symmetry, long- and even short-range orders with the formation of an irregular exotic atomic lattice with 3-, 4-, 5-, 6-, 7-, and even 8-member rings. Relaxation of both $C_{54}$ and $C_{54}H_{30}$ clusters leads to complete decomposition of finite regular 2D rectangular ladderane lattice which satisfies TI conditions with the formation of either graphene flakes or irregular 0D finite carbon clusters with breakdown of the topology and both long- and short-range orders.

The internal forces of ladderane are strong enough to overcome the C–C bonds regrouping potential barriers and completely disintegrate the lattice with breakdown of translation symmetry and topology of the regular crystalline lattice. By definition, the phonon dispersion calculations are performed for the lattices which satisfy the TI conditions of TI subspace of normal coordinates *x* which belongs to Hilbert complete active space of normal coordinates, $\boldsymbol{x} \in \boldsymbol{X}$. In the case of ladderane, the $\boldsymbol{X}$ space of infinite dimension includes aperiodic normal coordinates of disintegrated fragments of ladderane as well as finite TI $\boldsymbol{x}$, $\boldsymbol{X} \ni \boldsymbol{x}$, of regular ladderane lattice. Because of it, the constrained extremum localized in TI $\boldsymbol{x}$ subspace is just a regular point on Hilbert potential energy surface located far away from any true global minimum/maximum. Consequent calculations of phonon dispersion law at constrained TI SS NC minimum could artificially return no imaginary modes witnessing false stability of the lattice.

Summarizing the paragraph, it was found that 2D ladderane lattice does not satisfy TCT mandatory restrictions due to dramatic departure of the elementary ladderane rectangular fragments either from triangular ($sp^2$ carbon), or tetrahedral ($sp^3$ carbon) structural units. Linear PBCs lead to artificial stabilization of ladderane keeping intact its rectangular lattice with perfect translation symmetry. The lift of translation symmetry restrictions leads to complete breakdown of the integrity, translational and local symmetries of the lattice with formation of localized fragments caused by strong internal structural stress. In PBCs, the energy minimization of the ladderane 2D lattice is reduced to constrained minimization in subspace of translationally invariant normal coordinates far away from any true global/local minimum/maximum of Hilbert complete space of normal coordinates, which leads to artificial localization of a regular point on true potential energy surface as a special point for which no imaginary phonon modes were detected. 2D ladderane lattice suffers severe topological instability which cannot be detected by regular phonon calculations due to strong departure of 2D rectangular lattice from a true global minimum or maximum.

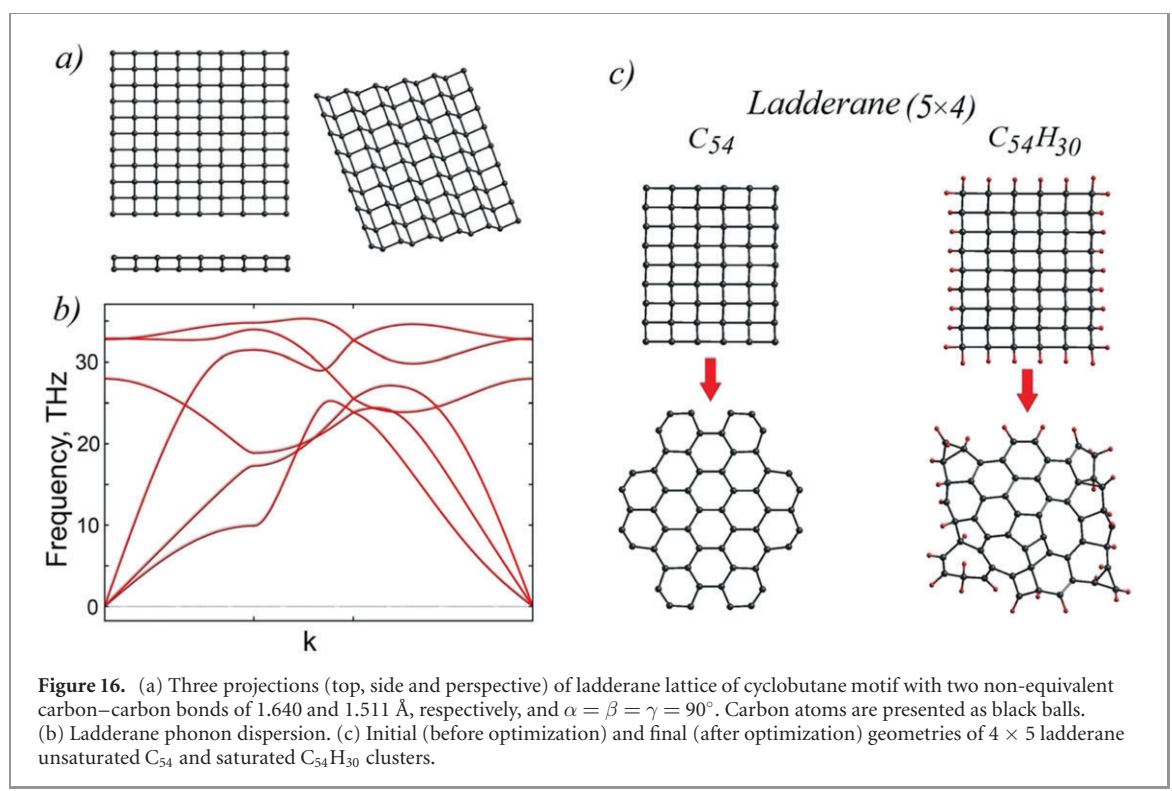

**Figure 16.** (a) Three projections (top, side and perspective) of ladderane lattice of cyclobutane motif with two non-equivalent carbon–carbon bonds of 1.640 and 1.511 Å, respectively, and $\alpha = \beta = \gamma = 90°$. Carbon atoms are presented as black balls. (b) Ladderane phonon dispersion. (c) Initial (before optimization) and final (after optimization) geometries of 4 × 5 ladderane unsaturated $C_{54}$ and saturated $C_{54}H_{30}$ clusters.

## 8. The mechanism of structural stabilization of 2D biphenylene lattice

All previous calculations of biphenylene lattice were performed for perfectly planar lattices applying symmetry restrictions either in cluster approach by $D_{2h}$ symmetry point group [27], or PBC [28, 122] *Pmmm* space group with six basis carbon atoms in the rectangular unit cell. The lattice constants of 2D biphenylene are $a = 3.76$ Å and $b = 4.52$ Å with 3 different C–C bonds, with the lengths of $l_1$, $l_2$, and $l_3$ equal to 1.45, 1.46, and 1.41 Å, respectively. No imaginary phonon modes were detected in previous phonon dispersion calculations [122].

To analyze topological stability of 2D biphenylene lattice (figure 17(a)) its parent constituting structural units, namely cyclobutadiene $C_4H_4$ (figure 17(b)) cyclooctatetraene $C_8H_8$ (figures 17(c) and (d)) and biphenylene $(C_6H_4)_2$ (figure 17(e)) should be considered first. Molecular structure of cyclobutadiene, cyclooctatetraene both ground state tub and planar transition state conformers and biphenylene were calculated using *ab initio* B3LYP/6-31G* and PW PBE approaches. Cyclobutadiene $C_4H_4$ belongs to [*n*]-annulene family (actually, it is the smallest [4]-annulene) and is characterized by its antiaromatic $\pi$-system with four $\pi$-electrons. It is a correlated molecule in which the Jahn–Teller effect distorts the carbon 4-member carbon cycle converting triplet carbon square to singlet carbon rectangule with two non-equivalent C–C bonds [123]. At the MR-CCSD(T) level of theory the ground state rectangular singlet is characterized by 1.483 and 1.291 Å long and short C–C bond lengths, respectively, whereas exited triplet square conformer has four equivalent 1.386 Å C–C bonds [124].

Antiaromatic cyclooctatetraene $C_8H_8$ (figure 17(d)) is another member of [*n*]-annulene ([8-annulene]) family. Its ground state tub-shaped conformer [125, 126] is characterized by two C–C–C non-equivalent angles of 121.1° and 117.6°. Anionic form $[C_8H_8]^{-2}$ is an aromatic molecule with a perfectly planar structure of $D_{8h}$ symmetry with eight $\angle$C–C–C equal to 135°. The neutral transition state conformer $C_8H_8$, cyclooctatetraene-TS (figure 17(c)), has a planar structure as well with reduced symmetry from $D_{8h}$ point group to $D_{4h}$ caused by the Jahn–Teller effect.

The free-standing $C_{12}H_8$ biphenylene unit (figure 17(e)) is a polycyclic molecule [127, 128] with two aromatic benzene rings connected through strongly antiaromatic central rectangular cyclobutadiene fragment [129]. Since the central antiaromatic $C_4$ unit keeps its planar structure due to structural limitations, the whole molecule also keeps its planar structure as well. In the unit cell of periodic lattice, every single $C_4$ fragment of 2D biphenylene displays an almost square structure with two types of C–C bonds, equal to 1.4572 and 1.4536 Å, respectively, which is very close to PW PBE bond length of 1.444 Å of $C_4H_4$ cyclobutadiene.

Every single $C_6$ hexagonal fragment in 2D biphenylene lattice has two types of $\angle$C–C–C angles equal to 110.0° (two angles) and 125.0° (four angles), which do not match neither perfect aromatic $C_6H_6$ (120.0°) nor parent PW PBE biphenylene (115.7°, 2 angles; 122.3°, 2 angles; and 122.0°, 2 angles). The C–C bond lengths (1.457 Å, 2 bonds; 1.405 Å, 4 bonds) of 2D biphenylene $C_6$ fragments are also significantly differed from parent molecular byphenilene with 4 types of C–C bonds, equal to 1.427 Å (2 bonds), 1.376 Å (4 bonds), 1.393 Å (2 bonds), and 1.420 Å (4 bonds).

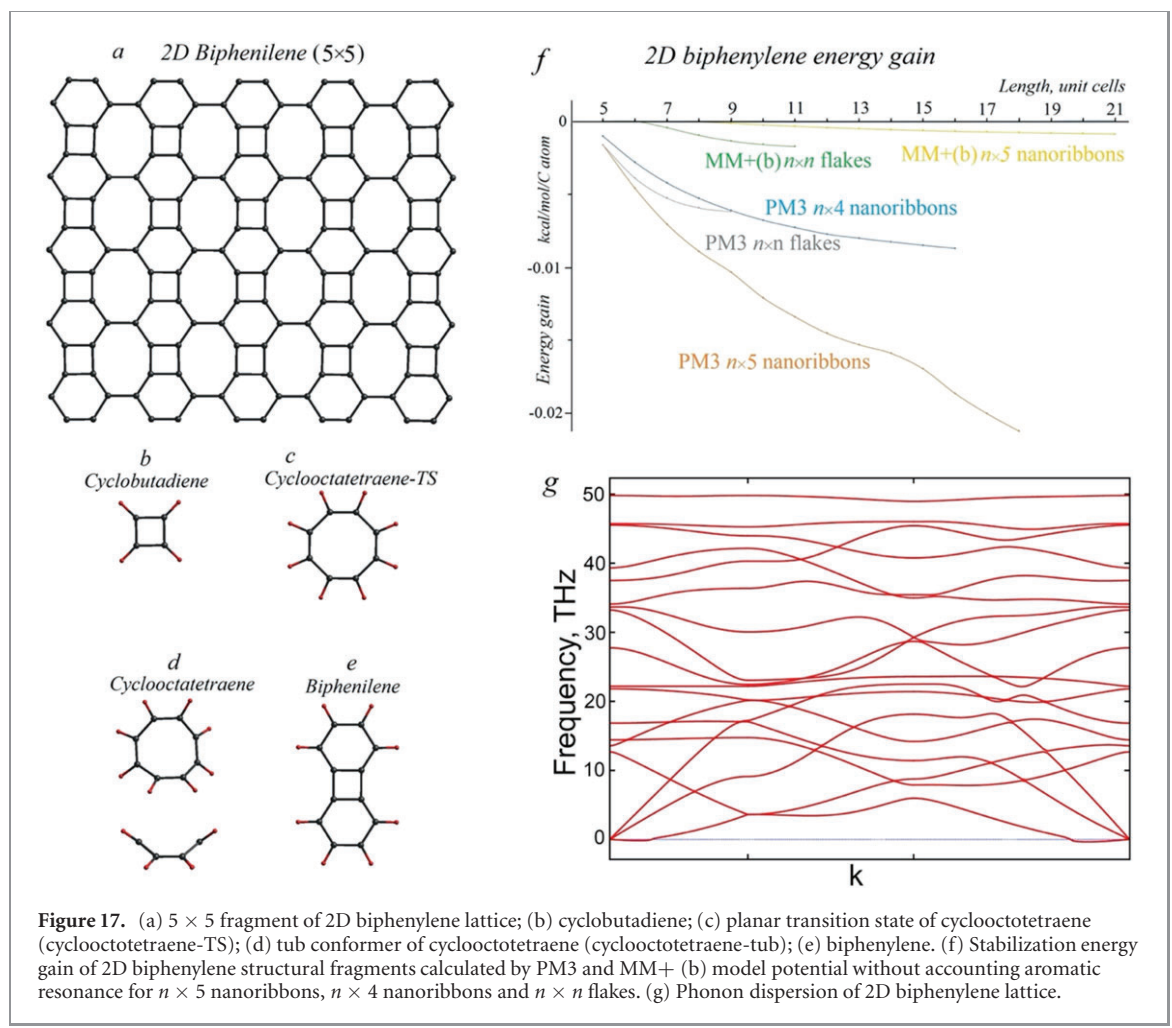

**Figure 17.** (a) 5 × 5 fragment of 2D biphenylene lattice; (b) cyclobutadiene; (c) planar transition state of cyclooctotetraene (cyclooctotetraene-TS); (d) tub conformer of cyclooctotetraene (cyclooctotetraene-tub); (e) biphenilene. (f) Stabilization energy gain of 2D biphenylene structural fragments calculated by PM3 and MM+ (b) model potential without accounting aromatic resonance for $n \times 5$ nanoribbons, $n \times 4$ nanoribbons and $n \times n$ flakes. (g) Phonon dispersion of 2D biphenylene lattice.

In 2D biphenylene, the most dramatic changes are revealed in $C_8$ fragments, which are very different from either cyclooctatetraene ground state non-planar $D_{2d}$ tub conformer, or transition state planar $D_{4h}$ one. $C_8$ of 2D biphenylene has 3 types of C–C bonds of 1.445 Å (2 bonds), 1.454 Å (2 bonds), and 1.405 Å (2 bonds) with 2 types of ∠C–C–C angles equal to 125.0° (4 angles) and 145.0° (4 angles), respectively, whereas tub conformer has non-zero torsion angle of 54.7°, 8 equivalent ∠C–C–C angles of 127.3° and two types of C–C bonds of 1.346 (4 bonds) and 1.469 Å (4 bonds). Planar TS conformer has 8 equivalent ∠C–C–C angles of 135.0° and two types of C–C bonds of 1.350 (4 bonds) and 1.472 Å (4 bonds).

Overall, that 2D biphenylene lattice does not match all cyclobutadiene $C_4$, benzene $C_6$, cyclooctatetraene $C_8$, or biphenylene $C_{12}H_8$ conformers. It could be considered as a correlated lattice for which different types of correlations, aromatic and antiaromatic resonances may cause dramatic structural distortions of constituting fragments, the lattice symmetry and topology.

The structure and topology of different 2D biphenylene fragments like $n \times n$ flakes and $n \times m$ ribbons were studied using model potential, PM3 semiempirical (figure SI1-13) and *ab initio* B3LYP/6-31G* methods. Stabilization bending energies of the fragments calculated by PM3 semiempirical and two types of model potentials with (MM+(a)) and without (MM+(b)) taking into account the aromatic resonance [66, 130]) are presented in (figure 17(f)).

PM3 structural optimization without symmetry restrictions of $2 \times 2$, $3 \times 2$, $3 \times 3$, and $4 \times 2$ flakes returned perfectly planar $D_{2h}$ atomic lattices without imaginary modes in vibration spectra. The Hessian calculations of a set of perfectly planar flakes optimized at PM3 level of theory returned numerous imaginary modes, namely: $5 \times 5$ flake 1 imaginary mode of 7.45i cm$^{-1}$; $6 \times 6$ flake 1 imaginary mode of 8.50i cm$^{-1}$; $7 \times 7$ flake 1 imaginary mode of 7.77i cm$^{-1}$; $9 \times 9$ flake 3 imaginary modes 19.26i, 14.01i, 7.06i cm$^{-1}$; $10 \times 10$ flake 2 imaginary modes 5.44i, 2.19i cm$^{-1}$. PM3 Hessian calculations of curved 5 × 5, 6 × 6, 4 × 5, 5 × 3, 5 × 4, 6 × 4, 7 × 4, and 6 × 7 flakes revealed no imaginary modes in vibration spectra. Since the curved 2D biphenylene flakes demonstrate rather extensive curvature radii (∼100–200 Å, see below) without breakdown of any chemical bonds and atomic regroupings with change of atomic coordination numbers and hybridization states, one can expect that the mistakes in energy determination of planar and curved conformers are the same which makes possible to compare the total energies of the conformers. The results of PM3 Hessian calculations demonstrate the transition nature of extended planar biphenylene flakes between two equivalent curved global minima.

All dependencies of PM3 bending stabilization energies upon dimension of $n \times 5$ and $n \times 4$ nanoribbons and $n \times n$ flakes of biphenylene lattice (figure 17(f)) demonstrate clear $1/x$ patterns. It is necessary to note that some deviations from $1/x$ pattern for $n \geqslant 14$ for $n \times 5$ nanoribbon is caused by a lack of accuracy of structural minimization of extended ribbons with the number of atoms in the lattices >450. The energy gain of square-shaped flakes converged to −0.006 kcal/mol/C atom which is smaller than the experimental values of vdW energy [112]. PM3 structural optimization of $n \times 4$ and $n \times 5$ nanoribbons reveals −0.01 and −0.02 kcal/mol/C atom stabilization energies, respectively, which is comparable with the lowest limit of vdW binding. Structural minimization of $n \times 4$ and $n \times 5$ nanoribbons revealed structural curvature with radii of $R_{n\times3} = 87.266$ Å and $R_{n\times4} = 85.139$ Å, respectively (figure SI1-13). The curvature radii of 42.761 × 36.287 × 4.042 Å 10 × 10 flake in $x$, $y$ and diagonal $x+y$ directions are found to be $R^x_{n\times n} = 172.5$ Å, $R^y_{n\times n} = 200.5$ Å, and $R^{xy}_{n\times n} = 167.8$ Å, respectively, with a maximum deviation from the plane equal to 4.042 Å.

The MM+ algorithm [130] allows one to separate different stretching and torsion components to the total energy of carbon-based atomic lattices with active $\pi$-system. Let us denote a set of parameters of the MM+ approach by taking into account the contribution of aromatic resonance as MM+(a). To elucidate the role of aromatic resonance in structural stabilization, the MM+(a) simulations were accompanied by structural optimization without taking into account the aromatic resonance using a set of MM+ parameters denoted by MM+(b). The MM+(b) stabilization energies of $n \times n$ flakes and $n \times 5$ nanoribbons are presented in figure 17(f). Since the aromatic resonance almost completely stabilizes the planar topology of 2D biphenylene lattices, the $n \times n$ and $n \times 5$ MM+(a) curves have almost constant stabilization energy values which are very close to 0, so they are not displayed in figure 17(f).

Like PM3 stabilization energies, the $n \times n$ and $n \times 5$ MM+(b) ones also demonstrate $1/x$ patterns with much smaller saturation limits of −0.002 and −0.001 kcal/mol/C atom. Once again, these values are much smaller of both vdW energies [112] and the accuracy of the MM+ approach [66, 130]. The curvature radii of the largest curved MM+(b) C$_{726}$H$_{66}$ 11 × 11 flake with 47.544 × 38.987 × 2.944 Å dimensions of C$_{2v}$ symmetry are equal to $R^x_{n\times n} = 294.9$ Å, $R^y_{n\times n} = 288.4$ Å, and $R^{xy}_{n\times n} = 306.9$ Å, which are significantly larger corresponding PM3 ones (see above). The Hessian calculations of MM+(b) $n \times n$ flakes revealed no imaginary modes for all curved conformers. In contrast, every single MM+(b) $n \times n$ planar conformer reveals one imaginary mode in the range of 7.92i–8.14i cm$^{-1}$.

The $n \times 5$ MM+(b) nanoribbons reveal a distinctive curved shape (figure SI1-14). In particular, the longest $n \times 5$ C$_{630}$H$_{94}$ nanoribbon of 91.862 Å length has $R_{n\times4} = 196.2$ Å curvature radius with the largest departure from $xy$ plain equal to 10.337 Å. As one can see, the curvature radius of $n \times 5$ MM+(b) nanoribbons is significantly larger than the PM3 one of $R_{n\times4} = 85.139$ Å (see above).

Taking into account the aromatic resonance contribution by applying of MM+(a) set of parameters leads to localization of almost perfect planar structures of $n \times n$ flakes with maximum departure of carbon atoms from the plane by 0.532 Å, and R.M.S. of D$_{2h}$ symmetry group of 0.010 (R.M.S. of C$_{2v}$ symmetry group is 0.000) for the largest curved C$_{726}$H$_{66}$ 11 × 11 flake with 47.363 × 38.784 Å dimensions. The 7 × 7 and 9 × 9 MM+(a) curved and planar flakes revealed no imaginary frequencies in Hessian calculations.

The MM+(a) set of parameters also leads to localization of almost planar $n \times 5$ nanoribbons with stabilization energies very close to 0, which is much less the accuracy of MM+ approach (not presented in figure 17(f) because of almost perfect coincidence of the curve with $X$ axis). The maximum departure of carbon atoms of the longest (91.316 Å) 20 × 4 nanoribbon from parent planar structure is 1.276 Å with curvature radius equal to 1633.0 Å and D$_{2h}$ R.M.S. equal to 0.274. It is necessary to note, that the maximum departure of carbon atoms for 15 × 5 nanoribbon is equal to just 0.054 Å with D$_{2h}$ R.M.S. equal to 0.012.

In table 2 structural and energetic characteristics of $n \times n, n = 3–9, k \times 5, k = 7–15$, and $5 \times l, l = 7–15$ 2D biphenylene fragments of finite dimensions calculated using *ab initio* B3LYP/6-31G* approach are presented. Following the results of PM3 simulations in which the curvature radius was estimated to be 85 Å (see above), an extended *ab initio* B3LYP/6-31G* geometry search was performed starting from curvature radius of 30 Å. It is necessary to note that independently of the starting geometry, geometry relaxation without symmetrical restrictions returned the same type of slightly corrugated structures. To reveal the stabilization bending energies and symmetry effects, perfectly planar conformers were calculated as well using the same *ab initio* B3LYP/6-31G* approach.

**Table 2.** Size (in unit cells and Å), chemical formula, stabilization energies (kcal/mol/C atom), departure from plane (Å), $D_{2h}$ R.M.S. and nature of planar and curved conformers of $n \times n$, $n = 3–9$, $k \times 5$, $k = 7–15$, and $5 \times l$, $l = 7–15$ 2D biphenylene finite fragments calculated using *ab initio* B3LYP/6-31G* method[a].

| Dimensions, octagonal fragments | Dimensions, Å | Chemical formula | Stabilization energy, kcal/mol/C atom | Departure from plane, Å | $D_{2h}$ R.M.S. | Nature of planar conformer | Nature of curved conformer |
|---|---|---|---|---|---|---|---|
| 3 × 3 | 11.885 × 10.017 | $C_{54}H_{18}$ | 0 | 0 | 0 | N.C.A. | N.C.A. |
| 4 × 4 | 16.373 × 13.821 | $C_{96}H_{24}$ | $2 \times 10^{-6}$ | 0.005 | 0.001 | N.C.A. | N.C.A. |
| 5 × 5 | 20.881 × 17.628 | $C_{150}H_{30}$ | $9 \times 10^{-6}$ | 0.012 | 0.002 | N.C.A. | N.C.A. |
| 6 × 6 | 25.364 × 21.439 | $C_{236}H_{36}$ | $6 \times 10^{-6}$ | 0.005 | 0.001 | N.C.A. | N.C.A. |
| 7 × 7 | 29.799 × 25.299 | $C_{294}H_{42}$ | $6 \times 10^{-6}$ | 0.019 | 0.004 | TS | TS |
| 8 × 8 | 34.278 × 29.094 | $C_{384}H_{48}$ | $2 \times 10^{-5}$ | 0.019 | 0.004 | TS | TS |
| 9 × 9 | 38.941 × 32.731 | $C_{486}H_{54}$ | — | — | — | TS | — |
|  |  |  | $6 \times 10^{-6}$ | 0.017 | 0.004 | N.C.A. | N.C.A. |
| 7 × 7[b] | 29.799 × 17.638 | $C_{210}H_{38}$ | $1 \times 10^{-5}$ | 0.018 | 0.004 | — | N.C.A. |
| 9 × 5 | 38.655 × 17.683 | $C_{270}H_{46}$ | $1 \times 10^{-4}$ | 0.015 | 0.068 | N.C.A. | N.C.A. |
| 11 × 5 | 47.558 × 17.683 | $C_{330}H_{54}$ | $7 \times 10^{-5}$ | 0.070 | 0.012 | TS | N.C.A. |
| 13 × 5 | 56.466 × 17.680 | $C_{390}H_{62}$ | $5 \times 10^{-5}$ | 0.068 | 0.013 | TS | N.C.A. |
| 15 × 5 | 65.359 × 17.690 | $C_{450}H_{70}$ | $7 \times 10^{-5}$ | 0.062 | 0.012 | TS | N.C.A. |
|  |  |  | $7 \times 10^{-6}$ | 0.015 | 0.003 | N.C.A. | N.C.A. |
| 5 × 7[b] | 21.078 × 25.079 | $C_{210}H_{34}$ | $9 \times 10^{-3}$ | 0.165 | 0.043 | — | N.C.A. |
| 5 × 9 | 20.875 × 32.918 | $C_{270}H_{38}$ | $1 \times 10^{-5}$ | 0.034 | 0.006 | TS | N.C.A. |
| 5 × 11 | 20.877 × 40.558 | $C_{330}H_{42}$ | $2 \times 10^{-5}$ | 0.032 | 0.006 | TS | N.C.A. |
| 5 × 12 | 20.874 × 44.380 | $C_{360}H_{44}$ | $5 \times 10^{-5}$ | 0.014 | 0.003 | TS | N.C.A. |
|  |  |  | $5 \times 10^{-5}$ | 0.016 | 0.003 | N.C.A. | TS |
| 5 × 13[c] | 20.873 × 48.195 | $C_{390}H_{46}$ | $-2 \times 10^{-4}$ | — | — | TS | — |
| 5 × 14 | 20.866 × 50.018 | $C_{420}H_{48}$ | $5 \times 10^{-5}$ | 0.018 | 0.004 | TS | TS |
| 5 × 15 | 20.873 × 55.831 | $C_{450}H_{50}$ | $4 \times 10^{-5}$ | 0.015 | 0.003 | TS | N.C.A. |

[a] TS = transition state; N.C.A. = 'no conclusive answer'.
[b] Two curved conformers were located.
[c] Two planar optimization trajectories were located.

All 3 types of fragments revealed the bending energies of 0.0–9 × 10$^{-3}$ kcal/mol/C atom, which is much lower the accuracy of the DFT B3LYP approach [68, 69, 131–135], for which the dissociation energies for the G2 set are determined with an accuracy of 3.5 kcal/mol. Calculated maximum of absolute departure of carbon atoms from the structural plane of 2D biphenylene fragments reaches 0.165 Å for one of 5 × 7 $C_{210}H_{34}$ conformers of 21.078 Å × 25.079 Å dimension with $D_{2h}$ symmetry R.M.S. 0.068 for 9 × 5 $C_{270}H_{46}$ nanoribbon, which provide unequivocal proof of visible distortion of parent planar $D_{2h}$ symmetry of all fragments except 3 × 3 $C_{54}H_{18}$ flake, for which internal total mechanical stress is not strong enough and cannot overcome $\pi$-electronic resonance making the lattice perfectly planar. It is necessary to note, that DFT B3LYP accuracy in the determination of interatomic distances is well below 0.001 Å (see references above).

Anizotropic nature of 2D biphenylene lattice leads to visibly different departures from planar structures for $k \times 5$ and $5 \times l$ nanoribbons in perpendicular directions (table 2) equal to 0.070 and 0.165 Å and $D_{2h}$ R.M.S. deviation of 0.068 and 0.043, respectively. Because of negligibly small stabilization energies of all considered fragments all of which are characterized by different structural shifts in perpendicular directions, one can conclude that 2D biphenylene lattice itself is characterized by very shallow potential energy surface in the vicinity of global minimum, which is close to perfectly planar structure. Since the basic structural units of 2D biphenylene, namely cyclobutadiene and cyclooctotetraene are tensed correlated antiaromatic structures, the entire 2D crystalline lattice of biphenylene could be regarded as a strongly correlated structure [136] which is characterized by both dynamical [137] and non-dynamical (see, for example, [138]) contributions.

The geometry optimization procedures are based on the search of zero gradient on total potential energy surfaces [139]. Rigorously, the outcome of geometry optimization should be accompanied by Hessian calculations at the extremum point to prove whether or not it is a local or global minimum rather than saddle point on complete potential energy surface. In the case of absence of imaginary modes in the calculated vibration spectrum, one can conclude that the localized geometry is a minimum, otherwise, it is proved that it is a saddle point. In general, for extended clusters (hundreds and thousands of atoms) the Hessian calculations are incredibly complex and unreliable making impossible to reveal the true nature of localized extreme points on potential energy surfaces.

For most cases, the optimization trajectories follow the minimum energy path on potential energy surfaces which couple the minimization of the gradient with a smooth decrease of the total energy in the vicinity of an extremal point (figure SI1-15), which can be either a minimum or a saddle point. Let us denote the first type of trajectories as 'no conclusive answer' (N.C.A.) trajectories since one should run additional Hessian calculations to reveal the true nature of the localized structure. For all N.C.A. trajectories, which couple the local extremum of potential energy surface with gradient minimum, the starting point has higher energy in respect to final localized point. In particular, even the localized extremal point is a saddle point, the N.C.A. trajectory returns a decline of energy in respect to the optimization steps. In general, no conclusive deductions about the nature of the localized point can be made based on just an analysis of the N.C.A.-type optimization trajectory.

Another special outcome of optimization procedure couples the gradient minimization with a raise in total energy (figure SI1-15). By definition, a trajectory along the raising energy pathway may only lead from a lower energy point on the potential energy surface to a saddle point with higher energy, or global or local maximum. Such sort of trajectories could be denoted as TS trajectories since they may lead to transition states of different orders with higher energies in respect to the starting point. In general, the TS-type optimization trajectory is a clear and unequivocal indication of non-ground state nature of any localized extremum point on potential energy surface as an outcome of optimization procedure.

Step-by-step extension of biphenylene 2D $l \times m$ clusters leads to decreasing the influence of cluster boundaries on the flake lattices improving the quality of initial structural guess. Based on TCT analysis (see above) and negligibly small energy differences between planar and non-planar conformers (table 2), one can expect that perfectly planar flakes may be a transition state on potential energy surfaces among multiple almost energy degenerated conformers of undetermined nature. The *ab initio* B3LYP 6-31G* optimization trajectories are presented in tables 2 and SI1-4. One can see that for extended perfectly planar clusters starting from $n \times n, n \geqslant 7, k \times 5, k \geqslant 11$ and $5 \times l, l \geqslant 9$ the planar conformers were localized within typical TS-type optimization trajectories, which might indicate their transition state character. Even the lattices of some curved conformers ($7 \times 7$ $C_{294}H_{42}$, $8 \times 8$ $C_{384}H_{48}$, $15 \times 5$ $C_{450}H_{70}$, $5 \times 13$ $C_{390}H_{46}$, and $5 \times 14$ $C_{420}H_{48}$) were localized through the TS-trajectories, which might reveal their possible TS nature.

Coexistence of multiple degenerated or close to degeneracy (energy differences 0–$10^{-3}$ kcal/mol/C atom, table 2) correlated transition states, local/global minima or maxima of the same symmetry and very close structural parameters of 2D biphenylene conformers contradicts to fundamental quantum adiabatic theorem [140], making impossible to exist stationary quantum state of the lattice. Following the adiabatic theorem and correlation effects within the lattice (see discussion above), the 2D biphenylene can be considered as a strongly-correlated non-adiabatic 2D crystal. Nevertheless, following the recent consideration [141] of closed quantum systems at final temperatures, in the basis of the instantaneous eigenstates of time-dependent Hamiltonian 2D biphenylene could exist as a correlated mixed quasi-Gibbs quantum state.

The phonon dispersion of perfectly planar 2D biphenylene lattice calculated by the PW PBE PBC method is presented in figure 17(g). An imaginary band with very small absolute amplitude is detected in the vicinity of Γ point. Taking into account the TCT-based structural analysis above one could conclude that the band may be at least partially attributed to a soft mode in the vicinity of global minimum as well as a result of numerical inaccuracy in calculations of phonon dispersion law.

Based on TCT analysis, analysis of possible role of electronic correlations and the effects of non-adiabaticity, it was found that 2D biphenylene could be considered as a strongly correlated non-adiabatic low-dimensional crystal. It was shown, that the aromatic resonance and electronic correlations almost cancel internal structural stress caused by distorted correlated antiaromatic lattice fragments, namely cyclobutadiene and cyclooctotetraene. The structural effects and stabilization binding energy of 2D biphenylene lattice strongly depend upon the method used to simulate the structure. In particular, the PM3 approach [67], which significantly overestimates the force constants, predicts the formation of biphenylene rings with ~85 Å radius. It is necessary to note that even at PM3 level of theory the stabilization bending energy is smaller than the vdW energy making 2D biphenylene lattice possible to be synthesized atop of some kind of support. Using model potential simulations with and without taking into account the aromatic resonance it was found that aromatic resonance almost completely compensates accumulated structural stress. *ab initio* B3LYP/6-31G* calculations of extended $k \times l, k, l = 3$–15 also demonstrate negligibly small stabilization bending energies with soft modes in the vicinity of the global minimum. Based on adiabatic theorem consideration, it was shown that free-standing low-dimensional biphenylene lattice can exist at final temperatures as a quasi-Gibbs mixed quantum state. Experimental synthesis of 2D biphenylene by bottom-up approach [28] on solid-state support is in perfect agreement with theoretical results obtained by TCT analysis accompanied by model potential and electronic structure simulations.

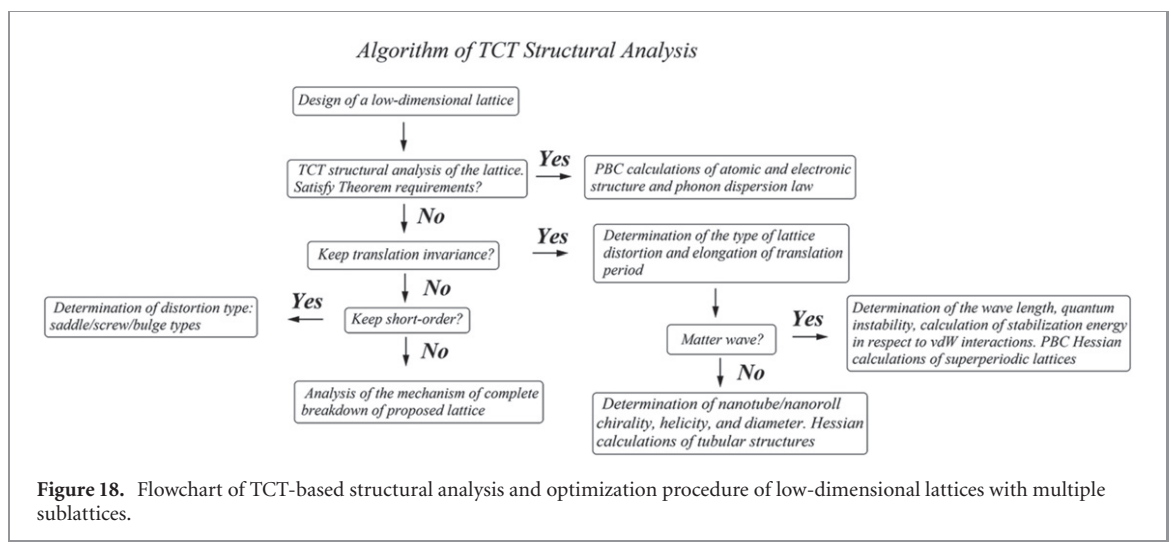

**Figure 18.** Flowchart of TCT-based structural analysis and optimization procedure of low-dimensional lattices with multiple sublattices.

## 9. TCT analysis of topological and quantum stability of low-dimensional lattices

Following TCT treatment coupled with numerical simulations, a flowchart (figure 18) for TCT structural analysis of low-dimensional lattices might be proposed. At the first step, after a theoretical design of low-dimensional crystalline lattice, one should verify whether or not relaxed free-standing structural units can perfectly fill in the 2D crystalline space. If the fragments perfectly fill in the lattice, PBC calculations of atomic and electronic structure and phonon dispersion can be performed and analyzed. For this particular case, the phonon calculations can be used to prove the stability of proposed low-dimensional materials. However, it is also suggested to carry out *ab-initio* molecular dynamic simulations to exclude false-predicted nanocrystals [142].

If the proposed lattice does not satisfy TCT requirements, rigorous study of the lattice stability should be performed. In this case, at the second step, conservation of TI conditions should be tested using, for example, electronic structure calculations of finite clusters. If the structural mismatch of the structural units leads to just multiplication of translation vectors in one or both directions, the type and characteristics of regular lattice deformations in a form of a structure wave should be determined (wavelength and amplitude, for example). For regular periodic patterns, the quantum stability of the lattice should be considered and stabilization bending energy should be compared with the energy of van-der-Waals interactions or external pressure to determine the ways to stabilize the proposed lattice by some sort of supports, external pressure, or formation of finite flakes to avoid lattice quantum instability.

Periodicity of the lattice can be eliminated just in one direction with the formation of a nanotube or a nanoroll. In this case, structural characteristics of 1D tubular forms (diameter, chirality, helicity) should be determined.

A complete breakdown of TI conditions can be either achieved in the case of torsion deformation of 2D lattice, by breakdown of the symmetry of the lattice, formation of saddle structures, or in the case of entire decomposition of the closest chemical environment of each single atom. In the case of conservation of the closest environment, small deformed flakes of proposed materials could be somehow synthesized. In this case, the phonon calculations can not be considered as a final and solid proof of structural stability of the lattice because of decomposition of TI conditions.

And finally, in the case of very poor initial guess of a lattice, even short-range chemical environment of any constituted atoms may be completely compromised by internal stress with a total breakdown of periodicity. In this case, the phonon calculations can not be considered as final and solid proof of structural stability of the lattice as well and the mechanism of structural instability should be revealed.

The proposed algorithm (figure 18) can be used to study topological and quantum stability of novel perspective low-dimensional materials.

## 10. Conclusions

Short- and long-range orders and deformations of a number of 0D, 1D, and 2D carbon atomic lattices with multiple non-equivalent sublattices were theoretically analyzed and simulated using model potential, semiempirical and DFT approaches. It was found that non-equivalent sublattices may cause uncompensated structural stress and torque with consequent mechanical deformation of the whole lattice. The influence of the key 0D fragments, constituted by 4-, 5-, 6-, 7-, and 8-member rings, on the structure of 2D carbon hexagonal graphene lattice was studied and it was found that all considered cores may cause breakdown of its symmetry and perfectly planar topology.

Based on structural analysis of low-dimensional atomic lattices, a Topology Conservation Theorem was formulated and proved. Assuming that at least one force constant of a low-dimensional crystalline lattice is very small or equal to zero, which is true with high accuracy for one-unit cell thick lattices, it was found that in contrast to conventional 3D crystals, the lack of perfect filling of 2D active crystalline space may lead to either formation of non-planar lattices or even to decomposition of entire regular crystalline structure. It was shown that PBC imply linear translation symmetry restrictions and as a result, causes artificial stabilization of both 1D and 2D low-dimensional lattices prone to violation of perfectly linear or planar topology. In addition, two TCT corollaries were formulated and were shown to be true by comparison with known experimental facts.

All possible types of perfect planar topology violations of low-dimensional lattices with and without breaking of either long- or short-range crystalline orders were considered, namely: (i) spontaneous formation of structure waves or regular helices of different nature with either variable or constant wavelengths with multiplication of translational vectors; (ii) spontaneous formation of 1D nanotubes and rolls; (iii) spontaneous formation of aperiodic saddle structures keeping intact short-range local atomic environment; (iv) complete breakdown of both long- and short-range orders accompanied by a complete breakdown of structural integrity the lattice; (v) possible stabilization of non-planar lattices by any kind of substrates in case of stabilization bending energies smaller the van-der-Waals interactions.

It was shown that the breakdown of TI causes dramatic expansion of complete active space of normal coordinates $X$ for which the TI coordinates correspond to a low-dimensional subspace $x \in X$. Constrained geometry optimization of the lattice in restricted TI coordinates subspace $x$ may lead to localization of ordinary point on complete Hilbert space of normal coordinates as an extremum with consequent erroneous calculations of phonon dispersion law at this point. It was found that for low-dimensional crystalline lattices with multiple non-equivalent sublattices the absence of imaginary modes in theoretical phonon spectra cannot be used as a solid and final proof of their either stability or metastability and should be combined with TCT analysis and robust consideration of finite clusters.

In particular, it was found that notorious 2D *phC(m,n)* ($m;n=0;1$) lattices are prone to form superperiodic structure waves, screws or helices with variable or constant wavelength, 1D nanotubes or nanorolls, and aperiodic atomic lattices as well, with perfectly planar regular 2D lattices as transition states between symmetrically equivalent bent global minima. It was found that at all levels of theory the stabilization bending energies of *phC(m,n)* lattices are of the same order of magnitude or even larger than the van-der-Waals interactions, which makes it impossible to stabilize planar conformers on top of any kind of supports at finite temperatures.

The standing structure waves for finite-size flakes can be described by homogeneous D'Alembert second-order differential equations for which the atomic positions are well-determined. In contrast to it, infinite free-standing structure waves are described by inhomogeneous D'Alembert equation which has a solution of an infinite superposition of plane waves with non-zero external force, for which the positions of all atoms in the lattice are completely undetermined. It was shown that finite wave conformers can exist and can be synthesized, whereas infinite free-standing wave lattices are prone to lattice quantum instability effects. In particular, the lattice quantum instability should lead to the breakdown of wave periodicity by instantaneous structural transformations.

In contrast to previous publications and in consistency with TCT analysis and finite cluster considerations, the phonon dispersion of planar 2D *phC(0,1)* lattice is characterized by one imaginary mode. The one-wave-length supercell PW PBC PBE electronic structure calculations of both wave and planar conformers of *phC(0,1)* unequivocally confirm the global minimum and transition state nature of wave and planar conformers, respectively. The phonon dispersion laws of *phC(1,0)* and *phC(1,1)* lattices also revealed imaginary modes in consistency with TCT and finite cluster behavior. Despite the screw behavior of elongated clusters and violation of TCT mandatory requirements, the phonon dispersion law of 2D *phC(0,0)* $R_{5,7}$ Haeckelite does not reveal any imaginary modes, which was interpreted as a result of the breakdown of lattice TI and phonon calculations performed at constrained minimum in TI coordinates subspace of normal coordinates. It was speculated that for free-standing 1D *phC(0,0)* nanoribbons which follow screw pattern, the lattices should suffer quantum instability.

It was found that at all levels of theory the Hessian calculations of finite *phC(m,n)* flakes require incredibly high accuracy of minimization of the lattices, with the gradient stopping criterium better than $10^{-5}$ kcal/mol/Å. Nevertheless, some obvious mistakes in calculations of vibration spectra made by different codes for extended low-dimensional clusters were detected.

It was shown that NVT AIMD PBC algorithm artificially confines periodic lattices inside MD simulation box of constant dimensions which prevents true structural relaxation and thermal equilibration in form of structural contraction or stretching with breakdown of topology and symmetry of low-dimensional crystals. For example, in the case of formation of structural waves one should perform AIMD simulations of 0D finite clusters to test stability of their topology and symmetry.

Using *ab initio* DFT calculations, the stabilization bending energy of 2D penta-graphene was estimated to be 1.9 kcal/mol/atom which is at least as twice as large as the vdW energy. It was shown that mathematically free-standing penta-graphene cannot form any regular 0D, 2D, or 3D geometrical solids like spherical fullerenes, tori, corkscrew tubes, or triple periodic minimal surfaces. Strong bending force constants cause formation of either 0D small saddle-shaped flakes, 1D nanotube of 11.96/14.27 Å inner/outer diameter, or complex aperiodic 1D nanorolls of 65–70 Å external diameters. It was shown that PBC symmetry restrictions artificially stabilize planar 2D penta-graphene, preventing breakdown of planar topology of the lattice. Constrained PBC minimization of planar 2D penta-graphene in translation symmetry subspace leads to artificial localization of a regular point on complete active space natural coordinates potential energy surface as a global minimum with a lack of imaginary modes in phonon dispersion. It was concluded that phonon dispersion cannot be used as final and solid proof of either stability or metastability of 2D penta-graphene lattice.

The TCT analysis of ladderane unequivocally indicated topological instability of the lattice due to dramatic departure of its rectangular structural units from either triangular $sp^2$, or tetrahedral $sp^3$ carbon atom coordinations. It was shown that imposing of translation symmetry leads to artificial stabilization of ladderane keeping intact its rectangular periodic lattice. In contrast to PBC calculations of infinite 2D ladderane, the PW PBE cluster optimization revealed its structural instability with complete loss of periodicity, structural integrity and even local symmetry of carbon atoms making up the initial lattice. It was shown that complete active space of ladderane normal coordinates of infinite dimensionality includes translation symmetry subspace, in which constrained minimization of the structure returns artificially stable geometry. The phonon dispersion law obtained at constrained minimum returns no imaginary modes because of significant departure of ladderane translation symmetry subspace from complete active space of natural coordinates true minimum, which corresponds to an infinite irregular ensemble of asymmetric carbon clusters of different nature. It was concluded that the phonon spectrum of ladderane cannot be used as final and solid proof of its structural stability/metastability.

Using TCT analysis followed by model potential simulations and electronic structure calculations of 2D biphenylene it was shown that the lattice does not satisfy the mandatory requirements of the theorem. It was found that stabilization of planar 2D biphenylene topology is caused by aromatic resonance and correlation effects which almost completely compensate internal mechanical stress caused by structural inconsistency of the fragments. Based on simulations of atomic and electronic structure it was found that stabilization energy of 2D biphenylene in order of magnitude is lower than the vdW interactions which allows the lattices to be synthesized at any kind of solid support. It was shown that free-standing biphenylene is a correlated non-adiabatic low-dimensional crystal that can exist at final temperatures as a quasi-Gibbs mixed quantum state.

Following the results of theoretical consideration and numerical simulations of 2D carbon lattices it was shown that TCT imposes rigorous mandatory limitations on topology, symmetry and structure of low-dimensional lattices with multiple sublattices. Based on TCT theorem, a flowchart of TCT structural analysis was proposed. The algorithm can be used to develop advanced methods to analyze structure and topology of low-dimensional crystalline lattices.

## Acknowledgments

PA acknowledges the National Research Foundation of the Republic of Korea for the support under the Grant NRF 2021R1A2C1010455. AK acknowledges Olle Engkvist Byggmästare foundation for the support under Contract No. 212-0178. The authors express their gratitudes to Profs. Drs A Oganov, H Ågren, A S Fedorov, Z Wang, L F C Pereira, G V Baryshnikov, L A Chernozatonskii, and B Yakobson for very kind and fruitful discussions. The authors thank the Swedish National Infrastructure for Computing (SNIC 2021-3-22) at the National Supercomputer Center of Linköping University (Sweden) partially funded by the Swedish Research Council through Grant Agreement No. 2018-05973.

## Data availability statement

The data that support the findings of this study are available upon reasonable request from the authors.

## ORCID iDs

Pavel V Avramov  https://orcid.org/0000-0003-0075-4198

Artem V Kuklin  https://orcid.org/0000-0002-9371-6213

## References

[1] Landau L D and Lifshitz E M 2005 *Course of Theoretical Physics Course of Theoretical Physics* 3rd edn (Amsterdam: Elsevier) part 1 section 137 pp 432–6
[2] Mermin N D 1968 Crystalline order in two dimensions *Phys. Rev.* **176** 250–4
[3] Nelson D R, Piran T and Weinberg S (ed) 2004 *Statistical Mechanics of Membranes and Surfaces* (Singapore: World Scientific)
[4] Nelson D R and Peliti L 1987 Fluctuations in membranes with crystalline and hexatic order *J. Phys. France* **48** 1085–92
[5] Le Doussal P and Radzihovsky L 1992 Self-consistent theory of polymerized membranes *Phys. Rev. Lett.* **69** 1209–12
[6] Ōsawa E 1970 Superaromaticity *Kagaku (Chemistry)* **25** 854–63 (In Japanese)
[7] Bochvar D A and Galpern E G 1973 Hypothetical systems: carbododecahedron, s-icosahedron, and carbo-s-icosahedron *Dokl. Acad. Nauk SSSR* **209** 610–2
[8] Kroto H W, Heath J R, O'Brien S C, Curl R F and Smalley R E 1985 $C_{60}$: buckminsterfullerene *Nature* **318** 162–3
[9] Novoselov K S, Geim A K, Morozov S V, Jiang D, Zhang Y, Dubonos S V, Grigorieva I V and Firsov A A 2004 Electric field effect in atomically thin carbon films *Science* **306** 666–9
[10] Wallace P R 1947 The band theory of graphite *Phys. Rev.* **71** 622–34
[11] Semenoff G W 1984 Condensed-matter simulation of a three-dimensional anomaly *Phys. Rev. Lett.* **53** 2449–52
[12] Zunger A 1978 Self-consistent LCAO calculation of the electronic properties of graphite: I. The regular graphite lattice *Phys. Rev.* B **17** 626
[13] Weinert M, Wimmer E and Freeman A J 1982 Total-energy all-electron density functional method for bulk solids and surfaces *Phys. Rev.* B **26** 4571
[14] Bostwick A, Ohta T, Seyller T, Horn K and Rotenberg E 2007 Quasiparticle dynamics in graphene *Nat. Phys.* **3** 36–40
[15] Iijima S 1991 Helical microtubules of graphitic carbon *Nature* **354** 56–8
[16] Hamada N, Sawada S-I and Oshiyama A 1992 New one-dimensional conductors: graphitic microtubules *Phys. Rev. Lett.* **68** 1579
[17] Mintmire J W, Dunlap B I and White C T 1992 Are fullerene tubules metallic? *Phys. Rev. Lett.* **68** 631–4
[18] Blase X, Benedoict L X, Shirley E L and Louie S G 1994 Hybridization effects and metallicity in small radius carbon nanotubes *Phys. Rev. Lett.* **72** 1878
[19] Avramov P V, Kudin G E and Scuseria K 2003 Single wall carbon nanotubes density of states: comparison of experiment and theory *Chem. Phys. Lett.* **370** 597–601
[20] Ouyang M, Huang J-L, Cheung C L and Lieber C M 2001 Energy gaps in 'metallic' single-walled carbon nanotubes *Science* **292** 702–5
[21] Rubio A, Corkill J L and Cohen M L 1994 Theory of graphitic boron nitride nanotubes *Phys. Rev.* B **49** 5081–4
[22] Blase X A, Rubio A, Louie S G and Cohen M L 1994 Stability and band gap constancy of boron nitride nanotubes *Europhys. Lett.* **28** 335–40
[23] Chopra N G, Luyken R J, Cherrey K, Crespi V H, Cohen M L, Louie S G and Zettl A 1995 Boron nitride nanotubes *Science* **269** 966–7
[24] Golberg D, Costa P M F J, Mitome M and Bando Y 2009 Properties and engineering of individual inorganic nanotubes in a transmission electron microscope *J. Mater. Chem.* **19** 909–20
[25] Chernozatonskii L A, Sorokin P B, Kvashnin A G and Kvashnin D G 2009 Diamond-like $C_2H$ nanolayer, diamane: simulation of the structure and properties *JETP Lett.* **90** 134–8
[26] Bakharev P V *et al* 2020 Chemically induced transformation of chemical vapour deposition grown bilayer graphene into fluorinated single-layer diamond *Nat. Nanotechnol.* **15** 59–66
[27] Karausha N N, Baryshnikov G V and Minaev B F 2014 DFT characterization of a new possible graphene allotrope *Chem. Phys. Lett.* **612** 229–33
[28] Fan Q *et al* 2021 Biphenylene network: a nonbenzenoid carbon allotrope *Science* **372** 852–6
[29] Boustani I 1997 New quasi-planar surfaces of bare boron *Surf. Sci.* **370** 355–63
[30] Mannix A J *et al* 2015 Synthesis of borophenes: anisotropic, two-dimensional boron polymorphs *Science* **350** 1513–6
[31] Liu Y Y, Zhou W X, Tang L M and Chen K Q 2014 An important mechanism for thermal rectification in graded nanowires *Appl. Phys. Lett.* **105** 203111
[32] Liu Y-Y, Zhou W-X and Chen K-Q 2015 Conjunction of standing wave and resonance in asymmetric nanowires: a mechanism for thermal rectification and remote energy accumulation *Sci. Rep.* **5** 17525
[33] Chen X-K, Xie Z-X, Zhou W-X, Tang L-M and Chen K-Q 2016 Phonon wave interference in graphene and boron nitride superlattice *Appl. Phys. Lett.* **109** 023101
[34] Togo A and Tanaka I 2015 First principles phonon calculations in materials science *Scr. Mater.* **108** 1–5
[35] Zhang S, Zhou J, Wang Q, Chen X, Kawazoe Y and Jena P 2015 Penta-graphene: a new carbon allotrope *Proc. Natl Acad. Sci. USA* **112** 2372–7
[36] Wang Z H, Zhou X F, Zhang X M, Zhu Q, Dong H F, Zhao M W and Oganov A R 2015 Phagraphene: a low-energy graphene allotrope composed of 5–6–7 carbon rings with distorted Dirac cones *Nano Lett.* **15** 6182–6
[37] Rahaman O, Mortazavi B, Dianat A, Cuniberti G and Rabczuk T 2017 Metamorphosis in carbon network: from penta-graphene to biphenylene under uniaxial tension *FlatChem* **1** 65–73
[38] Cranford S W 2016 When is 6 less than 5? Penta- to hexa-graphene transition *Carbon* **96** 421–8
[39] Avramov P, Demin V, Luo M, Choi C H, Sorokin P B, Yakobson B and Chernozatonskii L 2015 Translation symmetry breakdown in low-dimensional lattices of pentagonal rings *J. Phys. Chem. Lett.* **6** 4525–31
[40] Kuklin A V, Ågren H P V and Avramov P V 2020 Structural stability of single-layer $PdSe_2$ with pentagonal puckered morphology and its nanotubes *Phys. Chem. Chem. Phys.* **22** 8289–95
[41] Ewels C P, Rocquefelte X, Kroto H W, Rayson M J, Briddon P R and Heggie M I 2015 Predicting experimentally stable allotropes: instability of penta-graphene *Proc. Natl Acad. Sci. USA* **112** 15609–12

[42] Pereira L F C, Mortazavi B, Makaremic M and Rabczuk T 2016 Anisotropic thermal conductivity and mechanical properties of phagraphene: a molecular dynamics study *RSC Adv.* **6** 57773–9
[43] https://wolf.ifj.edu.pl/phonon/
[44] Parlinski K, Li Z Q and Kawazoe Y 1997 First-principle determination of the soft mode in cubic $ZrO_2$ *Phys. Rev. Lett.* **78** 4063
[45] http://computingformaterials.com/
[46] Gale J D 1997 GULP: a computer program for the symmetry-adapted simulation of solids *Faraday Trans.* **93** 629–37
[47] Gale J D and Rohl A L 2003 The general utility lattice program (GULP) *Mol. Simul.* **29** 291–341
[48] Podlivaev A I and Openov L A 2016 Possible nonplanar structure of phagraphene and its thermal stability *JETP Lett.* **103** 185–9
[49] Terrones H, Terrones M, Hernandez E, Grobert N, Charlier J-C and Ajayan P M 2000 New metallic allotropes of planar and tubular carbon *Phys. Rev. Lett.* **84** 1716–9
[50] Rocquefelte X, Rignanese G-M, Meunier V, Terrones H, Terrones M and Charlier J-C 2004 How to identify haeckelite structures: a theoretical study of their electronic and vibrational properties *Nano Lett.* **4** 805–10
[51] Thomas S, Jung H, Kim S, Jun B, Lee C H and Lee S U 2019 Two-dimensional haeckelite h567: a promising high capacity and fast Li diffusion anode material for lithium-ion batteries *Carbon* **148** 344–53
[52] Lusk M T and Carr L D 2008 Nanoengineering defect structures on graphene *Phys. Rev. Lett.* **100** 175503
[53] Stone A J and Wales D J 1986 Theoretical studies of icosahedral $C_{60}$ and some related species *Chem. Phys. Lett.* **128** 501–3
[54] Crespi V H, Benedict L X, Cohen M L and Louie S G 1996 Prediction of a pure-carbon planar covalent metal *Phys. Rev.* B **53** 303
[55] Weller T, Ellerby M, Saxena S S, Smith R and Skipper N 2005 Superconductivity in the intercalated graphite compounds $C_6Yb$ and $C_6Ca$ *Nat. Phys.* **1** 39–41
[56] Emery N, Herold C, d'Astuto M, Garcia V, Bellin C, Mareche J F, Lagrange P and Loupias G 2005 Superconductivity of bulk $CaC_6$ *Phys. Rev. Lett.* **95** 087003
[57] Csanyi G, Pickard C J, Simons B D and Needs R J 2007 Graphite intercalation compounds under pressure: a first-principles density functional theory study *Phys. Rev.* B **75** 085432
[58] Yin H C *et al* 2019 Stone–Wales graphene: a two-dimensional carbon semimetal with magic stability *Phys. Rev.* B **99** 041405(R)
[59] Deza M, Fowler P W, Shtogrin M and Vietze K 2000 Pentaheptite modifications of the graphite sheet *J. Chem. Inf. Comput. Sci.* **40** 1325–32
[60] He L, Guo S, Lei J, Sha Z and Liu Z 2014 The effect of Stone–Thrower–Wales defects on mechanical properties of graphene sheets—a molecular dynamics study *Carbon* **75** 124–32
[61] Kotakoski J, Krasheninnikov A V, Kaiser U and Meyer J C 2011 From point defects in graphene to two-dimensional amorphous carbon *Phys. Rev. Lett.* **106** 105505
[62] Shirodkar S N and Waghmare U V 2012 Electronic and vibrational signatures of Stone–Wales defects in graphene: first-principles analysis *Phys. Rev.* B **86** 165401
[63] Bets K V and Yakobson B I 2009 Spontaneous twist and intrinsic instabilities of pristine graphene nanoribbons *Nano Res.* **2** 161–6
[64] Shenoy V B, Reddy C D, Ramasubramaniam A and Zhang Y W 2008 Edge-stress-induced warping of graphene sheets and nanoribbons *Phys. Rev. Lett.* **101** 245501
[65] Sen D, Das B K, Saha S, Roy R, Mitra A and Chattopadhyay K K 2019 $sp^3$ bonded two-dimensional allotrope of carbon: a first-principles prediction (Carbon)146 pp 430–7
[66] Allinger N L J 1977 Conformational analysis. 130. MM2. A hydrocarbon force field utilizing V1 and V2 torsional terms *J. Am. Chem. Soc.* **99** 8127–34
[67] Stewart J J P 1990 MOPAC: a semiempirical molecular orbital program *J. Comput. Aided Mol. Des.* **4** 1–103
[68] Becke A D 1988 Density-functional exchange-energy approximation with correct asymptotic behavior *Phys. Rev.* A **38** 3098–100
[69] Lee C, Yang W and Parr R G 1988 Development of the Colle–Salvetti correlation-energy formula into a functional of the electron density *Phys. Rev.* B **37** 785–9
[70] Hehre W J, Random L, Schleyer P v R and Pople J A 1986 *Ab Initio Molecular Orbital Theory* (New York: Wiley)
[71] Rassolov V A, Pople J A, Ratner M A and Windus T L 1998 *J. Chem. Phys.* **109** 1223–9
[72] Schmidt M W *et al* 1993 General atomic and molecular electronic structure system *J. Comput. Chem.* **14** 1347–63
[73] Grimme S 2006 Semiempirical GGA-type density functional constructed with a long-range dispersion correction *J. Comput. Chem.* **27** 1787–99
[74] Kresse G and Furthmüller J 1996 Efficient iterative schemes for *ab initio* total-energy calculations using a plane-wave basis set *Phys. Rev.* B **54** 11169–86
[75] Kresse G and Hafner J 1993 *Ab initio* molecular dynamics for liquid metals *Phys. Rev.* B **47** 558–61
[76] Blöchl P E 1994 Projector augmented-wave method *Phys. Rev.* B **50** 17953–79
[77] Perdew J P, Burke K and Ernzerhof M 1996 *Phys. Rev. Lett.* **77** 3865
Perdew J P, Burke K and Ernzerhof M 1997 *Phys. Rev. Lett.* **78** 1396 (erratum)
[78] Monkhorst H J and Pack J D 1976 Special points for Brillouin-zone integrations *Phys. Rev.* B **13** 5188–92
[79] Momma K and Izumi F 2011 VESTA 3 for three-dimensional visualization of crystal, volumetric and morphology data *J. Appl. Crystallogr.* **44** 1272–6
[80] Nosé S 1984 A unified formulation of the constant temperature molecular dynamics methods *J. Chem. Phys.* **81** 511–9
[81] Hoover W G 1985 Canonical dynamics: equilibrium phase-space distributions *Phys. Rev.* A **31** 1695–7
[82] Avramov P V, Fedorov D G, Sorokin P B, Sakai S, Entani S, Ohtomo M, Matsumoto Y and Naramoto H 2012 Intrinsic edge asymmetry in narrow zigzag hexagonal heteroatomic nanoribbons causes their subtle uniform curvature *J. Phys. Chem. Lett.* **3** 2003–8
[83] Goldberg M 1935 The isoperimetric problem of polyhedra *Tohoku Math. J. First Series* **40** 226–36
[84] Goldberg M 1937 A class of multi-symmetric polyhedra *Tohoku Math. J. First Series* **43** 104–8
[85] Schwarz H A 1890 *Gesammeite Mathematische Abhandlungen* (Berlin: Springer)
[86] Schoen A H 1969 Quasi-regular saddle polyhedral *Not. Am. Math. Soc.* **16** 97–8
[87] Nitsche J C C 1985 *Vorlesüngen über Minimalflächen* (Berlin: Springer)
[88] Mackay A L and Terrones H 1991 Diamond from graphite *Nature* **352** 762
[89] Lenosky T, Czonze X, Teter M P and Elser V 1992 Energetics of negatively curved graphytic carbon *Nature* **355** 333–5
[90] Vanderbilt D and Tersoff J 1992 Negative-curvature fullerene analog of $C_{60}$ *Phys. Rev. Lett.* **68** 511–3
[91] Terrones H and Terrones M 1997 Quasiperiodic icosahedral graphite sheets and high-genus fullerenes with nonpositive Gaussian curvature *Phys. Rev.* B **55** 9969–74

[92] Terrones H and Terrones M 1998 Fullerenes and nanotubes with non-positive Gaussian curvature *Carbon* **36** 725–30
[93] Barth W E and Lawton R G 1971 The synthesis of corannulene *J. Am. Chem. Soc.* **93** 1730–45
[94] Shen M, Ignatyev I S, Xie Y and Schaefer H F III 1993 [7]Circulene: a remarkably floppy polycyclic aromatic $C_{28}H_{14}$ isomer *J. Phys. Chem.* **97** 3212–6
[95] Yamamoto K, Harada T, Nakazaki M, Nakao T, Kai Y, Harada S and Kasai N 1983 Synthesis and characterization of [7]circulene *J. Am. Chem. Soc.* **105** 7171
[96] Martin J M L, Franqois J P and Gijbels R 1991 A critical comparison of MINDO/3, MNDO, AM1, and PM3 for a model problem: carbon clusters C2–C10. An ad hoc reparametrization of MNDO well suited for the accurate prediction of their spectroscopic constants *J. Comput. Chem.* **12** 52–70
[97] Wood E 1964 *Bell System Technical Journal* 43 (Hoboken, NJ: Wiley) pp 541–59
[98] Milošević I, Popović Z, Volonakis G, Logothetidis S and Damnjanović M 2007 Electromechanical switch based on pentaheptite nanotubes *Phys. Rev.* B **76** 115406
[99] Damnjanović M, Popović Z, Volonakis G, Logothetidis S and Milošević I 2009 On the pentaheptite nanotubes *Mater. Manuf. Process.* **24** 1124–6
[100] Cornell W D *et al* 1995 A second generation force field for the simulation of proteins, nucleic acids, and organic molecules *J. Am. Chem. Soc.* **117** 5179–97
[101] Larach S and Shrader R E 1956 Electroluminescence from boron nitride *Phys. Rev.* **102** 582
[102] Elias D C *et al* 2009 Control of graphene's properties by reversible hydrogenation: evidence for graphane *Science* **323** 610–3
[103] Ketabi N, Tolhurst T M, Leedahl B, Liu H, Li Y and Moewes A 2017 How functional groups change the electronic structure of graphdiyne: theory and experiment *Carbon* **123** 1–6
[104] Jia Z, Li Y, Zuo Z, Liu H, Huang C and Li Y 2017 Synthesis and properties of 2D carbon—graphdiyne *Acc. Chem. Res.* **50** 2470–8
[105] Hu Y *et al* 2022 Synthesis of γ-graphyne using dynamic covalent chemistry *Nat. Synth.* **1** 449–54
[106] Liu X *et al* 2022 Constructing two-dimensional holey graphyne with unusual annulative π-extension *Matter* **5** 2306–18
[107] Li Y, Zhang J, Wang Q, Jin Y, Huang D, Cui Q and Zou G 2010 Nitrogen-rich carbon nitride hollow vessels: synthesis, characterization, and their properties *J. Phys. Chem.* B **114** 9429–34
[108] Lee J S, Wang X Q, Luo H M and Dai S 2010 *Adv. Mater.* **22** 1004
[109] Zhao W, Gebhardt J, Spath F, Gotterbarm K, Gleichweit C, Steinruck H-P, Gçrling A and Papp C 2015 Reversible hydrogenation of graphene on Ni(111)—synthesis of 'graphone' *Chem. Eur. J.* **21** 3347–58
[110] Prinz V Y, Seleznev V A, Gutakovsky A K, Chehovskiy A V, Preobrazhenskii V V, Putyato M A and Gavrilova T A 2000 Free-standing and overgrown InGaAs/GaAs nanotubes, nanohelices and their arrays *Physica* E **6** 828–31
[111] Kotakoski J, Meyer J C, Kurasch S, Santos-Cottin D, Kaiser U and Krasheninnikov A V 2011 Stone–Wales-type transformations in carbon nanostructures driven by electron irradiation *Phys. Rev.* B **83** 245420
[112] Atkins P and de Paula J 2006 *Physical Chemistry for the Life Sciences* (Oxford: Oxford University Press)
[113] Mishra S *et al* 2018 Tailoring bond topologies in open-shell graphene nanostructures *ACS Nano* **12** 11917–27
[114] Gerlich S, Eibenberger S, Tomandl M, Nimmrichter S, Hornberger K, Fagan P J, Tüxen J, Mayor M and Arndt M 2011 Quantum interference of large organic molecules *Nat. Commun.* **2** 263
[115] Meyer J C, Geim A K, Katsnelson M I, Novoselov K S, Booth T J and Roth S 2007 The structure of suspended graphene sheets *Nature* **446** 60–3
[116] Fasolino A, Los J H and Katsnelson M I 2007 Intrinsic ripples in graphene *Nat. Mater.* **6** 858
[117] Tapaszt L, Dumitrică T, Kim S J, Nemes-Incze P, Hwang C and Biró L P 2012 Breakdown of continuum mechanics for nanometre-wavelength rippling of graphene *Nat. Phys.* **8** 739
[118] Terrones M, Hsu W K, Hare J P, Kroto H W, Terrones H and Walton D R M 1996 Graphitic structures: from planar to spheres, toroids and helices *Phil. Trans. R. Soc.* A **354** 2025–54
[119] Kuzubov A A, Avramov P V, Tomilin F N and Ovchinnikov S G 2001 Theoretical investigation of toroidal forms of carbon and their endohedral derivatives with Li ions *Phys. Solid State* **43** 1982–8
[120] Schoen A H 1970 Infinite periodic minimal surfaces without self-intersections *NASA Technical Note, NASA, TN D-5541* NASA (Washington, DC)
[121] Hyde S T and O'Keeffe M 2017 At sixes and sevens, and eights, and nines: schwarzites $p^3$, $p = 7, 8, 9$ *Struct. Chem.* **28** 113–21
[122] Luo Y, Ren C, Xu Y, Yu J, Wang S and Sun M 2021 A first principles investigation on the structural, mechanical, electronic, and catalytic properties of biphenylene *Sci. Rep.* **11** 19008
[123] Senn P 1992 A simple quantum mechanical model that illustrates the Jahn–Teller effect *J. Chem. Educ.* **69** 819–21
[124] Balkova A and Bartlett R J 1994 A multireference coupled-cluster study of the ground state and lowest excited states of cyclobutadiene *J. Chem. Phys.* **101** 8972–87
[125] Johnson A W 1947 Cyclooctatetraene (Organic Chemistry) *Sci. Prog.* **35** 506–15
[126] Kaufman H S, Fankuchen I and Mark H 1948 Structure of cyclo-octatetraene *Nature* **161** 165
[127] Cava M P and Mitchell M J 1967 *Cyclobutadiene and Related Compounds Organic Chemistry* vol 10 (New York: Academic) pp 255–316
[128] Barton J W 1969 *Nonbenzenoid Aromatics* vol 1 ed J P Snyder (New York: Academic) pp 32–62
[129] Steiner E and Fowler P W 1996 Ring currents in aromatic hydrocarbons *Int. J. Quantum Chem.* **60** 609–16
[130] Hocquet A and Langgård M 1998 An evaluation of the MM+ force field *J. Mol. Model.* **4** 94–112
[131] Barone V 1994 Inclusion of Hartree–Fock exchange in the density functional approach. Benchmark computations for diatomic molecules containing H, B, C, N, O, and F atoms *Chem. Phys. Lett.* **226** 392–8
[132] Petersson G A, Malick D K, Wilson W G, Ochterski J W, Montgomery J A and Frisch M J 1998 Calibration and comparison of the Gaussian-2, complete basis set, and density functional methods for computational thermochemistry *J. Chem. Phys.* **109** 10570–9
[133] Wang X J, Wong L H, Hu L H, Chan C Y, Su Z and Chen G H 2004 Improving the accuracy of density-functional theory calculation: the statistical correction approach *J. Phys. Chem.* A **108** 8514–25
[134] Becke A D 1992 Density-functional thermochemistry: II. The effect of the Perdew–Wang generalized-gradient correlation correction *J. Chem. Phys.* **97** 9173–7
[135] Becke A D 1992 Density-functional thermochemistry: I. The effect of the exchange-only gradient correction *J. Chem. Phys.* **96** 2155–60

[136] Avramov P V, Fedorov D G, Irle S, Kuzubov A A and Morokuma K 2009 Strong electron correlations determine energetic stability and electronic properties of Er-doped Goldberg-type silicon quantum dots *J. Phys. Chem.* C **113** 15964–8
[137] McGuire J H 1997 *Electron Correlation Dynamics in Atomic Collisions* (Cambridge: Cambridge University Press)
[138] Cramer C J 2002 *Essentials of Computational Chemistry* (New York: Wiley)
[139] Levine I N 2014 *Quantum Chemistry* 7th edn (London: Pearson)
[140] Born M and Fock V A 1928 Beweis des Adiabatensatzes *Z. Phys.* **51** 165–80
[141] Il'in N, Aristova A and Lychkovskiy O 2021 Adiabatic theorem for closed quantum systems initialized at finite temperature *Phys. Rev.* A **104** L030202
[142] Malyi O I, Sopiha K V and Persson C 2019 Energy, phonon, and dynamic stability criteria of two-dimensional materials *ACS Appl. Mater. Interfaces* **11** 24876–84

*Topological and quantum stability of low-dimensional crystalline lattices with multiple nonequivalent sublattices*

*Supplementary Information 1* (SI1 Section)

Pavel V. Avramov[a,*], and Artem V. Kuklin[b]

[a] Department of Chemistry, College of Natural Sciences, Kyungpook National University, 80 Daehak-ro, Buk-gu, Daegu, 41566, South Korea

[b] Department of Physics and Astronomy, Uppsala University, Box 516, SE-751 20 Uppsala, Sweden

*E-mail: paul.veniaminovich@knu.ac.kr

**Table SI1-1**. Imaginary vibrational frequencies of planar $C_{40}H_{16}$ carbon flake calculated at the PM3 level of theory, in *cm$^{-1}$*.

| Closed-shell singlet | Open-shell singlet | Opened shell triplet |
|---|---|---|
| 1. 122.43*i* | 1. 126.76*i* | 1. 132.69*i* |
| 2. 61.09*i* | 2. 61.31*i* | 2. 68.21*i* |
| 3. 52.02*i* | 3. 52.14*i* | 3. 59.29*i* |
| 4. 30.56*i* | 4. 33.38*i* | 4. 35.37*i* |

**Table SI1-2**. PM3 imaginary vibrational frequencies of planar $C_{62}H_{20}$ (in *cm$^{-1}$*).

| Closed-shell singlet imaginary frequencies | Opened-shell singlet imaginary frequencies |
|---|---|
| 1. 141.08*i* | 1. 144.05*i* |
| 2. 112.02*i* | 2. 113.60*i* |
| 3. 30.13*i* | 3. 59.19*i* |
| 4. 24.29*i* | 4. 39.47*i* |
| 5. 23.83*i* | 5. 32.38*i* |
|  | 6. 22.65*i* |
|  | 7. 2.47*i* |

**Table SI1-3**. Vibrational frequencies of wave and planar $nA\times 1B$ phagraphene nanoribbons calculated at model potential, and PM3 levels of theory.

| | | | |
|---|---|---|---|
| Wave conformers | | | |
| Model potential | | | |
| Nanoribbon $phC(0,1)$ $nA\times 1B$ | Number of imaginary modes | Range of imaginary ($i$) modes, $cm^{-1}$ | Smallest real mode |
| $phC(0,1)$ $1A\times 1B$ | 0 | - | 77.91 |
| $phC(0,1)$ $2A\times 1B$ | 0 | - | 40.05 |
| $phC(0,1)$ $3A\times 1B$ | 0 | - | 20.80 |
| $phC(0,1)$ $4A\times 1B$ | 0 | - | 7.00 |
| $phC(0,1)$ $5A\times 1B$ | 0 | - | 4.88 |
| $phC(0,1)$ $6A\times 1B$ | 0 | - | 2.37 |
| $phC(0,1)$ $7A\times 1B$ | 0 | - | 1.70 |
| $phC(0,1)$ $8A\times 1B$ | 0 | - | 1.62 |
| $phC(0,1)$ $9A\times 1B$ | 0 | - | 1.61 |
| $phC(0,1)$ $10A\times 1B$ | 0 | - | 1.56 |
| $phC(0,1)$ $11A\times 1B$ | 0 | - | 1.90 |
| $phC(0,1)$ $12A\times 1B$ | 0 | - | 2.07 |
| $phC(0,1)$ $13A\times 1B$ | 1 | $4.74i$ (poor convergency) | 0.64 |
| $phC(0,1)$ $14A\times 1B$ | 0 | - | 1.65 |
| $phC(0,1)$ $15A\times 1B$ | 0 | - | 1.23 |
| $phC(0,1)$ $20A\times 1B$ | 2 | $12.01i$; $8.83i$ (poor convergency) | 0.90 |
| $phC(0,1)$ $25A\times 1B$ | 1 | $4.91i$ (poor convergency) | 1.26 |
| PM3 | | | |
| $phC(0,1)$ $1A\times 1B$ | 0 | - | 40.12 |
| $phC(0,1)$ $2A\times 1B$ | 0 | - | 20.31 |
| $phC(0,1)$ $3A\times 1B$ | 0 | - | 10.56 |
| $phC(0,1)$ $4A\times 1B$ | 0 | - | 6.16 |

| Nanoribbon phC(0,1) nA×1B | Number of imaginary modes | Range of imaginary modes, $cm^{-1}$ | Smallest real mode |
|---|---|---|---|
| phC(0,1) 5A×1B | 0 | - | 4.70 |
| phC(0,1) 6A×1B | 0 | - | 4.25 |
| phC(0,1) 7A×1B | 0 | - | 3.64 |
| phC(0,1) 8A×1B | 0 | - | 0.71* |
| phC(0,1) 9A×1B | 0 | - | 6.18 |
| phC(0,1) 10A×1B | 0 | - | 1.34* |
| phC(0,1) 11A×1B | 0 | - | 0.58* |
| phC(0,1) 12A×1B | 0 | - | 0.09* |
| phC(0,1) 13A×1B | 0 | - | 1.20* |
| phC(0,1) 14A×1B | 0 | - | 0.96* |
| phC(0,1) 15A×1B | 0 | - | 2.14* |
| phC(0,1) 16A×1B | 0 | - | 1.84* |
| phC(0,1) 17A×1B | 0 | - | 1.82* |
| Planar conformers ||||
| Model potential ||||
| Nanoribbon phC(0,1) nA×1B | Number of imaginary modes | Range of imaginary modes, $cm^{-1}$ | Smallest real mode |
| phC(0,1) 1A×1B | 1 | 68.50$i$ | 124.58 |
| phC(0,1) 2A×1B | 2 | 71.28$i$, 50.28$i$ | 55.12 |
| phC(0,1) 3A×1B | 4 | 73.08$i$ - 29.21$i$ | 63.13 |
| phC(0,1) 4A×1B | 5 | 73.83$i$ - 18.54$i$ | 48.27 |
| phC(0,1) 5A×1B | 7 | 74.24$i$ - 12.74$i$ | 41.05 |
| phC(0,1) 6A×1B | 8 | 74.49$i$ - 9.24$i$ | 34.79 |
| phC(0,1) 7A×1B | 10 | 74.66$i$ - 7.01$i$ | 30.23 |
| phC(0,1) 8A×1B | 11 | 74.79$i$ - 5.49$i$ | 24.16 |
| phC(0,1) 9A×1B | 12 | 74.88$i$ - 4.34$i$ | 4.55 |
| phC(0,1) 10A×1B | 14 | 74.95$i$ - 3.92$i$ | 16.17 |
| phC(0,1) 11A×1B | 15 | 75.13$i$ - 3.03$i$ | 13.57 |
| *phC*(0,1) *12A×1B* | 17 | 75.30*i* - 2.83*i* | 11.55 |
| *phC*(0,1) *13A×1B* | 18 | 75.40*i* - 4.16*i* | 9.95 |
| *phC*(0,1) *14A×1B* | 20 | 75.55*i* - 2.89*i* | 8.65 |
| *phC*(0,1) *15A×1B* | 21 | 75.58*i* - 2.81*i* | 7.64 |
| *phC*(0,1) *16A×1B* | 23 | 75.66*i* - 1.63*i* | 6.71 |
| *phC*(0,1) *17A×1B* | 24 | 75.65*i* - 1.34*i* | 5.99 |
| *phC*(0,1) *18A×1B* | 26 | 75.66*i* - 1.89*i* | 5.36 |
| *phC*(0,1) *19A×1B* | 27 | 75.75*i* - 1.09*i* | 4.84 |
| *phC*(0,1) *20A×1B* | 29 | 75.73*i* - 0.95*i* | 4.37 |
| *phC*(0,1) *21A×1B* | 30 | 75.77*i* - 2.39*i* | 3.98 |
| *phC*(0,1) *22A×1B* | 31 | 75.79*i* - 1.47*i* | 3.65 |
| *phC*(0,1) *23A×1B* | 33 | 75.78*i* - 0.80*i* | 3.47 |
| *phC*(0,1) *24A×1B* | 34 | 75.80*i* - 1.34*i* | 2.98 |
| *phC*(0,1) *25A×1B* | 36 | 75.83*i* - 1.26*i* | 2.83 |
| PM3 | | | |
| *phC*(0,1) *1A×1B* | 2 | 71.28*i*, 69.97*i* | 51.08 |
| *phC*(0,1) *2A×1B* | 4 | 111.28*i* - 45.43*i* | 38.44 |
| *phC*(0,1) *3A×1B* | 6 | 124.74*i* - 27.06*i* | 13.74 |
| *phC*(0,1) *4A×1B* | 9 | 129.97*i* - 16.22*i* | 32.29 |
| *phC*(0,1) *5A×1B* | 11 | 132.17*i* - 11.10*i* | 25.66 |
| *phC*(0,1) *6A×1B* | 14 | 133.31*i* - 8.04*i* | 21.12 |
| *phC*(0,1) *7A×1B* | 16 | 133.93*i* - 6.12*i* | 16.20 |
| *phC*(0,1) *8A×1B* | 18 | 134.27*i* - 4.76*i* | 12.84 |
| *phC*(0,1) *9A×1B* | 21 | 134.42*i* - 3.79*i* | 3.29 |
| *phC*(0,1) *10A×1B* | >20 | 134.96*i* | 3.21 |
| *phC*(0,1) *11A×1B*[**] | >20 | 134.82*i* | ~3 |
| *phC*(0,1) *12A×1B*[**] | >20 | 134.93*i* | ~3 |
| *phC*(0,1) *13A×1B*** | >20 | 135.01*i* | ˜3 |
| *phC*(0,1) *14A×1B*** | >20 | 135.5*i* | ˜3 |
| *phC*(0,1) *15A×1B*** | >20 | 135.16*i* | ˜3 |
| *phC*(0,1) *16A×1B*** | >20 | 135.23*i* | 3.55[*] |
| *phC*(0,1) *17A×1B*** | >20 | 135.26*i* | 3.13[*] |
| *phC*(0,1) *18A×1B*** | >20 | 135.28*i* | 2.82[*] |

[*] Low-frequency bending modes

[**] For which GAMESS code makes mistakes

**Table SI1-4.** Optimization protocols (Energies kcal/mol, gradients kcal/mol/Å, R.M.S. gradients kcal/mol/Å and screenshots at each optimization steps) for planar and curved conformers of some $n \times n$, $k \times 5$, and $5 \times l$ 2D biphenylene finite fragments calculated using *ab initio* B3LYP/6-31G* method. In screenshots, the optimized geometries are circled by red ovals, and minimal energies are circled by green ovals.

| 2D biphenylene flake | Planar conformer | Curved conformer |
|---|---|---|
| $7 \times 7$ $C_{294}H_{42}$ | 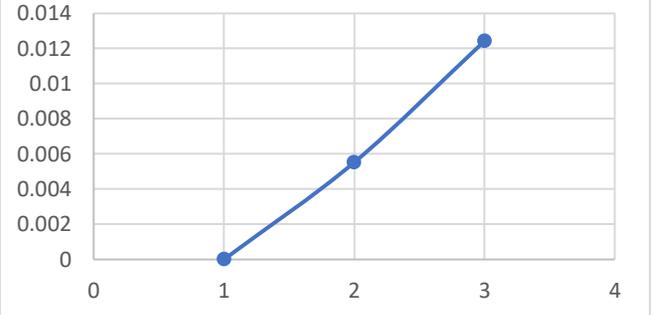 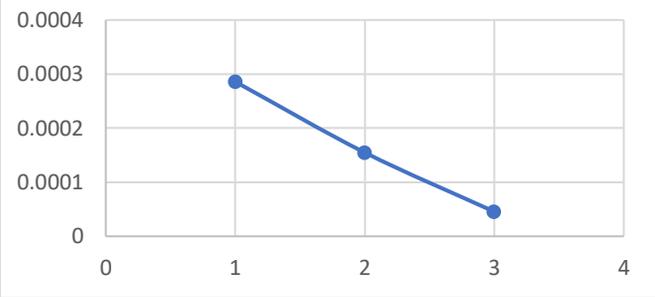 | 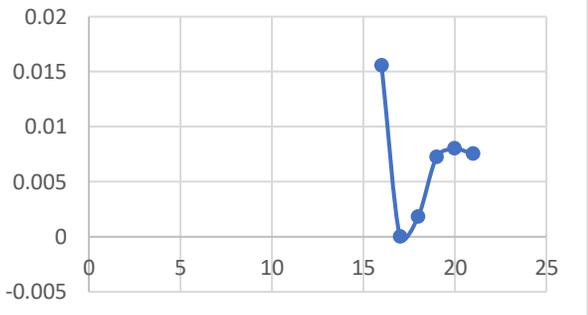 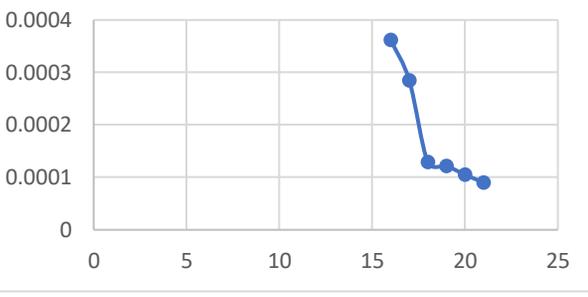 |
| | Planar conformer RMS Gradient, kcal/mol /A | Curved conformer RMS Gradient, kcal/mol/A |
| $8 \times 8$ $C_{384}H_{48}$ | Planar conformer Energy, kcal/mol | Curved conformer Energy, kcal/mol |

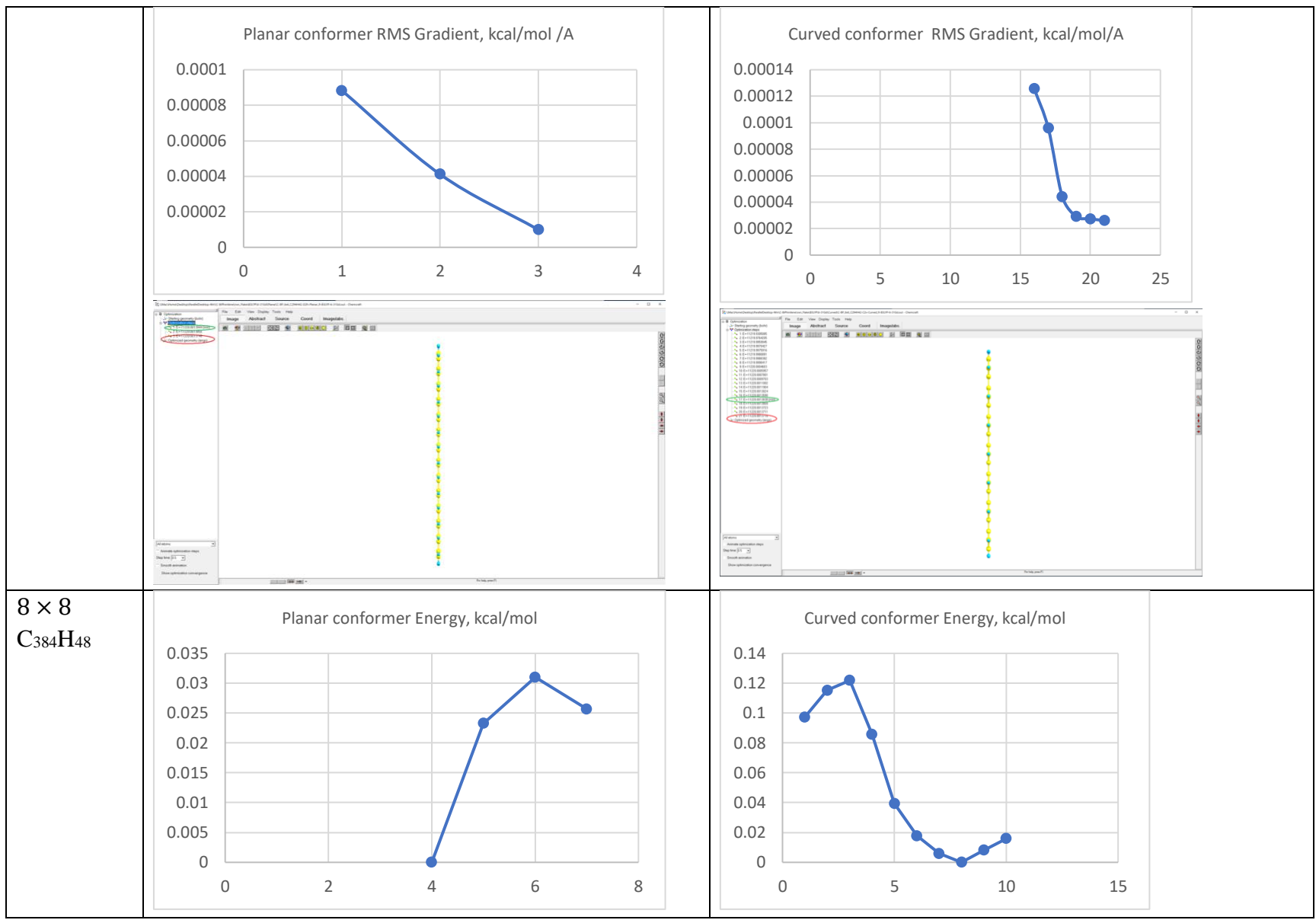

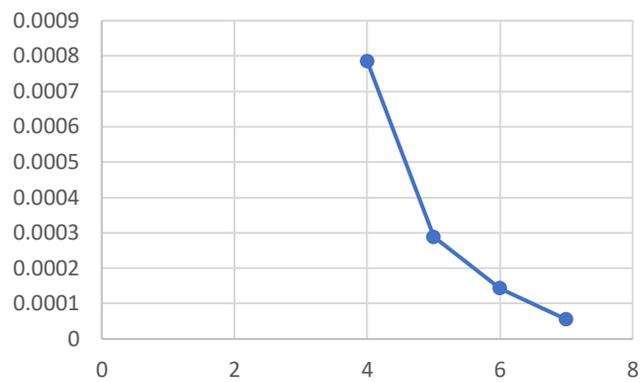
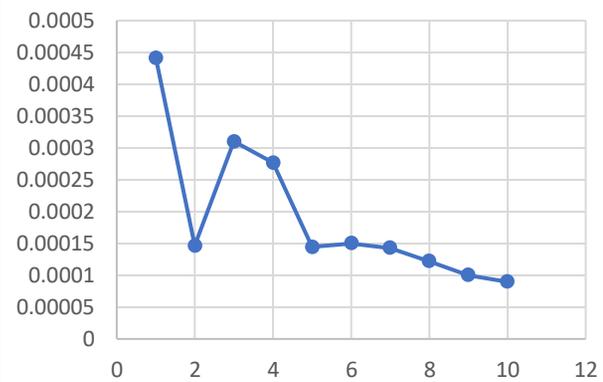
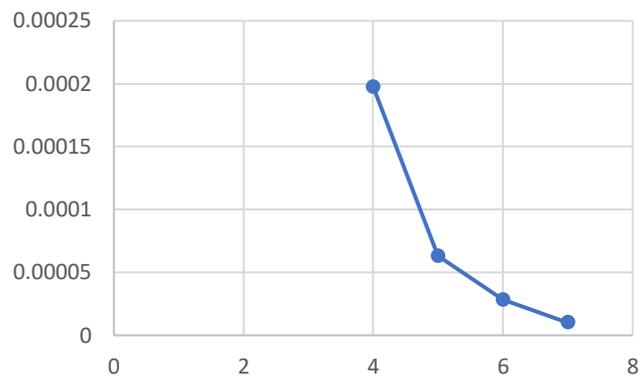
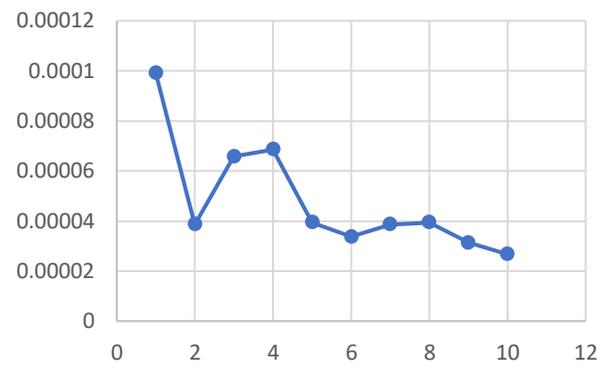

| | | |
|---|---|---|
| | 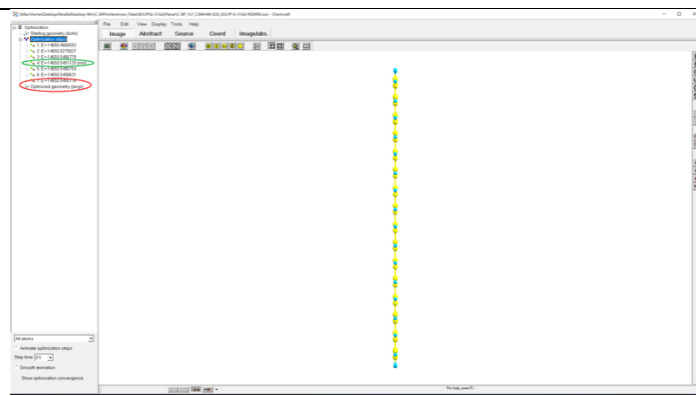 | 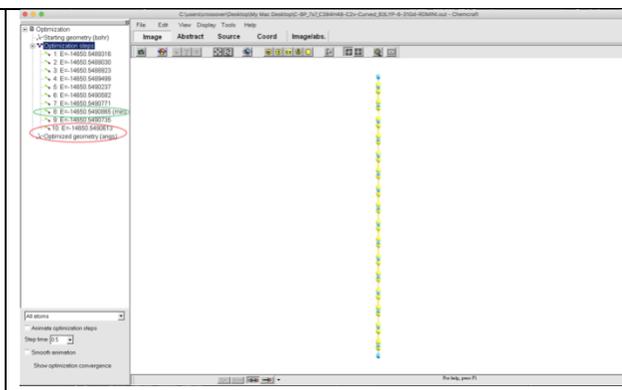 |
| $9 \times 9$ $C_{486}H_{54}$ | 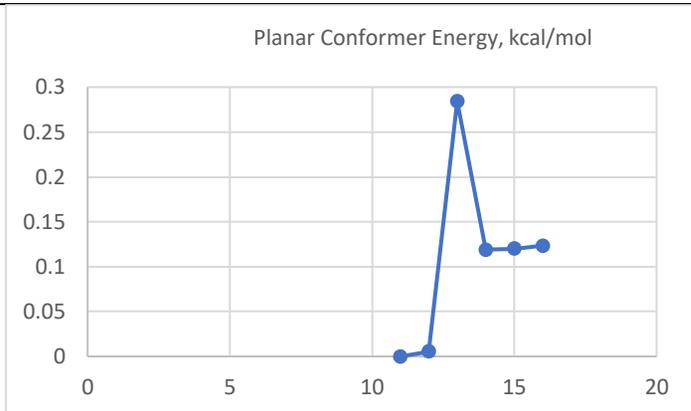<br>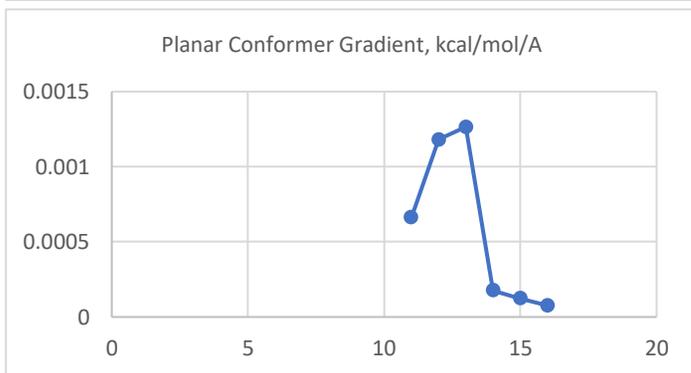 | -- |
| | 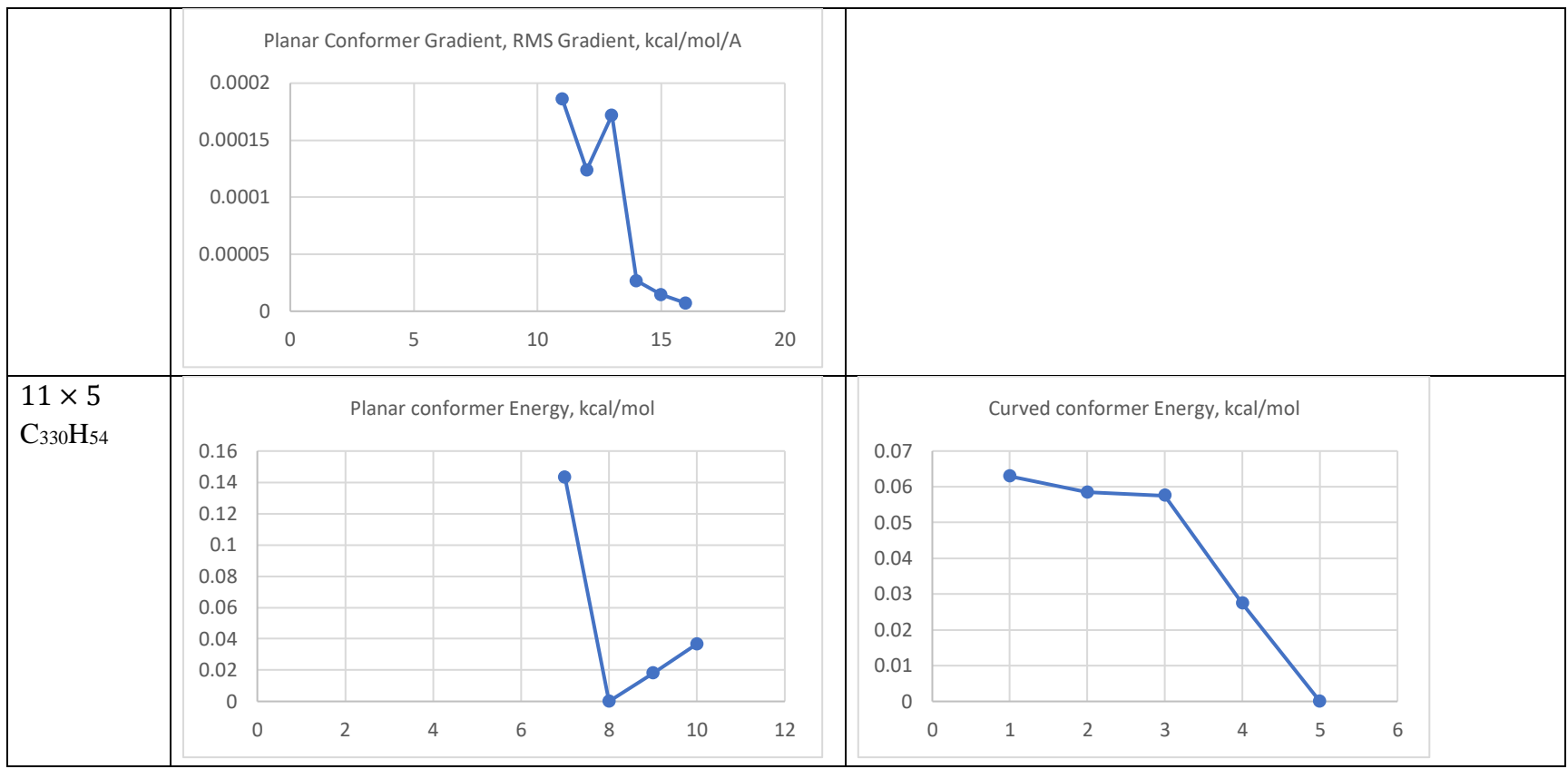 | |
| $11 \times 5$ $C_{330}H_{54}$ | | |

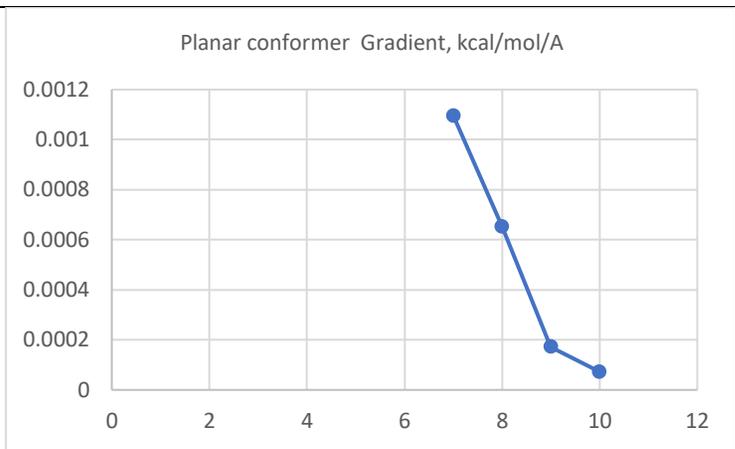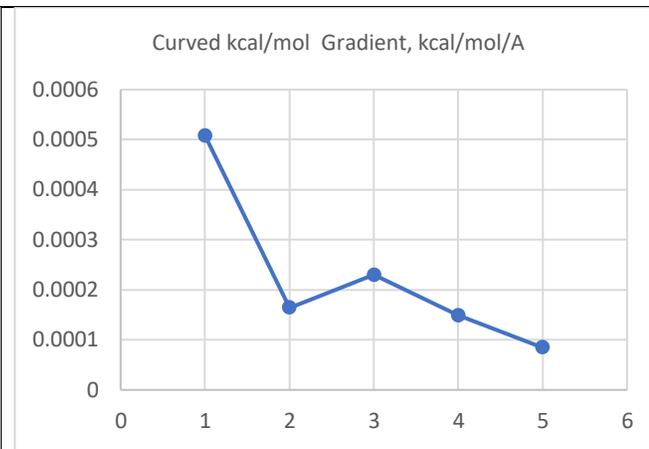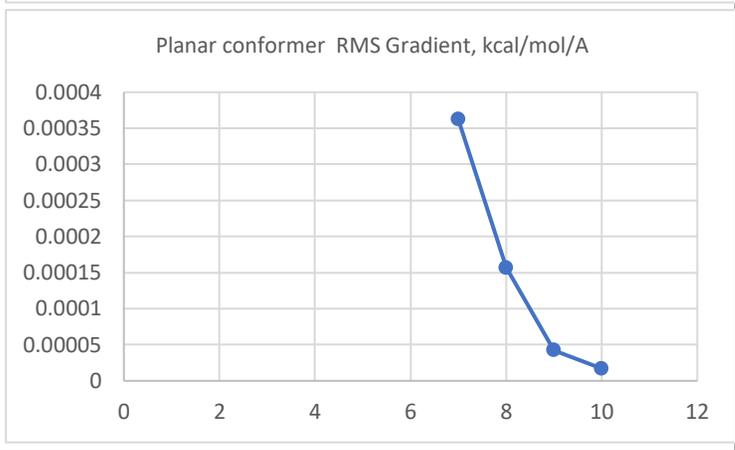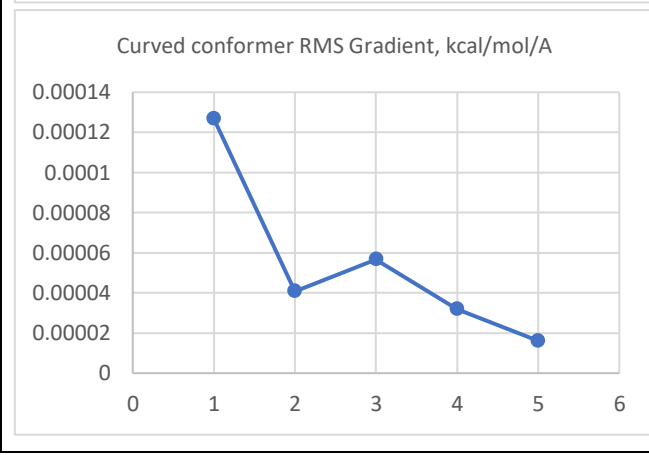

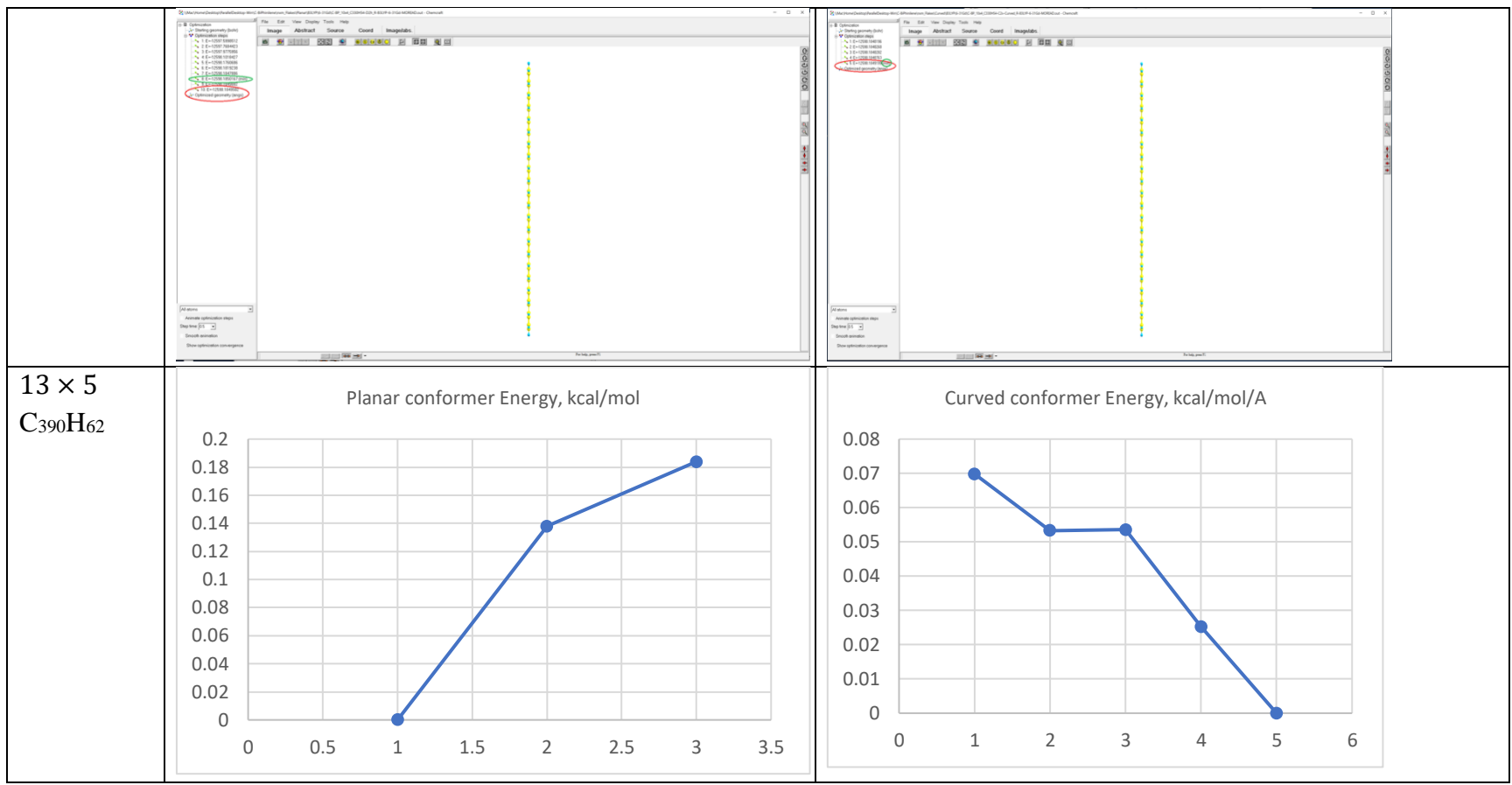

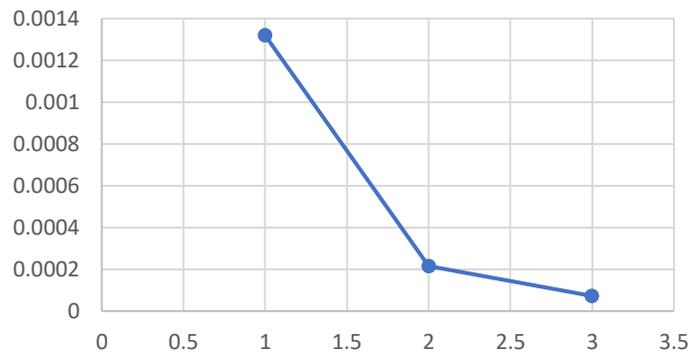
Planar conformer Gradient, kcal/mol/A

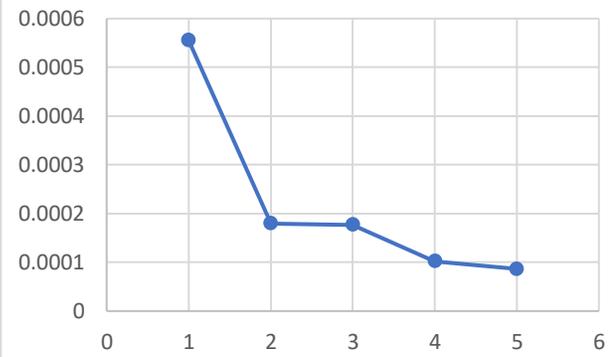
Curved conformer Gradient, kcal/mol/A

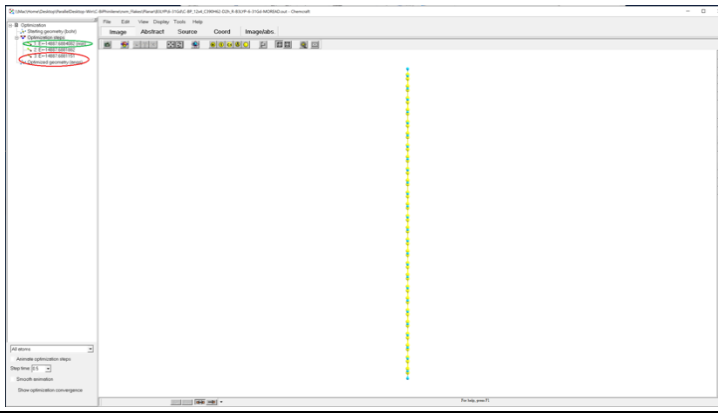

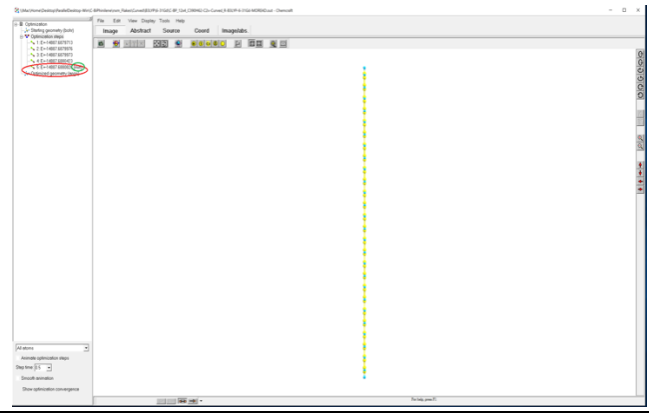

| 15 × 5<br>$C_{450}H_{70}$ | 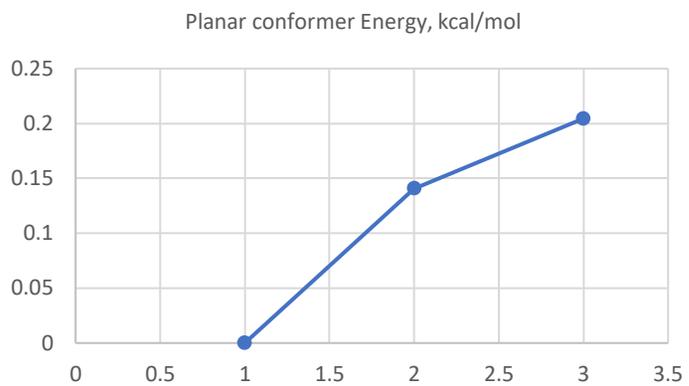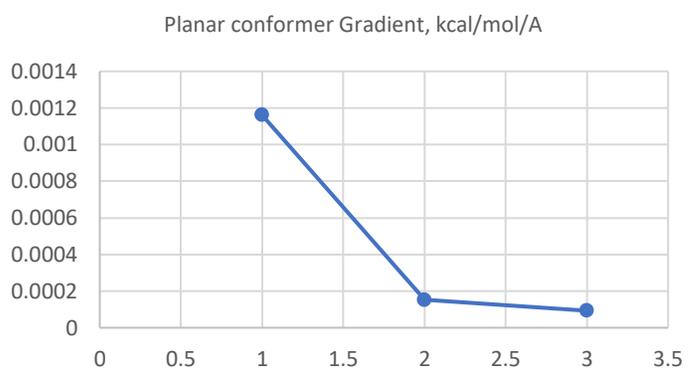 | 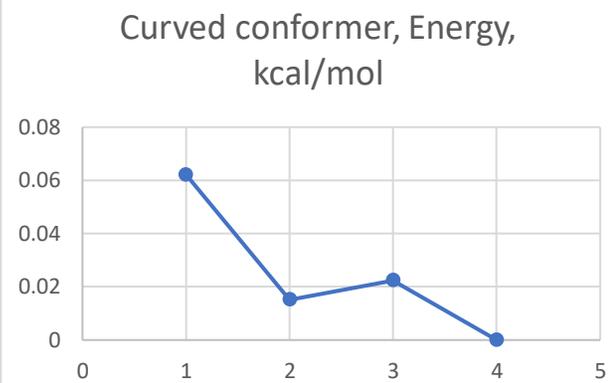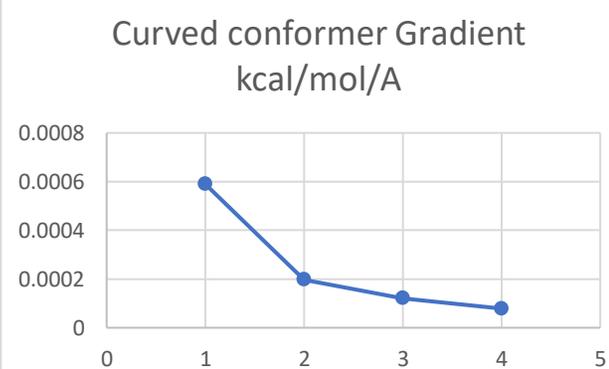 |
|---|---|---|
| | 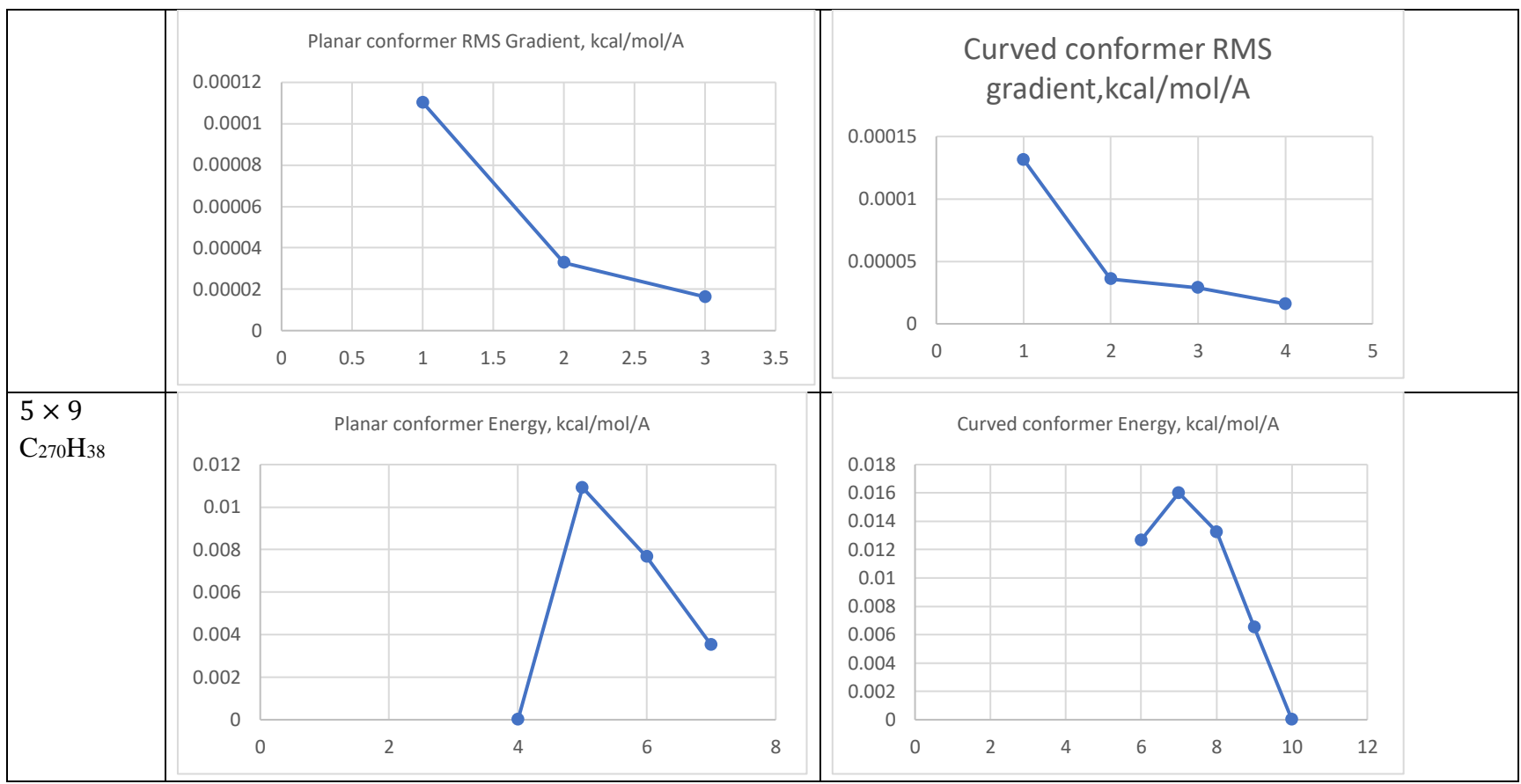 | |
| $5 \times 9$ $C_{270}H_{38}$ | | |

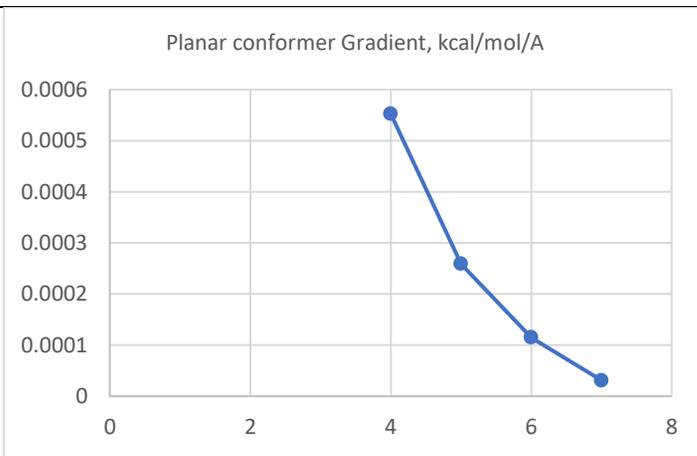
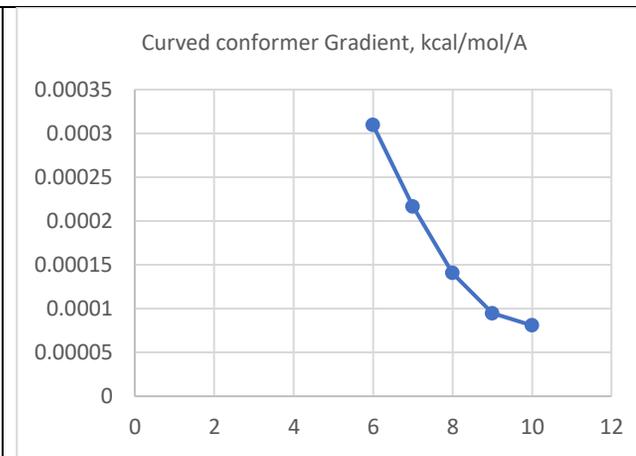
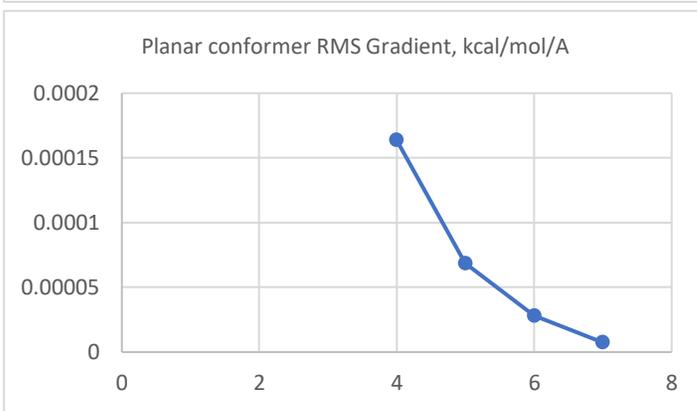
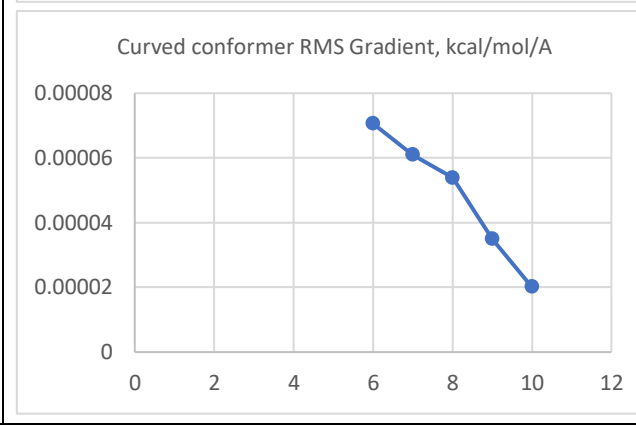

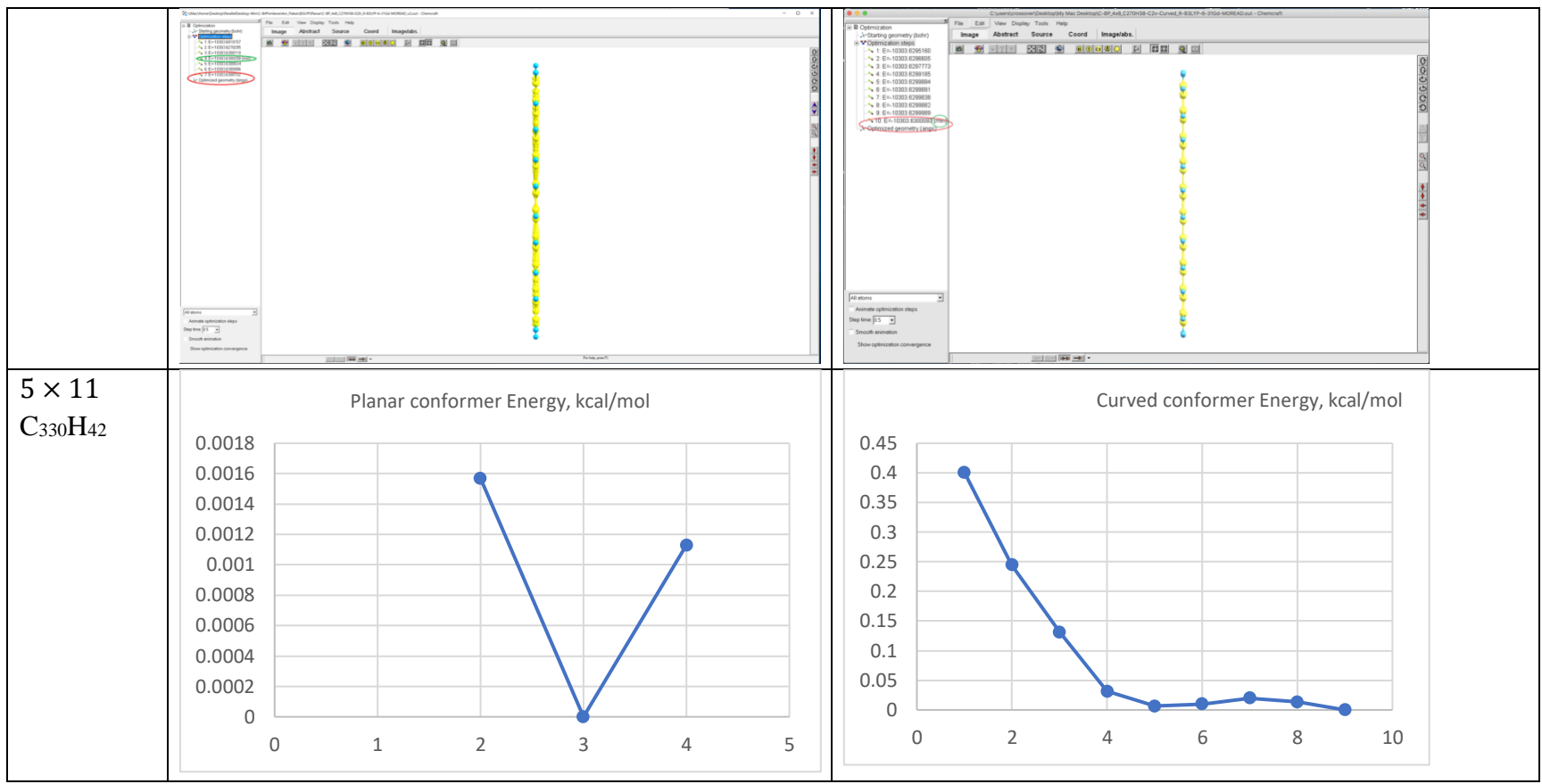
| | | |
|---|---|---|
| $5 \times 11$ $C_{330}H_{42}$ | Planar conformer Energy, kcal/mol | Curved conformer Energy, kcal/mol |

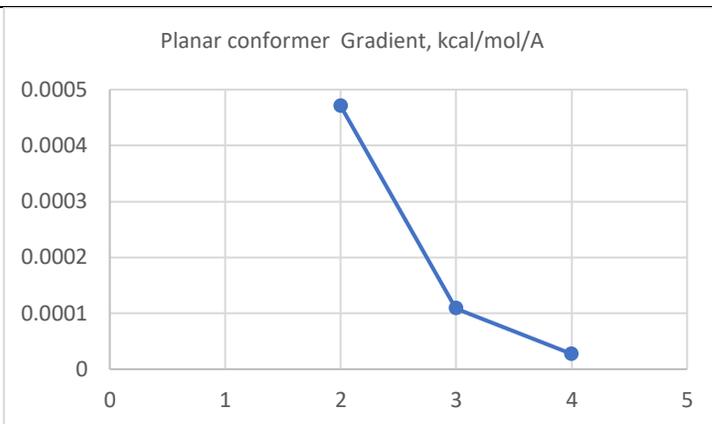
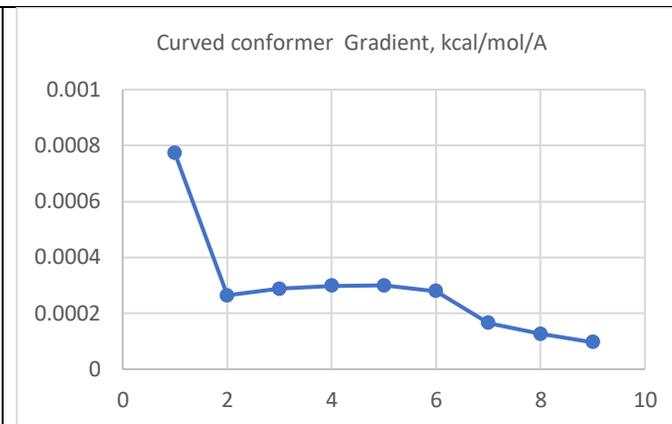
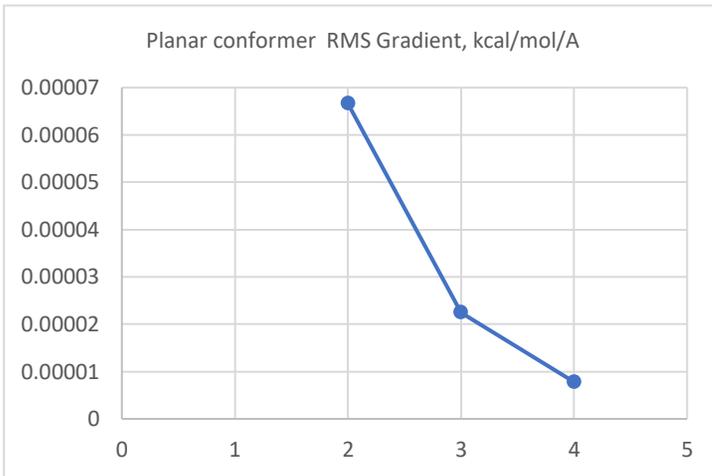
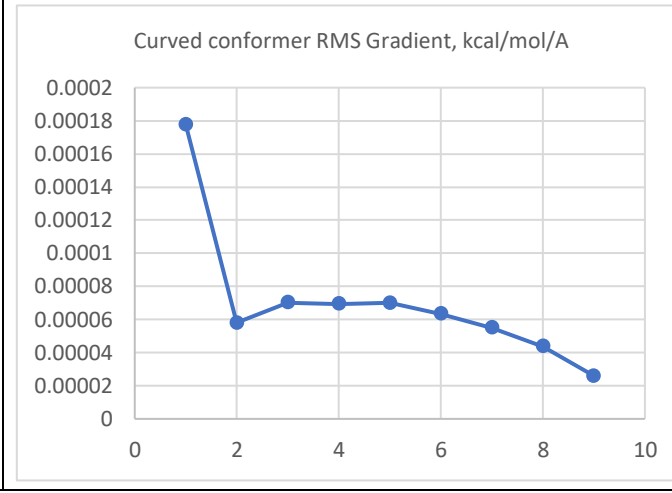

| | | |
|---|---|---|
| | 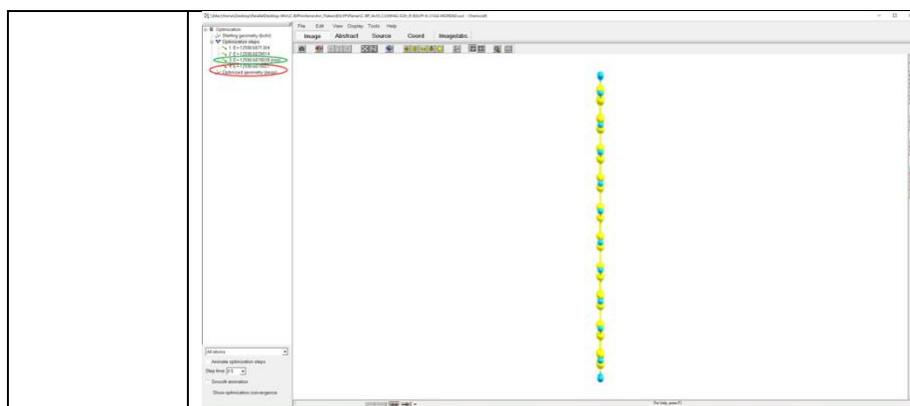 | 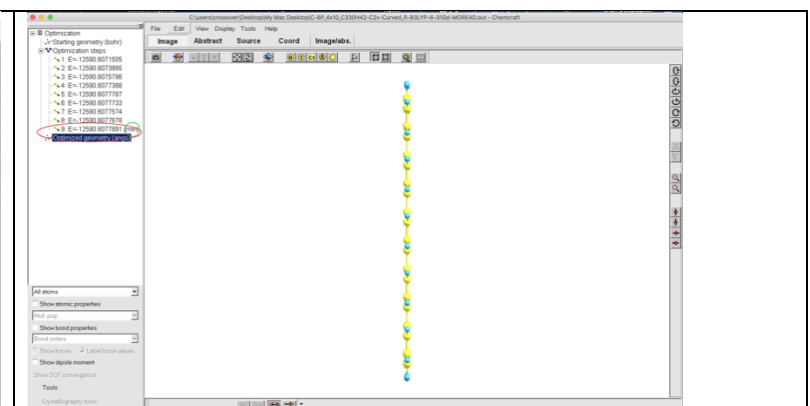 |
| $5 \times 12$ $C_{360}H_{44}$ | 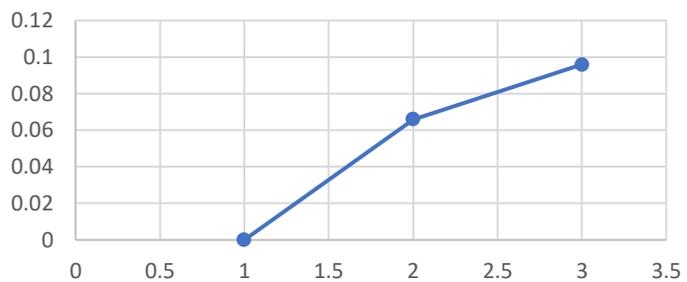 | 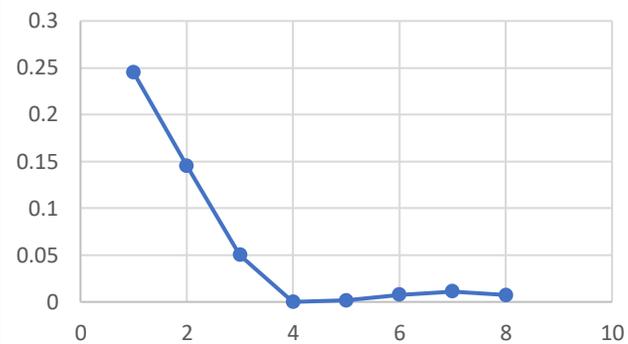 |

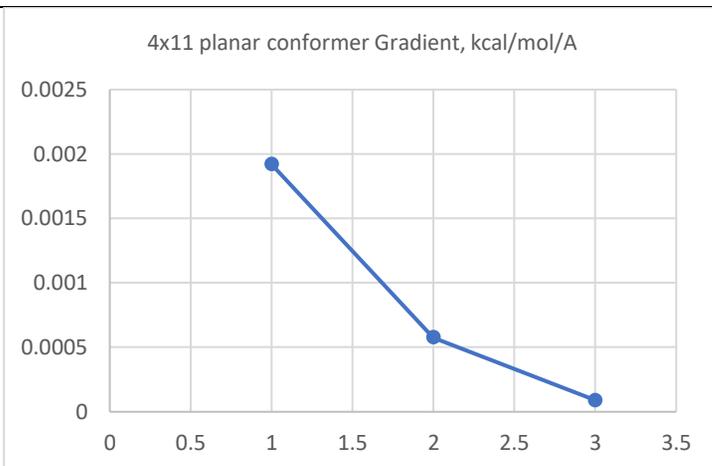
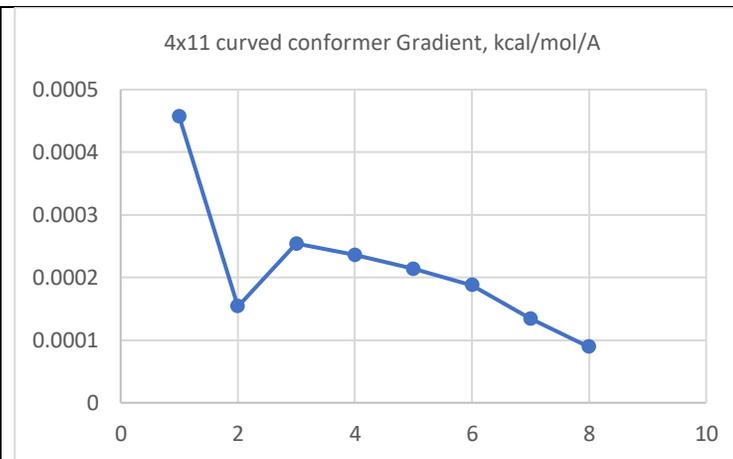
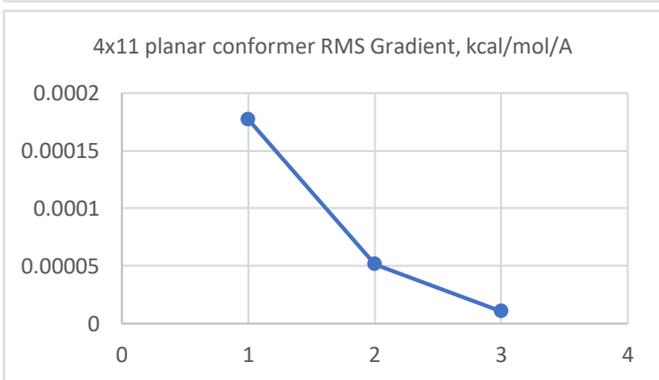
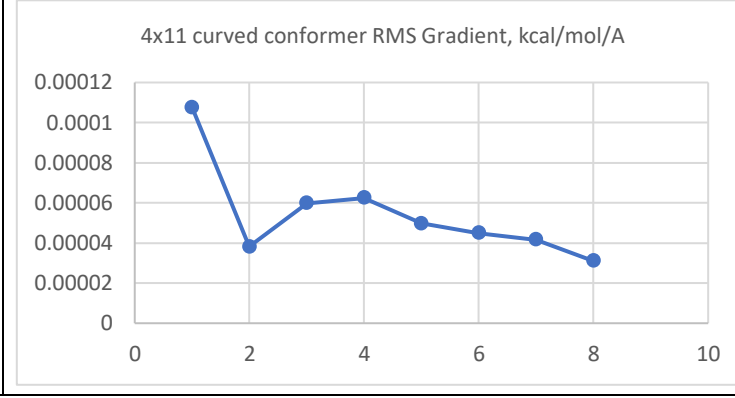

| | | |
|---|---|---|
| | 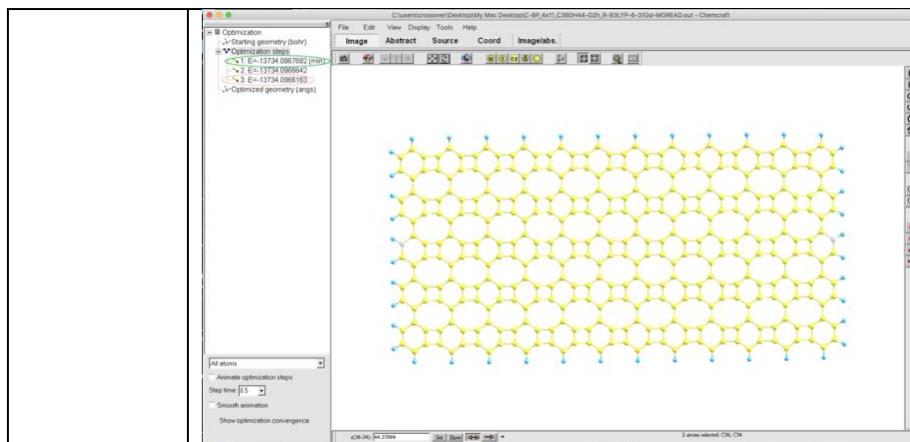 | 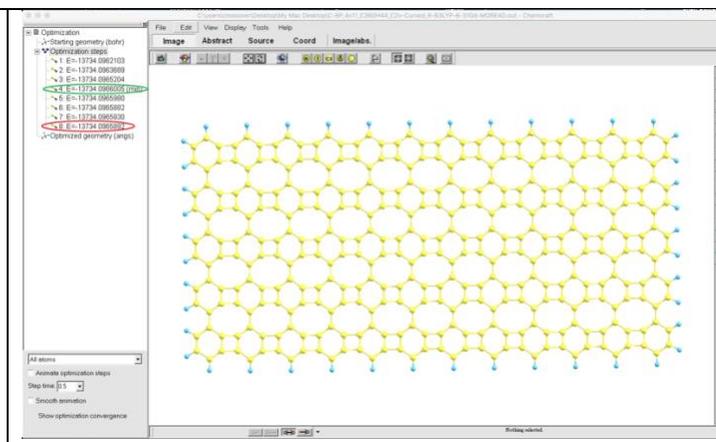 |
| 5 × 13<br>$C_{390}H_{46}$ | 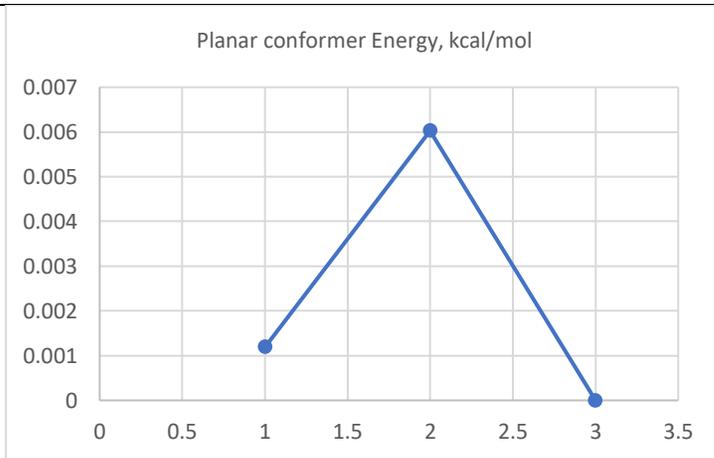 | 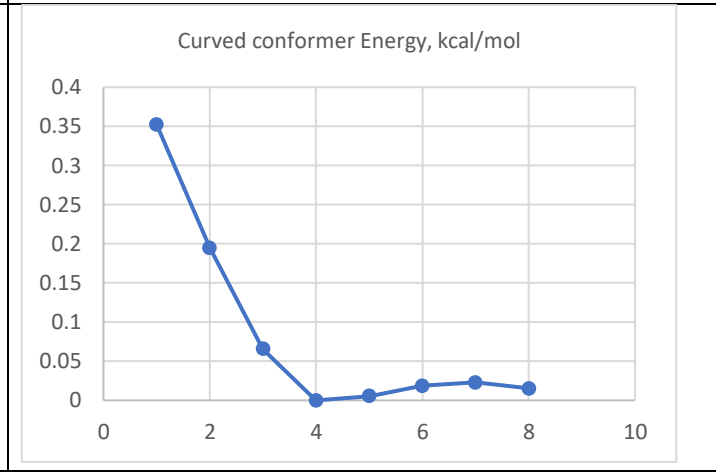 |

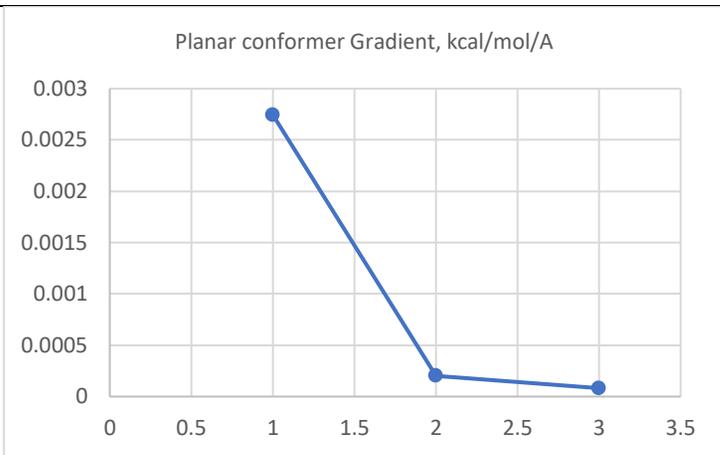
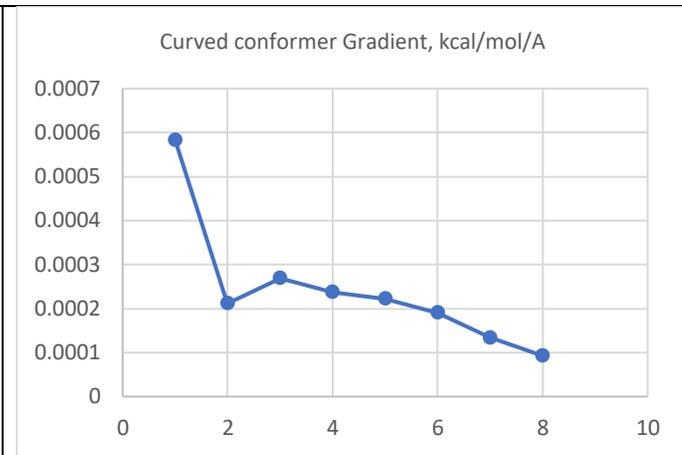
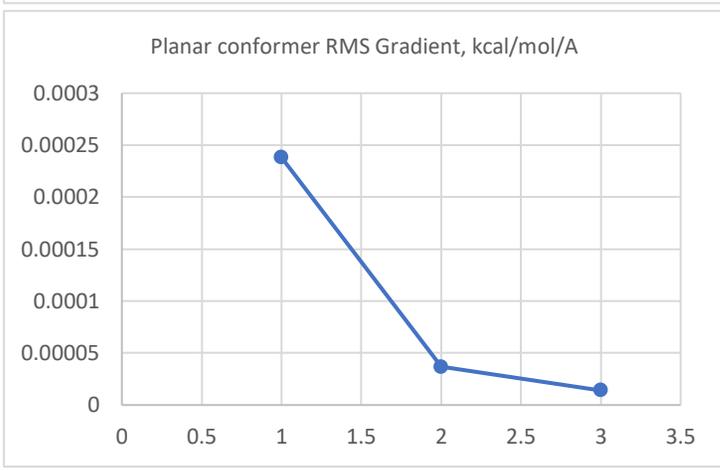
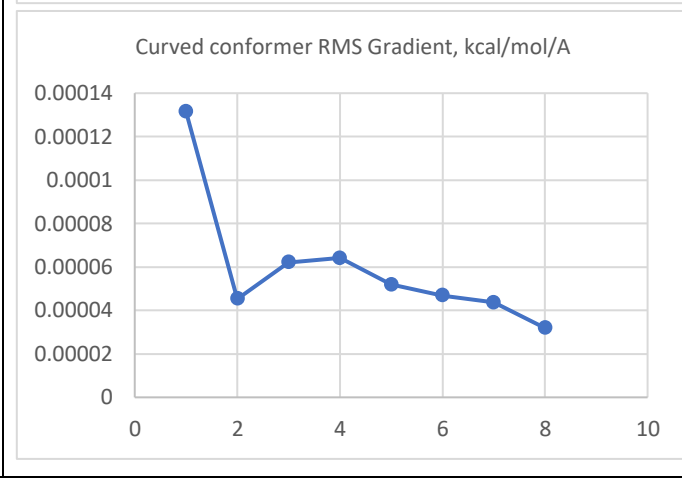

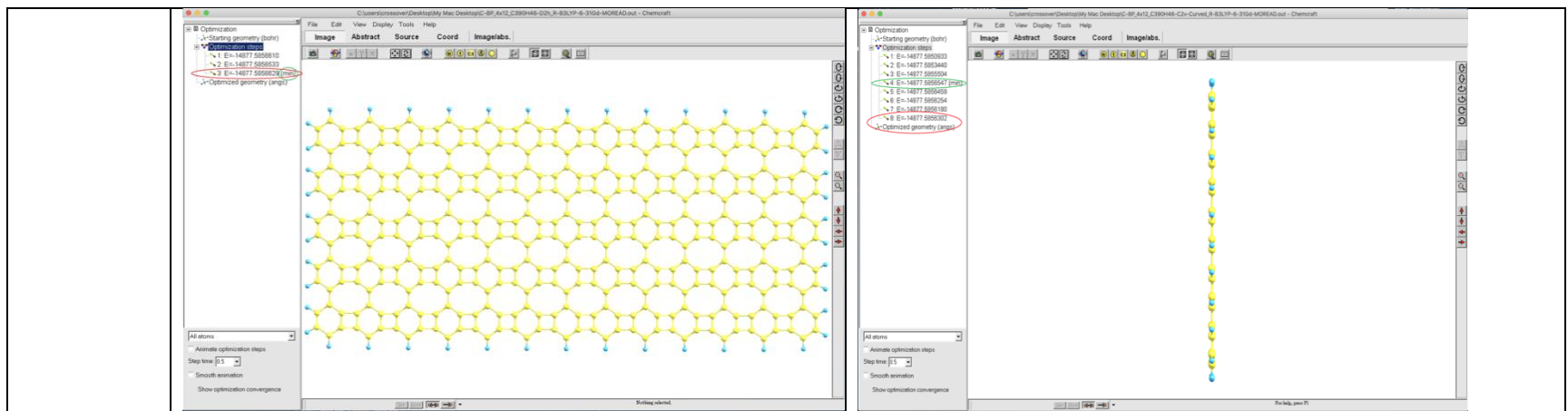

| $5 \times 13$ $C_{390}H_{46}$ | 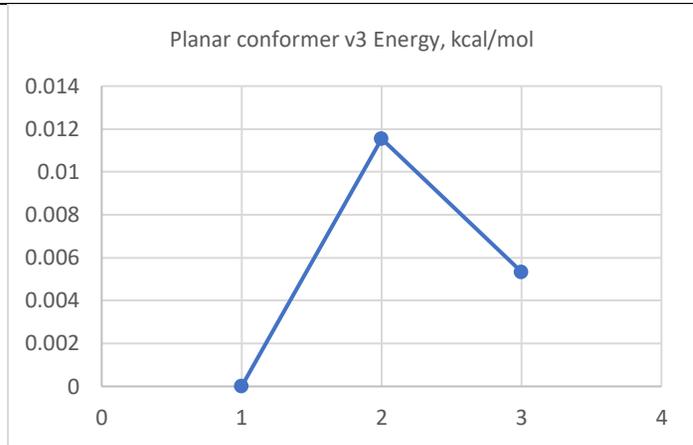 | |

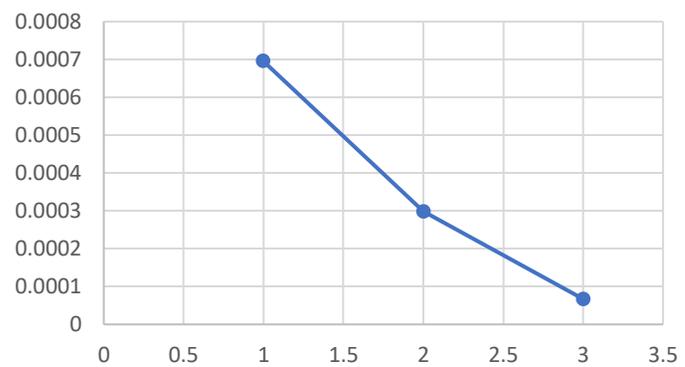

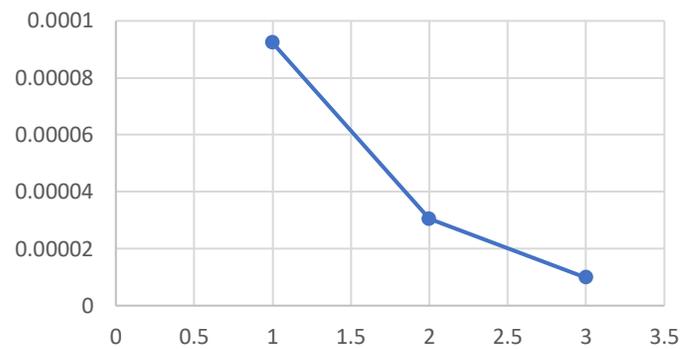

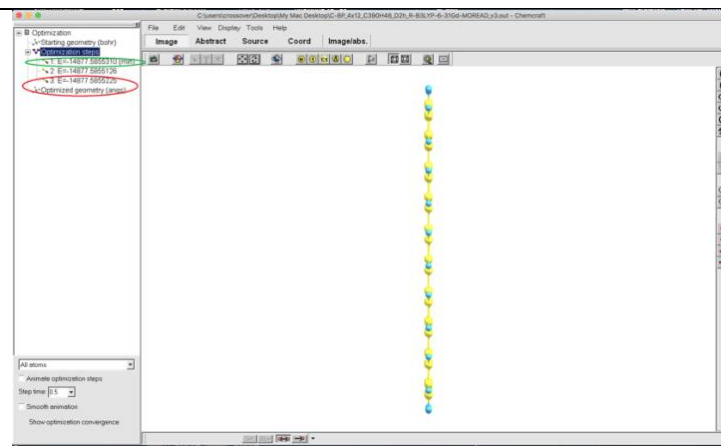

| $5 \times 14$ $C_{420}H_{48}$ | 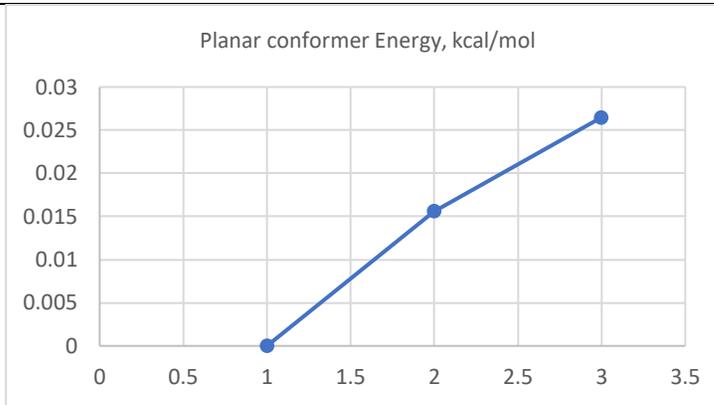 | 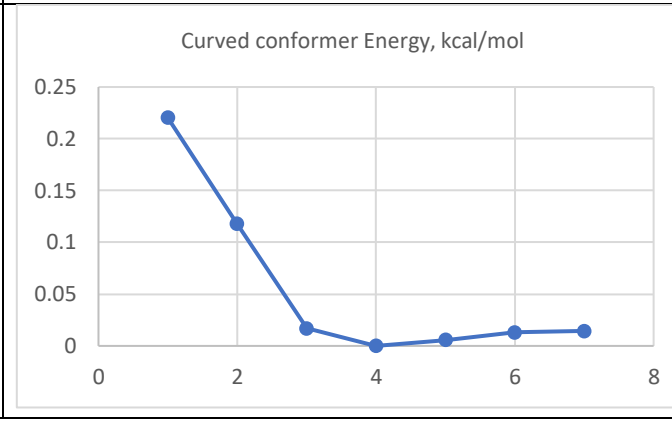 |

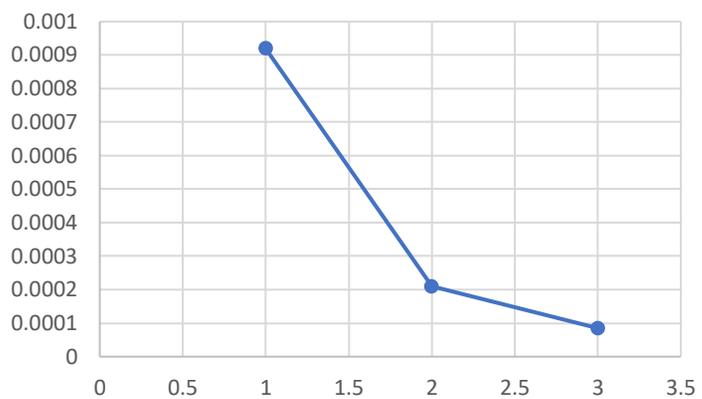
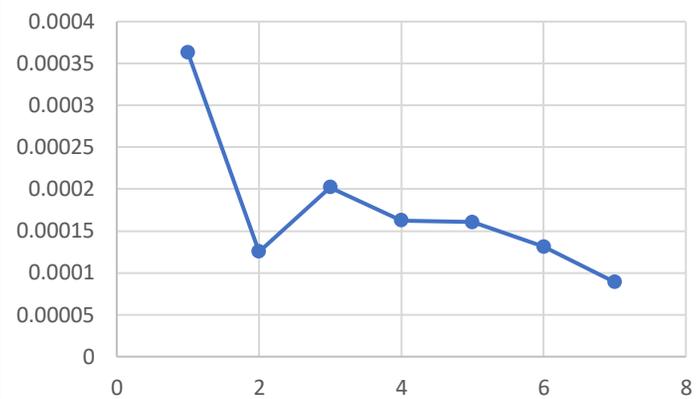
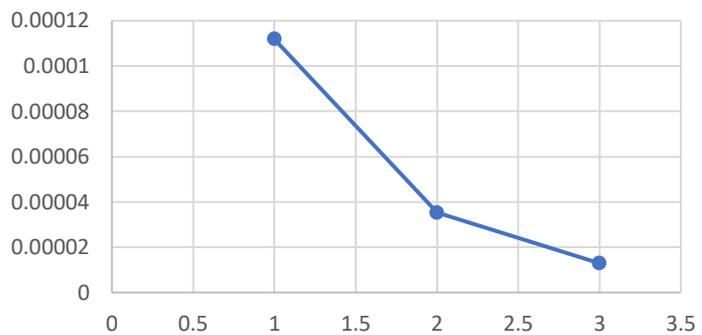
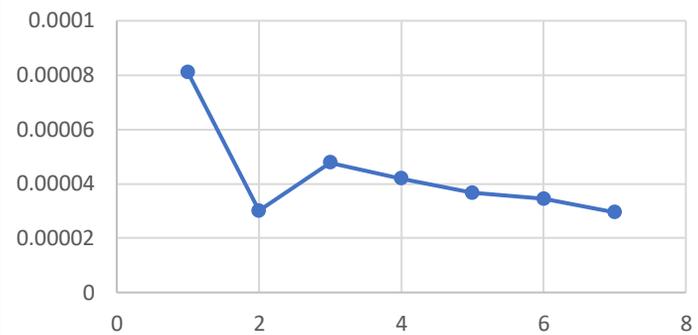

| 5 × 15 C$_{450}$H$_{50}$ | 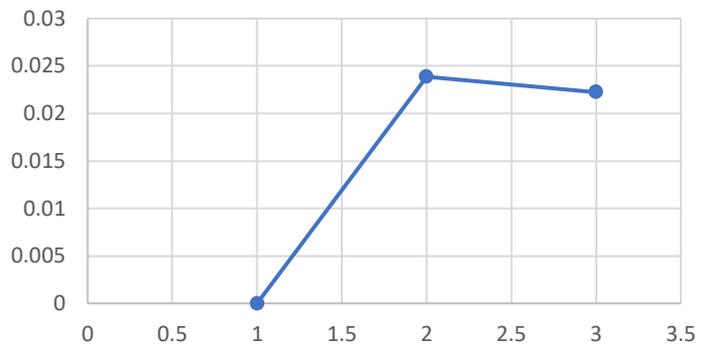 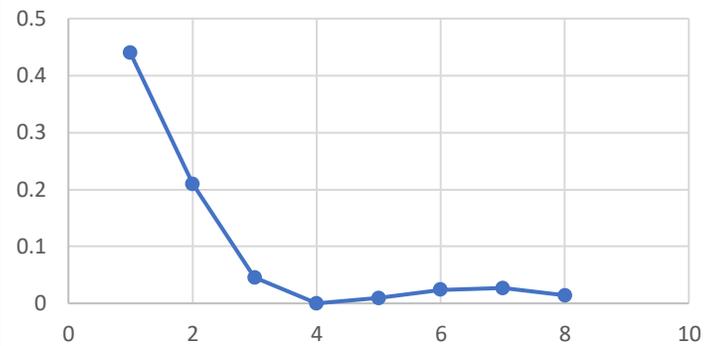 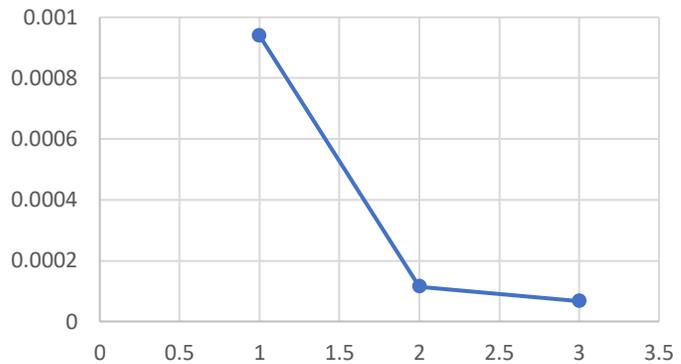 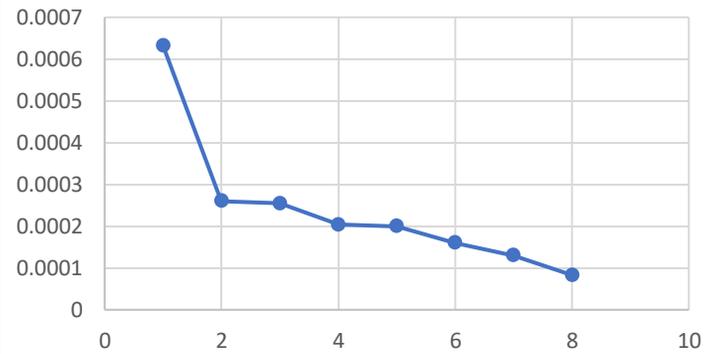 |
|---|---|

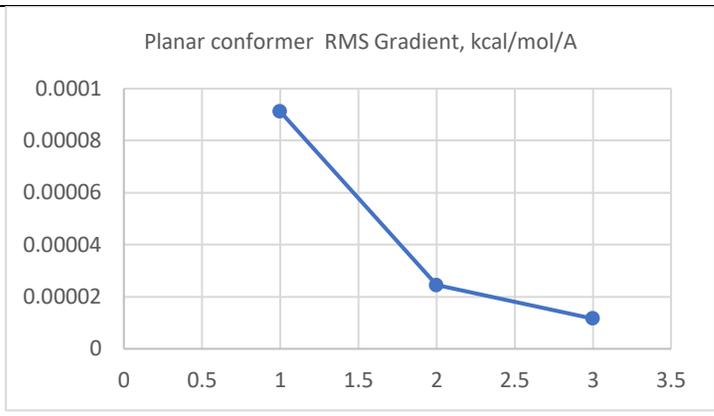 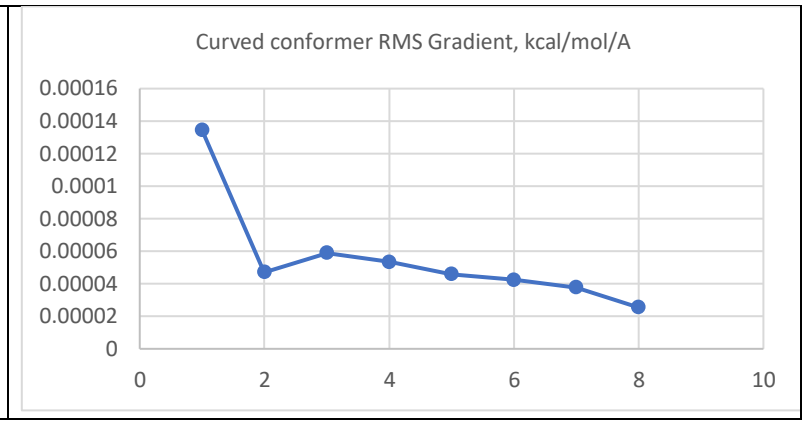

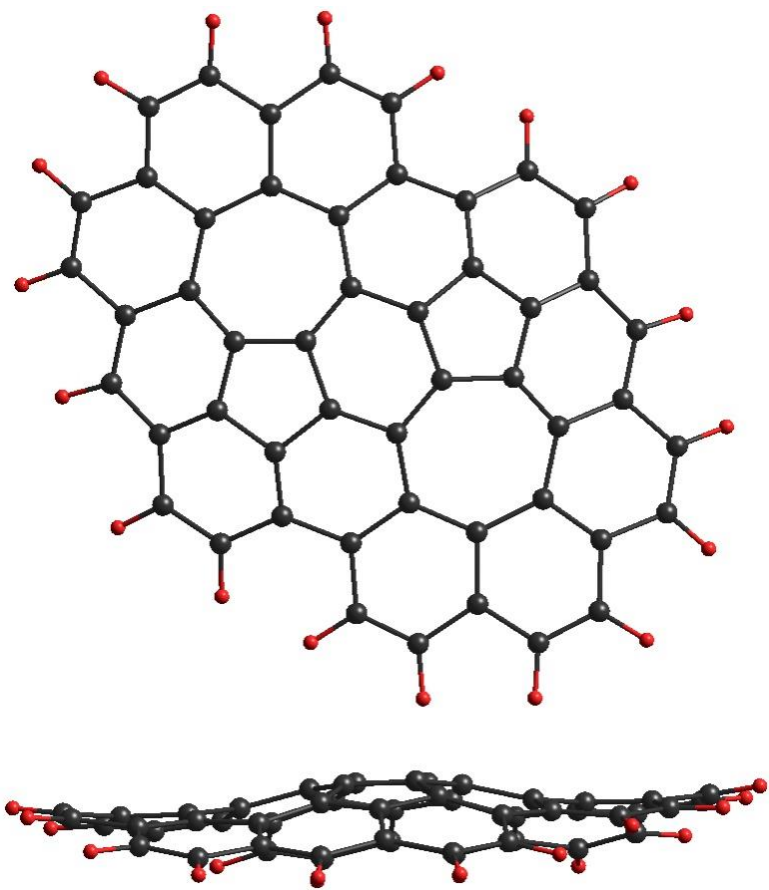

**Figure SI1-1.** Top and side views of minimized curved conformer of $C_{50}H_{15}$ cluster, which represents two 5/7 phagraphene cores embedded into $C_6$ environment. Optimization was performed at PM3 and B3LYP/6-31G* levels of theory with relative energies of planar transition state equal to 5.641 and 0.504 kcal/mol, respectively.

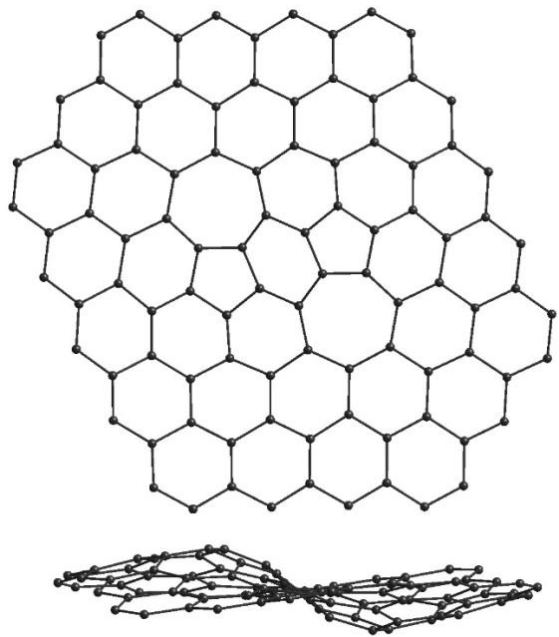

**Figure SI1-2.** Top and side views of carbon backbone of curved conformer of $C_{96}H_{24}$ cluster (ground state of $C_i$ or $C_s$ symmetry) which represents a phagraphene structural unit with double (5/7) core embedded into graphene lattice. The atomic structure was relaxed using both model potential and semiempirical PM3 optimizations.

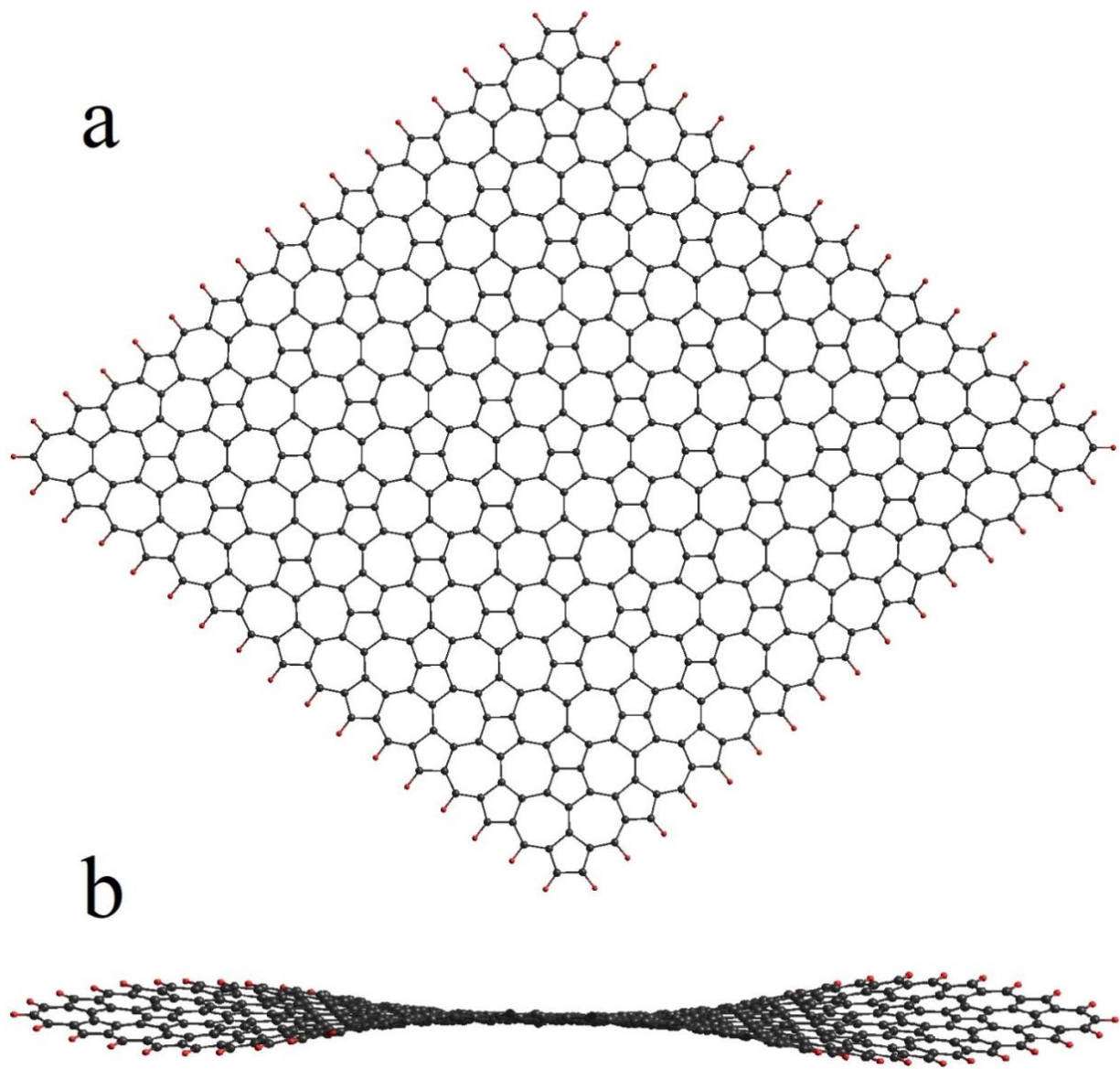

**Figure SI1-3.** Top (a) and side (b) views of PM3 8 × 8 *phC(0,0)* flake. Bending profile of the flake is clearly seen in the side projection.

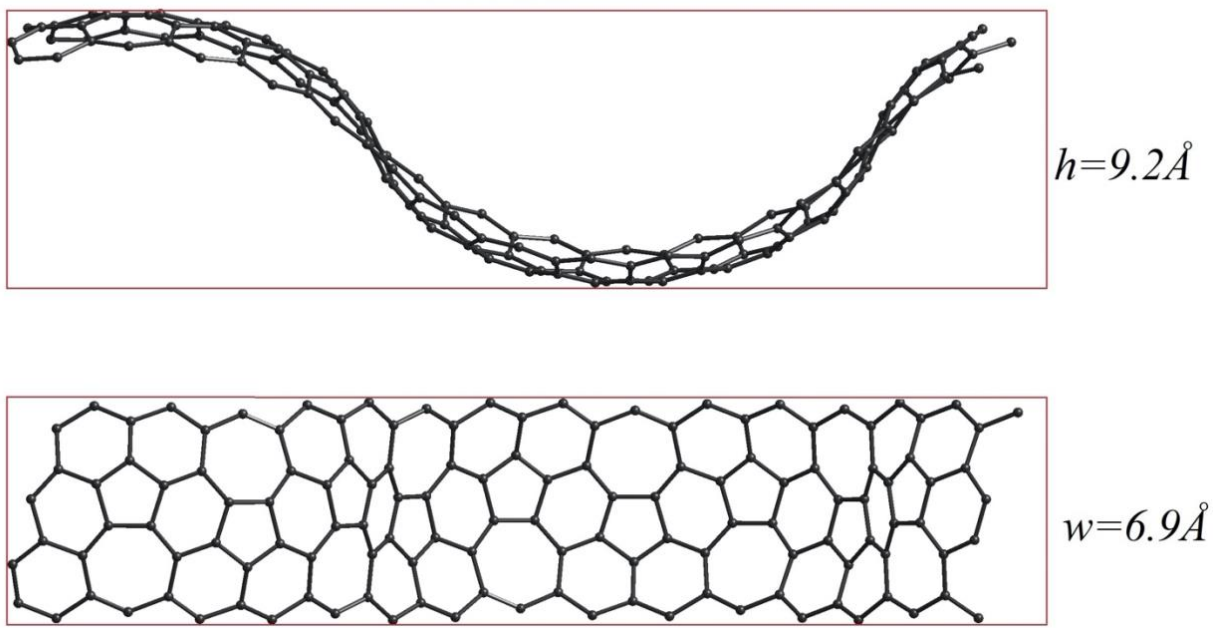

**Figure SI1-4.** Structure and volume of structure wave box of *phC(0,1) nA×1B* nanoribbon.

*The structure wave box dimensions for nA×4B phagraphene nanoribbons*
*l=115.8Å     1548 carbon atoms*

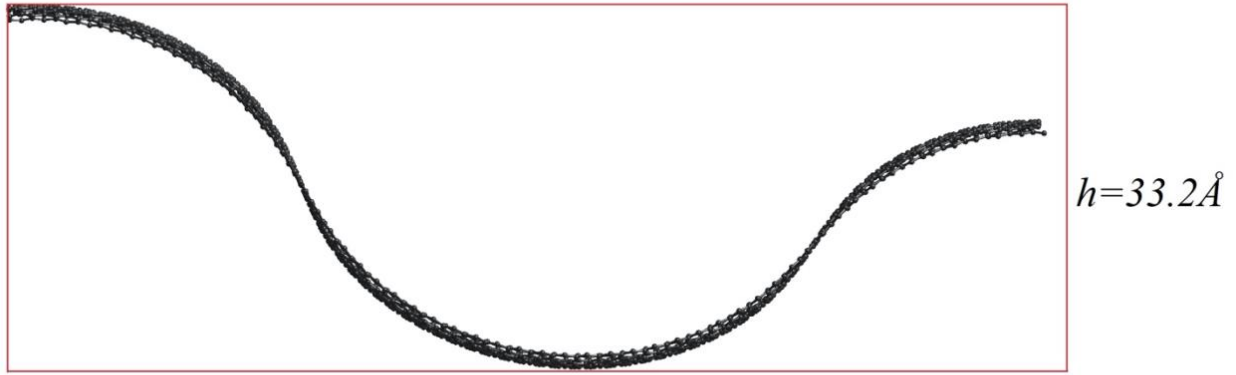

*h=33.2Å*

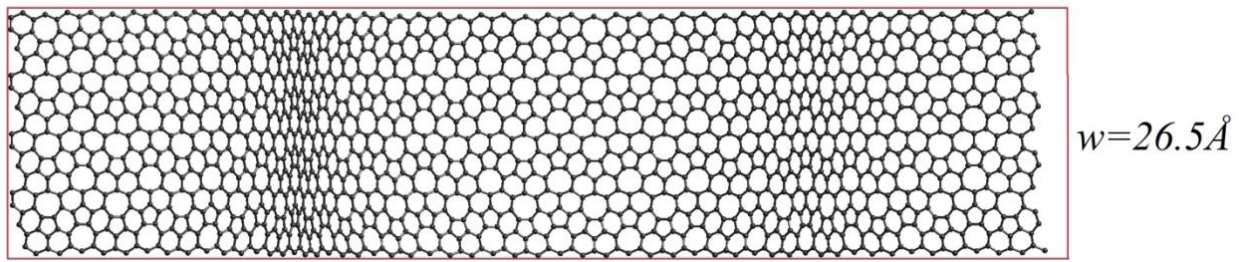

*w=26.5Å*

**Figure SI1-5.** Structure and volume of structure wave box of *phC(0,1) nA×4B* nanoribbon.

| a | b |
|---|---|
| 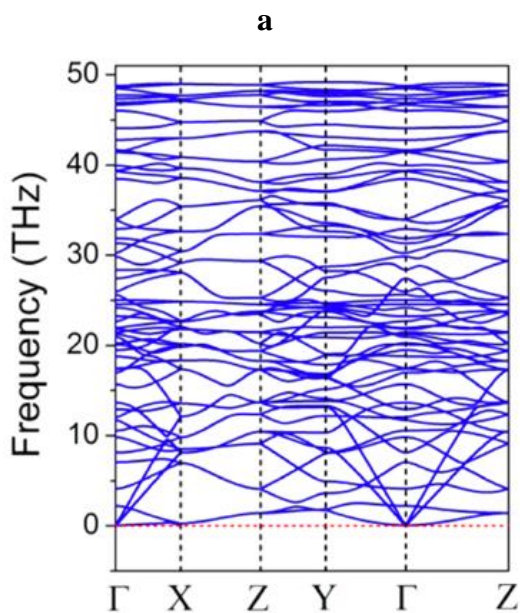 | 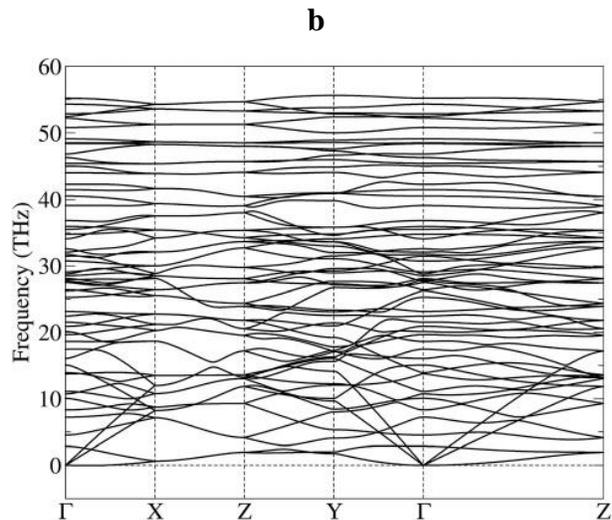 |

**Figure SI1-6a**. Phagraphene phonon dispersion [1] for 2×2 and 3×3 supercells calculated using PHONON code [3-5]. No imaginary frequencies were detected.

**Figure SI1-6b**. Phonon dispersion for phagraphene [2] modeled by Tersoff potential. No imaginary frequencies were detected. The spectrum was calculated using GULP [6, 7] code.

[1] Z. H. Wang, X. F. Zhou, X. M. Zhang, Q. Zhu, H. F. Dong, M. W. Zhao, and A. R. Oganov, Phagraphene: A Low-Energy Graphene Allotrope Composed of 5–6–7 Carbon Rings with Distorted Dirac Cones, Nano Lett. 15, 6182-6186 (2015).

[2] L.F.C. Pereira, B. Mortazavi, M. Makaremic, T. Rabczuk, Anisotropic thermal conductivity and mechanical properties of phagraphene: a molecular dynamics study, RSC Adv. 6, 57773-57779 (2016). DOI: 10.1039/C6RA05082D

[3] https://wolf.ifj.edu.pl/phonon/

[4] K. Parlinski, Z.Q.Li, Y.Kawazoe, First-principle determination of the soft mode in cubic ZrO2, Phys.Rev.Lett. 78, 4063 (1997).

[5] http://www.computingformaterials.com/

[6] J.D. Gale, GULP: A computer program for the symmetry-adapted simulation of solids, J. Chem. Soc., Faraday Trans., 93, 629–637 (1997).

[7] J.D. Gale, A. L. Rohl, The General Utility Lattice Program, Mol. Simul., 29, 291–341 (2003).

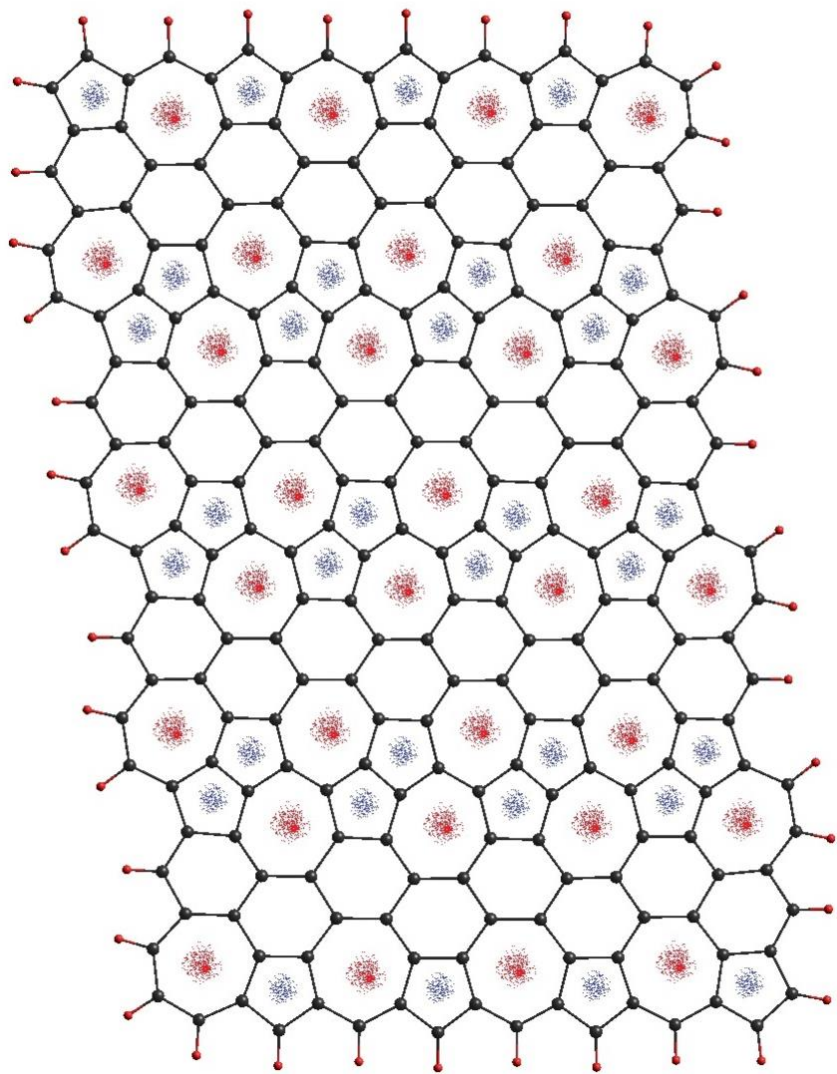

**Figure SI1-7**. A finite fragment ($C_{232}H_{42}$ flake) of planar $4 \times 4$ $phC(1,0)$ lattice. Heptagons are highlighted by red spray and pentagons are highlighted by blue spray. Carbon atoms are presented in black and hydrogen atoms are presented in red.

# phC(1,0) 6×6 Flake

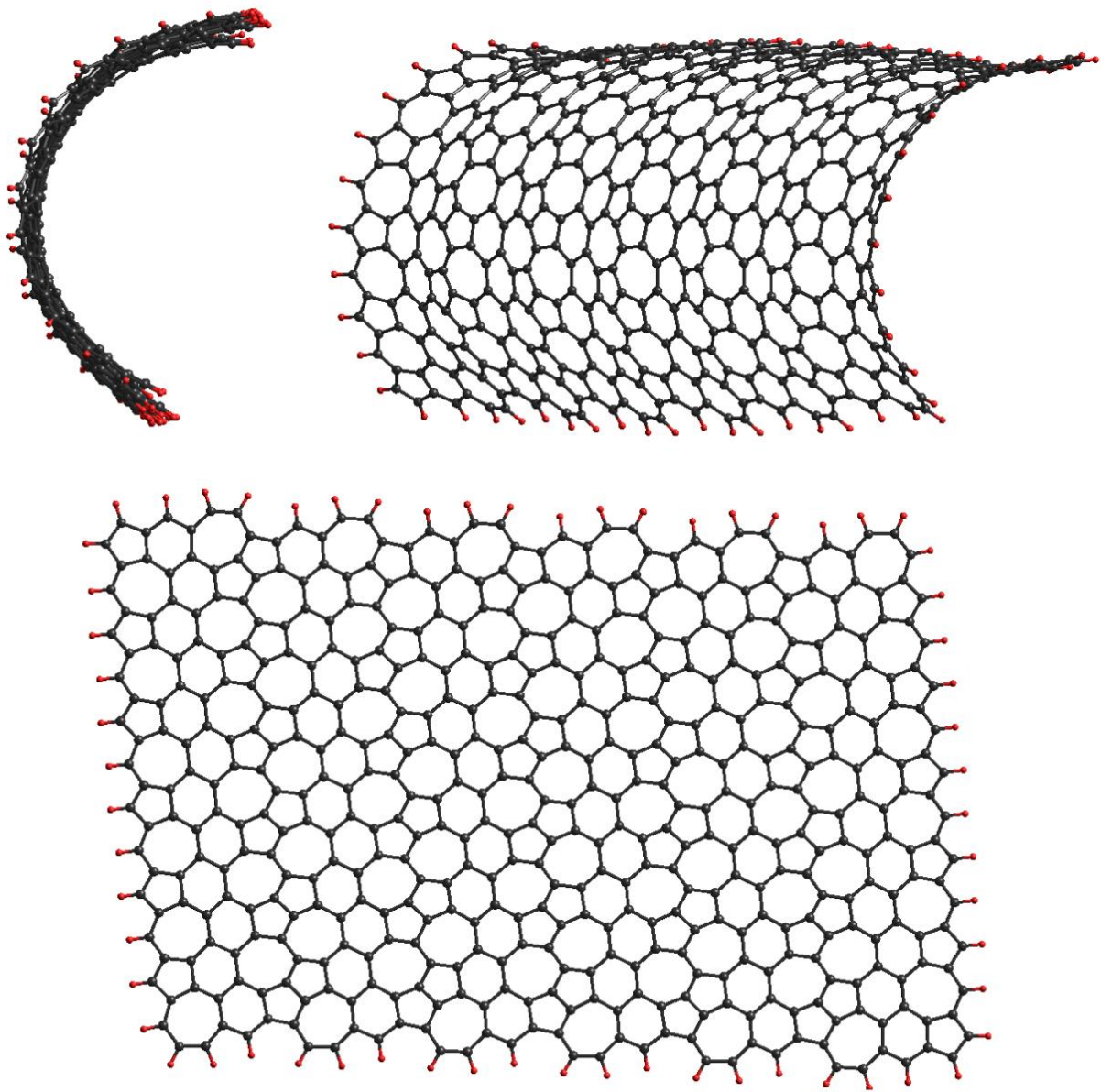

**Figure SI1-8**. Two projections of curved 6A × 6B *phC(1,0)* nanoflake (a fragment of self-assembled *phC(1,0)* chiral (13,1) nanotube) and planar 6A × 6B *phC(1,0)* nanoflake from top to bottom. Both structures were relaxed at both model potential and PM3 levels of theory. Carbon atoms are presented in black and hydrogen atoms are presented in red.

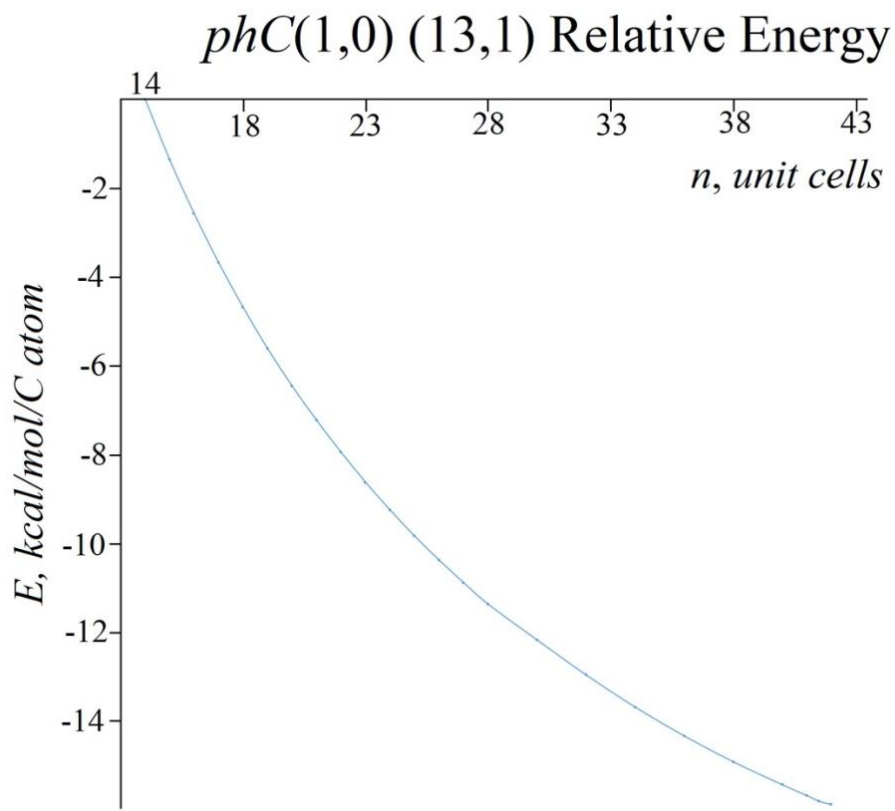

**Figure SI1-9**. The PM3 energy gain (kcal/mol/C atom) of final *phC(1,0)* (13,1) chiral nanotubes of different lengths with respect to the energy of the shortest completed *phC(1,0)* (13,1) nanotube constituted by 14 unit cells.

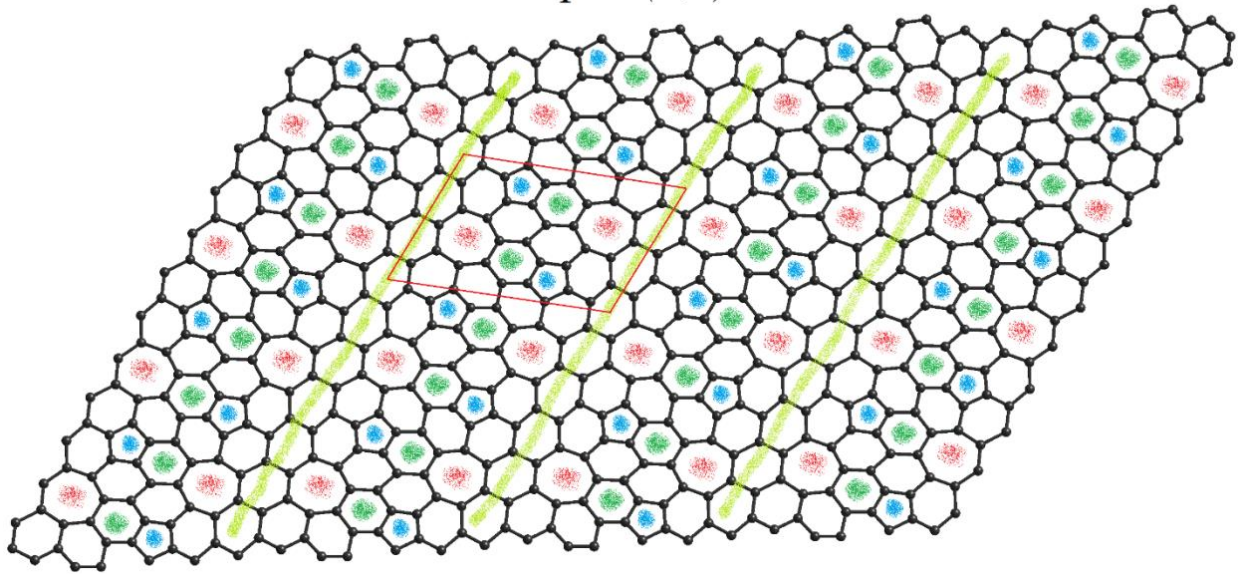

**Figure SI1-10**. a) $4A \times 4B$ fragment of planar 2D *phC(1,1)* lattice. Carbon atoms are presented in black. Pentagonal rings are highlighted in blue, heptagonal rings are highlighted in red, and hexagonal rings, which separate pentagonal and heptagonal fragments of the same unit cell, are highlighted in green. Lines of fused hexagons are highlighted in light green. A *phC(1,1)* unit cell is denoted by rhombus made by red lines.

## Penta-graphene flakes, rings, tubes and rolls

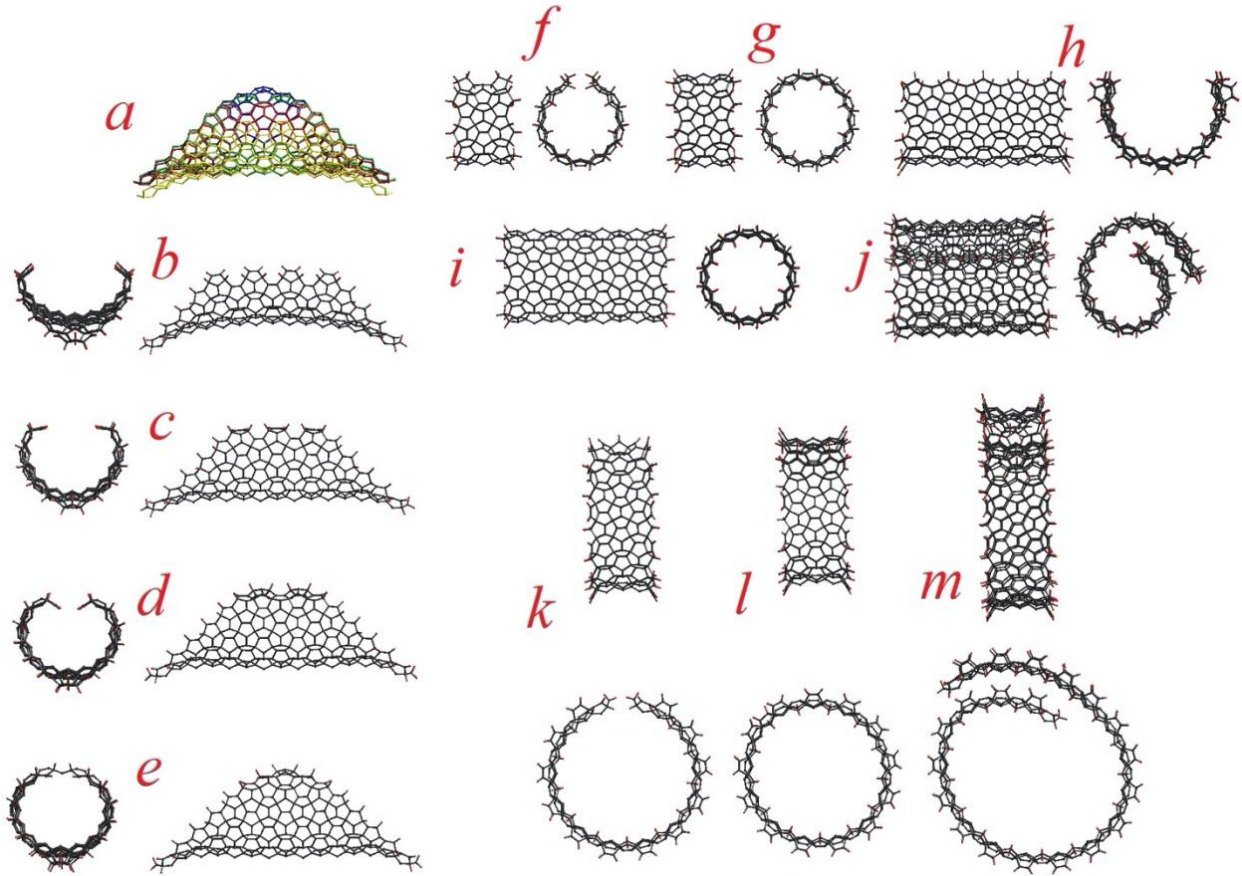

**Figure SI1-11**. Penta-graphene flakes, rings, tubes and rolls calculated using PM3 approach. *a*) Comparative side view of penta-graphene $na \times 8b$ ($n = 5 - 8$) flakes. $5a \times 8b$ $C_{360}$ flake is presented in yellow, $6a \times 8b$ $C_{396}$ flake is presented in red, $7a \times 8b$ $C_{420}$ flake is presented in green and locked $8a \times 8b$ $C_{430}$ flake is presented in blue. *b*) Two side views of $5a \times 8b$ $C_{360}$ flake. *c*) Two side views of $6a \times 8b$ $C_{396}$ flake. *d*) Two side views of $7a \times 8b$ $C_{420}$ flake. *e*) Two side views of $8a \times 8b$ $C_{430}$ flake. *f*) and *g*) Two side views of unlocked and locked of *b*-oriented ring. *h*) and *i*) Two side views of unlocked and locked of *b*-oriented nanotube fragment of 5 unit cell length. *j*) A roll formed by rolling of *a*-oriented nanoribbon. The main axis of the roll is parallel to *b*-direction. *k*) and *l*) Two side views of unlocked and locked of *a*-oriented ring. *m*) A roll formed by rolling of *b*-oriented nanoribbon. The main axis of the roll is parallel to *a*-direction.

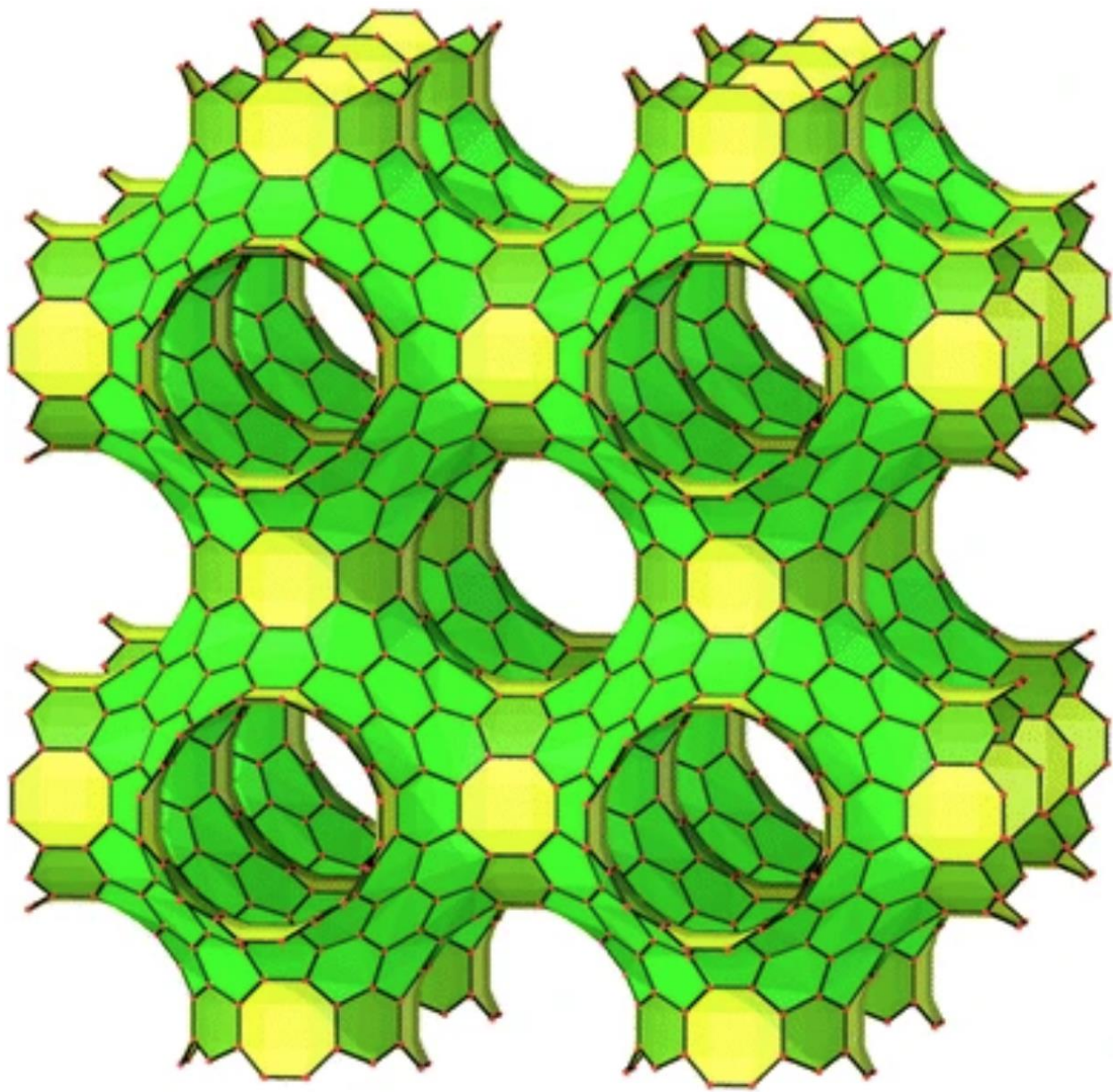

**Figure SI1-12**. Structure of the unit cell of simplest schwarzite [8, 9] constituted by hexagons and octagons.

[8] A.L. Mackay, H. Terrones, Diamond from graphite, Nature 352, 762 (1991).

[9] S.T. Hyde, M. O'Keeffe, At sixes and sevens, and eights, and nines: schwarzites p3, p=7, 8, 9, Struct. Chem. 28, 113–121 (2017). DOI 10.1007/s11224-016-0782-1.

# Penta-graphene rolls
## Model Potential
Zig-zag roll, 32 unit cells, $C_{700}H_{268}$

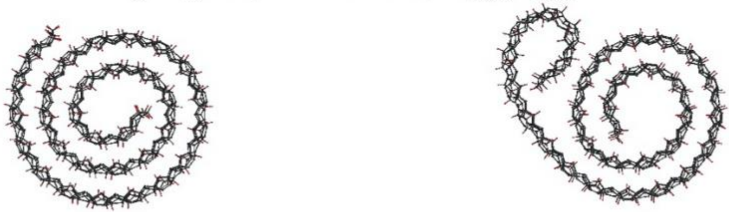

## Semiempirical PM3
Zig-zag roll, 32 unit cells, $C_{700}H_{268}$

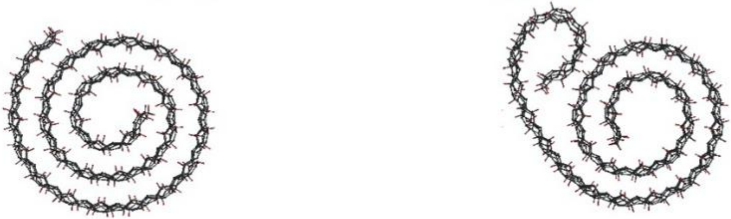

## Model Potential
Zig-zag roll, 162 unit cells  
$C_{3560}H_{1308}$

Zig-zag roll, 163 unit cell  
$C_{3582}H_{1316}$

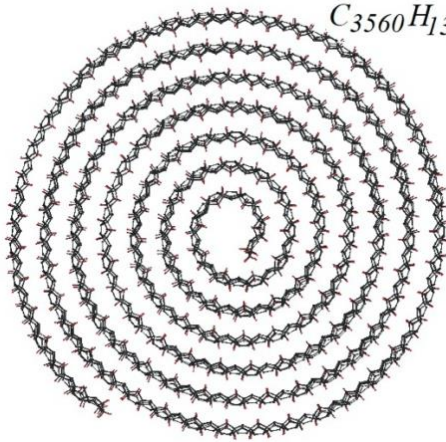 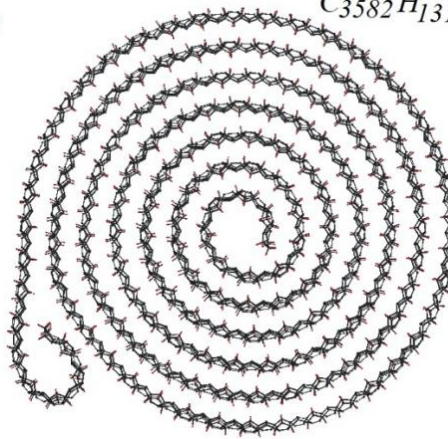

**Figure SI1-13**. Single- and double penta-graphene *b*-oriented rolls of 32 ($C_{700}H_{268}$), 162 ($C_{3560}H_{1308}$) and 163 ($C_{3582}H_{1316}$) unit cell length of different shapes calculated using model potential and semiempirical PM3 methods. The 163 ($C_{3582}H_{1316}$) spontaneously forms double coil structure.

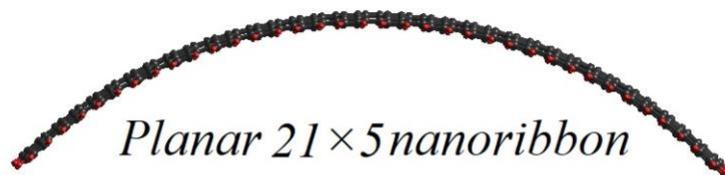
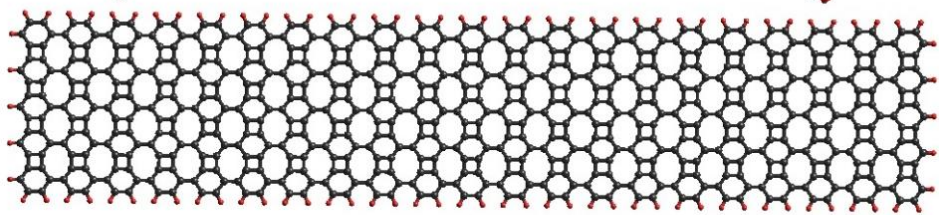
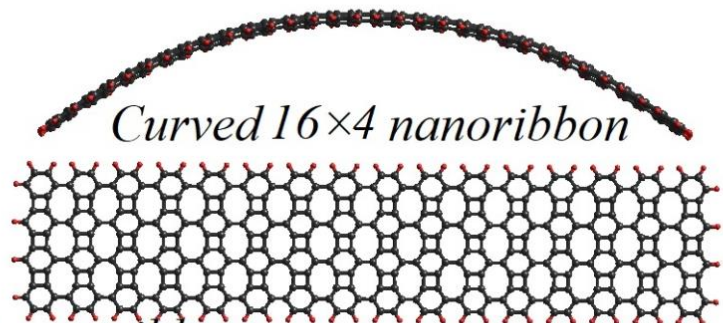
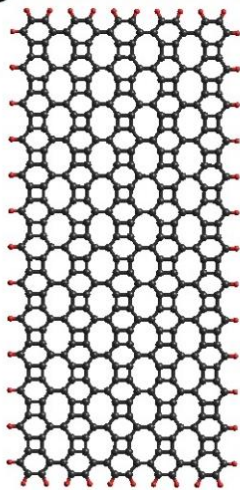
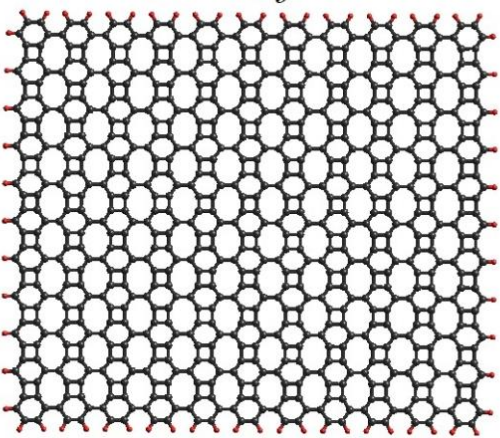

**Figure SI1-14**. Extended 2D biphenylene lattice finite-size fragments, from top to bottom: Curved 18×5 nanoribbon ($C_{540}H_{82}$); planar 21×5 nanoribbon ($C_{630}H_{94}$); side and top views of curved 16×4 nanoribbon ($C_{384}H_{72}$); 5×13 nanoribbon ($C_{390}H_{46}$); 11×11 flake ($C_{726}H_{66}$). Carbon atoms are depicted in black, hydrogen atoms are depicted in red. The lattice dimensions are presented in unit cells.

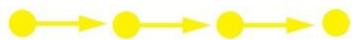
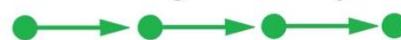
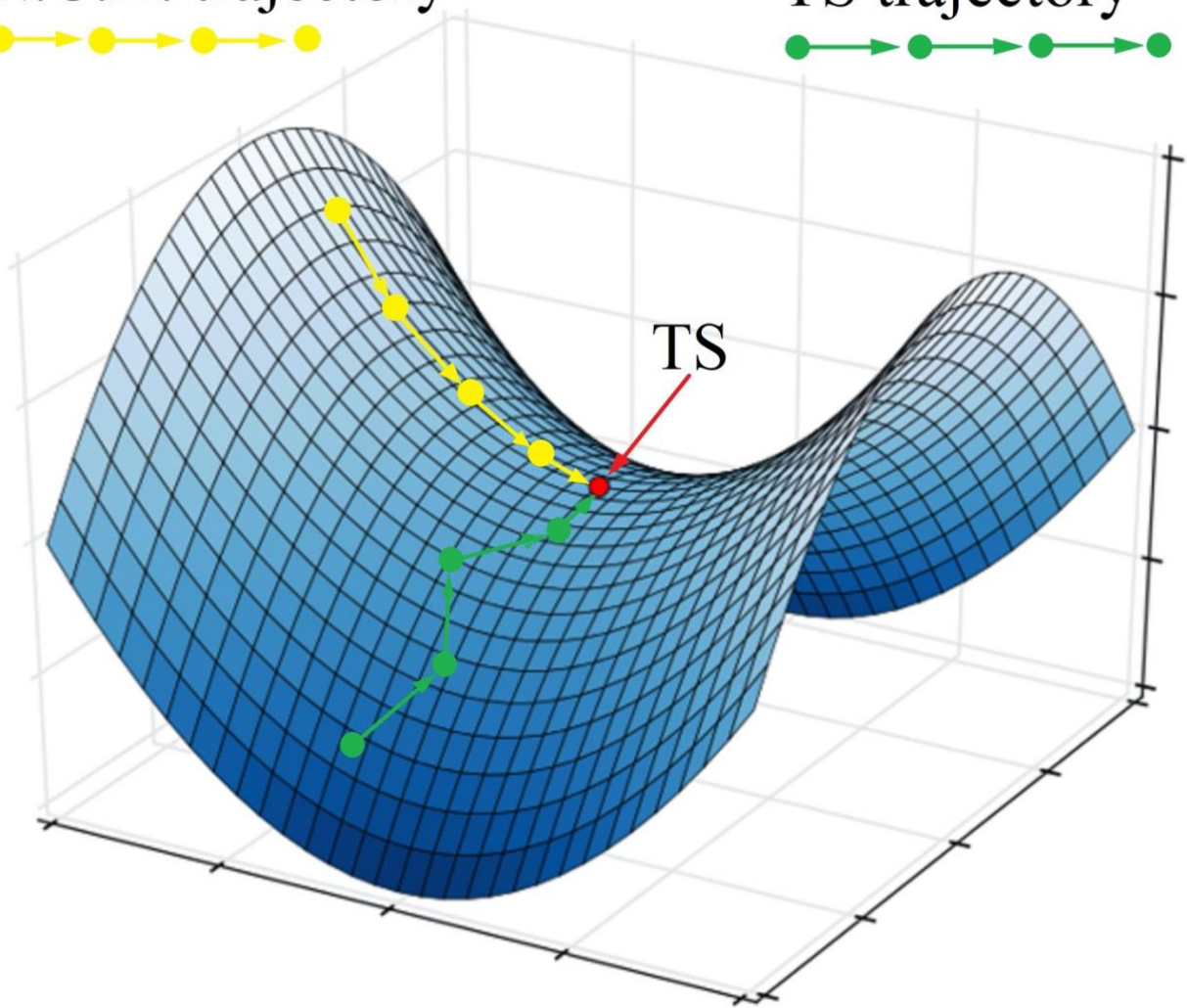

**Figure SI1-15**. No Conclusive Answer (N.C.A. trajectory, in yellow) and Transition State (TS trajectory, in green) optimization trajectories, which converge to the same transition state (TS, depicted as red dot at the saddle point). Depending on starting point, the total energy on potential energy surface may either decline (N.C.A. trajectory), or raise (TS trajectory) during optimization procedure. In other words, N.C.A. trajectory couples the gradient minimum with local minimum on optimization energy curve, whereas the TS trajectory couples the gradient minimum with local maximum on optimization energy curve. Based on N.C.A., additional Hessian calculations are needed to reveal whether or not the localized extreme point is minimum or maximum on potential energy curve. In contrast, TS-type trajectory is a clear indication of transition state character of the localized extreme point.

*Topological and quantum stability of low-dimensional crystalline lattices with multiple nonequivalent sublattices*

*Supplementary Information 2* (SI2 Section)

Pavel V. Avramov[a,*], and Artem V. Kuklin[b]

[a] Department of Chemistry, College of Natural Sciences, Kyungpook National University, 80 Daehak-ro, Buk-gu, Daegu, 41566, South Korea

[b] Department of Physics and Astronomy, Uppsala University, Box 516, SE-751 20 Uppsala, Sweden

*E-mail: paul.veniaminovich@knu.ac.kr

# THEORETICAL MODEL OF INTERNAL MECHANICAL STRESS OF 1D AND 2D LATTICES WITH MULTIPLE NONEQUIVALENT SUBLATTICES

### A. 1D h-BN zig-zag narrow nanoribbon case

A hypothetical 1D *h*-BN zig-zag narrow nanoribbon (*h*-BN ZNR) of one $B_3N_3$ hexagonal fragment width (Figure SI2-1) may be considered as the simplest case of a low-dimensional lattice with multiple non-equivalent sublattices. Without loss of generality one can introduce $a_1$, $a_2$, and $b_1$, $b_2$ nonequivalent sublattices for N and B basis atoms, respectively, associated with symmetrically nonequivalent $\boldsymbol{a}_{a1}$, $\boldsymbol{a}_{a2}$, $\boldsymbol{a}_{b1}$, $\boldsymbol{a}_{b2}$ translation vectors oriented along *X* direction. In particular, both external atomic rows have coordination numbers equal to 2 (or, in the case of hydrogen-substituted external atoms, 3, with symmetrically different environment), with different neighborhood, namely $a_1$ N atoms have 2 $b_2$ B neighbors while $b_1$ B atoms have 2 $a_2$ N neighbors. The coordination numbers of $a_2$ N and $b_2$ B atoms are equal to 3 with 2 $b_1$ and 1 $b_2$ boron atoms and 2 $a_1$ and 1 $a_2$ nitrogen neighbors, respectively. For simplicity all interatomic bond lengths can be considered equal to $R_{NB}$. Since the basis atoms, the boundary conditions, coordination numbers and types of environments of $a_1/a_2$ and $b_1/b_2$ N and B sublattices are nonequivalent, one can write:

$$\boldsymbol{a}_{a1} \neq \boldsymbol{a}_{a2} \neq \boldsymbol{a}_{b1} \neq \boldsymbol{a}_{b2}$$

The N $a_1$ and B $b_1$ sublattices with coordination numbers equal to 2 possess different force constants $Q_1$, $Q_2$ of vibrations along *X* direction because of different symmetry and environment of the edge N and B atoms even $a_2$-$b_1$-$a_2$ and $b_2$-$a_1$-$b_2$ angles and $R_{NB}$ length of all N-B bonds keep initial parameters of perfect 2D *h*-BN. The force constant $Q$ along the *X* direction and $R_{NB}$ length of B-N bonds for $a_2$ and $b_2$ sublattices with coordination numbers equal to 3 keep the initial $Q_i$ value of perfect 2D *h*-BN lattice. So, for $\boldsymbol{a}_{a2}$ and $\boldsymbol{a}_{b2}$ translation vectors and the force constants the following relationships can be written:

$$Q_1 \neq Q_2 \neq Q$$

$$\boldsymbol{a}_{a2} = \boldsymbol{a}_{b2}$$

In harmonic approximation, the stress energies $E_i$ caused by any deformation along $X$ can be written as:

$$E_i = Q_i \boldsymbol{a}_i^2 + B\boldsymbol{a}_i$$

where constant $B$ is the same for all sublattices. For perfect hexagonal lattice $|\boldsymbol{a}_{a1}| = |\boldsymbol{a}_{b1}| = \sqrt{3}R_{NB}$. Following the symmetry restrictions the following relationships can be written:

$$E_{a2} = Q|\boldsymbol{a}_{a2}|^2 + B|\boldsymbol{a}_{a2}| = Q|\boldsymbol{a}_{b2}|^2 + B|\boldsymbol{a}_{b2}| = E_{b2}$$

$$E_{a1} = Q_1|\boldsymbol{a}_{a1}|^2 + B|\boldsymbol{a}_{a1}| \neq Q_2|\boldsymbol{a}_{b1}|^2 + B|\boldsymbol{a}_{b1}| = E_{b1}$$

For the forces acting on $a_i$ and $b_i$ sublattices one can write:

$$F_{a2} = F_{b2} = (E_{a2})' = (E_{b2})' = 2Q\boldsymbol{a}_{a2} + B = 2Q\boldsymbol{a}_{b2} + B$$

$$F_{a1} \neq F_{b1}$$

$$F_{a1} = (E_{a1})' = 2Q_1\boldsymbol{a}_{a1} + B$$

$$F_{b1} = (E_{b1})' = 2Q_2\boldsymbol{a}_{b1} + B$$

Symmetrical non-equivalency of the forces acting on the boundary B and N atoms ($F_{a1} \neq F_{b1}$) creates mechanical stress and structural curvature of ultranarrow $h$-BN ZNR and other similar zig-zag heteroatomic nanoribbons [1].

For perfect hexagonal lattice the distance between $a_1$ and $b_1$ nodes are:

$$R_{a1-b1} = 2R_{NB} = 2R_{a2-b2},$$

so the torques $\tau_i$ caused by uncompensated forces acting on nonequivalent sublattices of $h$-BN ZNR in respect to the motionless center of the mass can be written as:

$$\tau_i = r \times F_i$$

For $a_2$, $b_2$ sublattices the $\alpha$ angles in respect to the $X$ direction and center of ezch single $C_3B_3$ fragment are equal to $1/6 \pi$ and $-1/6 \pi$, respectively (Figure SI2-1), so:

$$\tau_{a2} = R_{NB} \cdot |F_{a2}| \cdot sin(1/6 \pi) = {R_{NB}|F_{a2}|}/{2}$$

$$\tau_{b2} = R_{NB} \cdot |F_{b2}| \cdot sin(-1/6 \pi) = -{R_{NB}|F_{a2}|}/{2}$$

$$\tau_{b2} = -\tau_{a2}$$

and $a_2$, $b_2$ associated torques mutually compensate each other.

For $a_1$, $b_1$ sublattices of the opposite N and B atoms the $\alpha$ angles for are equal to $-\pi/2$ and $\pi/2$, respectively, (Figure SI2-1), and

$$|r_{a1}| = |r_{b1}| = R_{NB},$$

so, taking into account asymmetry of *sin* function:

$$\tau_{a1} = R_{NB} \cdot |F_{a1}| \cdot \sin(-\pi/2) = -R_{NB}|F_{a1}|$$

$$\tau_{b1} = R_{NB} \cdot |F_{b1}| \cdot \sin(\pi/2) = R_{NB}|F_{b1}|$$

The $\tau_{b1}$ and $\tau_{a1}$ are oriented in opposite directions. The total torque acting on the nanostructure is the sum of the torques associated with the sublattices. Since $F_{a1} \neq F_{b1}$

$$\tau_{a1} + \tau_{b1} = R_{NB}|F_{b1}| - R_{NB}|F_{a1}| = R_{NB}(|F_{b1}| - |F_{a1}|) \neq 0$$

and the torques do not compensate each other. Since both nonequivalent torques associated with $a_1$ and $b_1$ sublattices are perpendicular to the plane of *h*-BN ZNR lattice and oriented in opposite directions, they lead to displacements of the atoms with the formation of a cone fragment of one $B_3N_3$ width and breaking down the linear translation symmetry [1, 2].

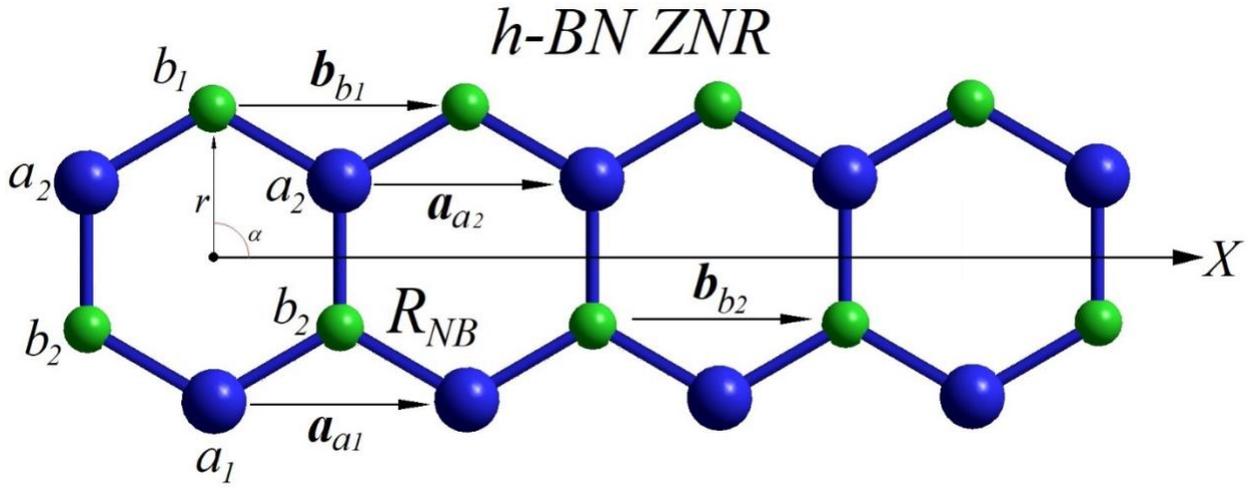

**Figure SI2-1**. A fragment of 1D *h*-BN zig-zag narrow nanoribbon of one $B_3N_3$ hexagonal fragment width, oriented along the *X*-axis. $a_1$, $a_2$, $b_1$, and $b_2$ are nonequivalent sublattices filled by N ($a_1$, $a_2$, in blue) and B ($b_1$, $b_2$, in green) atoms, respectively. $R_{NB}$ is the length of interatomic bonds, $\mathbf{a}_{a1}$, $\mathbf{a}_{a2}$, $\mathbf{a}_{b1}$, $\mathbf{a}_{b2}$ are translation vectors associated with $a_1$, $a_2$, $b_1$, $b_2$ sublattices, respectively. $\alpha$ is the angle between *X* direction and *r*, where *r* is an atomic radius vector.

## C. 2D graphene and h-BN cases

Both 2D graphene and *h*-BN crystalline lattices have two basis atoms in hexagonal unit cells associated with 2 translation vectors $a_1$ and $a_2$ with absolute values $|a_1| = |a_{12}| = \sqrt{3}R$, where $R$ is the length of either C-C or B-N chemical bonds. The angle between translation vectors, $\alpha = \pi/3$ or $\alpha = 2\pi/3$ can be chosen in two different ways depending on the choice of the unit cell. For the sake of simplicity let's consider the case of the first graphene sublattice (Figure 3) with $\alpha_1 = \pi/3$ and the force constant $Q$.

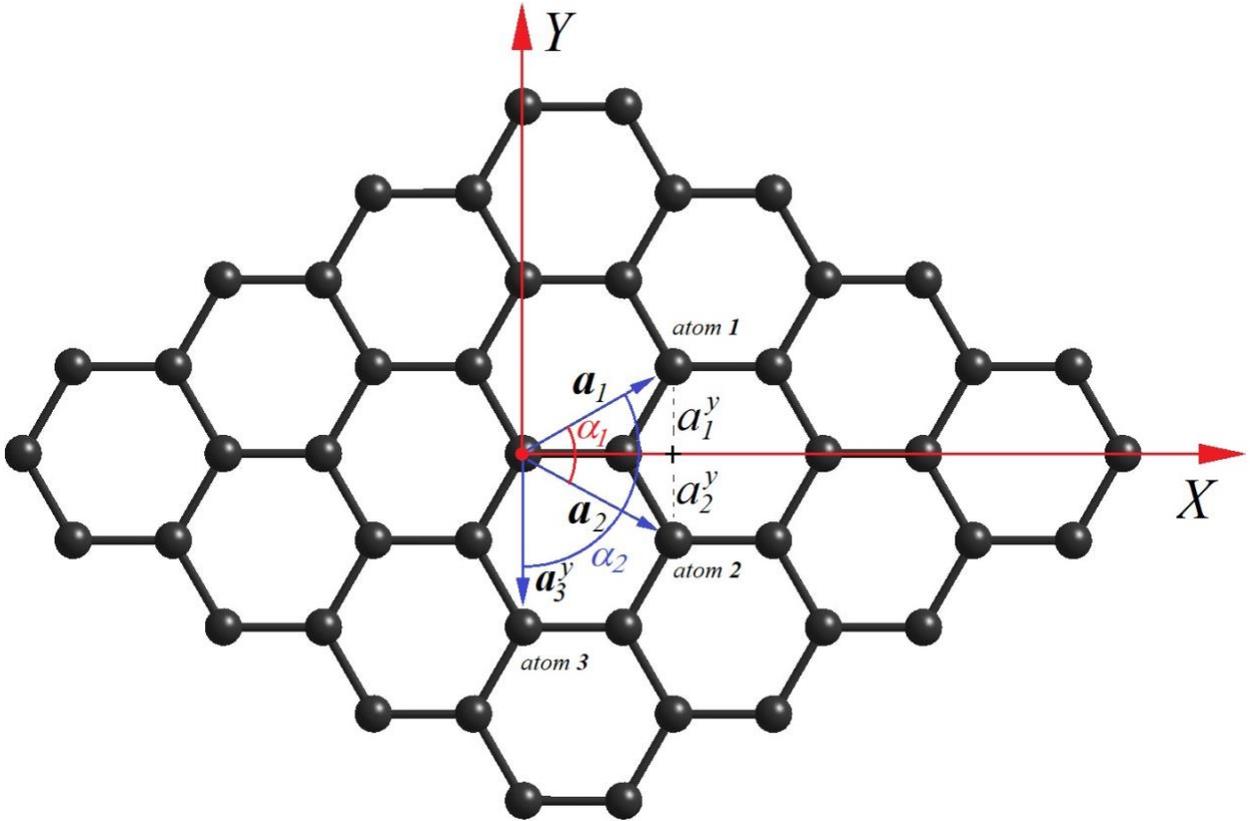

**Figure SI2-2**. 2D hexagonal crystalline lattice of graphene. $a_1$ and $a_2$ are linear translation vectors for which $|a_1| = |a_{12}| = \sqrt{3}R$, $a_1^y = \frac{\sqrt{3}}{2}R$, $a_2^y = -\frac{\sqrt{3}}{2}R$, and $a_3^y = -\sqrt{3}R$ are Y coordinates of atoms 1, 2 and 3, angles between translation vectors $\alpha_1 = \pi/3$, $\alpha_2 = 2\pi/3$. The X coordinates of atoms 1, 2 and 3, $a_1^x = \frac{3}{2}R$, $a_2^x = \frac{3}{2}R$, and $a_3^x = 0$ are not depicted in the figure.

Because of symmetry restrictions, the X coordinates of atoms 1 and 2, $\{a_1^x, a_1^y; a_2^x, a_2^y\}$, generated by translation vectors $a_1$ and $a_2$ are equal to each other with Y coordinates have opposite sign. In respect of the origin of the coordinate system (red dot in the center Figure SI2-2), the coordinates of atoms 1 and 2 are:

$$\left(\tfrac{3}{2}R, \tfrac{\sqrt{3}}{2}R\right)$$

$$\left(\tfrac{3}{2}R, -\tfrac{\sqrt{3}}{2}R\right)$$

In harmonic approximation the $X$ and $Y$ components of $\boldsymbol{a}_1, \boldsymbol{a}_2$ associated energies and forces are:

$$E^x_{a1} = E^x_{a2} = Q|a^x_1|^2 = Q|a^x_2|^2$$

$$E^y_{a1} = Q|a^y_1|^2$$

$$E^y_{a2} = Q|a^y_2|^2$$

$$F^x_{a1} = F^x_{a2} = 2Qa^x_1 = 2Qa^x_2 = 3QR$$

$$F^y_{a1} = 2Qa^y_1 = \sqrt{3}QR$$

$$F^y_{a2} = 2Qa^y_2 = -\sqrt{3}QR.$$

For the second equivalent choice of the unit lattice of graphene with $\alpha_2 = 2\pi/3$, the coordinates of the generated atoms 1 and 3 are $\{a^x_1, a^y_1;\ a^x_3, a^y_3\}$. In respect of the origin of the coordinate system, the coordinates of the atom 3 and the energy components are

$$\left(0, \sqrt{3}R\right)$$

$$E^x_{a3} = Q|a^x_3|^2 = 0$$

$$E^y_{a3} = Q|a^y_3|^2$$

Because of symmetry of the lattice, the energy components must be equal to each other:

$$E^x_{a1} = E^x_{a2} = E^x_{a3}$$

$$E^y_{a1} = E^y_{a2} = E^{xy}_{a3}$$

This conditions can be satisfied only all components are equal to 0, so, for pristine hexagonal lattices of graphene and *h*-BN, the *X* and *Y* components of forces mutually compensate each other and the total stress of the lattice is equal to 0. For the torques acting on atoms 1 and 2 in respect to the origin of the coordinate system one can write:

$$F_1 = F_2$$
$$\tau_1 = \boldsymbol{r} \times \boldsymbol{F}_1 = |\sqrt{3}R| F_1 \sin(\pi/6)$$
$$\tau_2 = \boldsymbol{r} \times \boldsymbol{F}_2 = |\sqrt{3}R| F_2 \sin(-\pi/6)$$
$$\tau_1 = -\tau_2$$

and the torques mutually compensate each other as well. For perfect 2D hexagonal lattices like graphene and *h*-BN the forces acting on atoms and associated structural stresses and torques mutually compensate each other, so they keep perfect linear translation symmetry and planar 2D structure.

## TOPOLOGICAL AND QUANTUM STABILITY OF 1D AND 2D *phC(n,m)* LATTICES

### B. *phC(0,1)* (phagraphene) lattice

*Dynamical stability of phC(0,1) lattice*

The impulse absolute value of the structural wave of *phC(0,1)* $nA \times 1B$ is $|p| = h/\lambda = \frac{6.6262}{3.4} \times \frac{10^{-34}}{10^{-9}} \times \frac{kg \cdot m^2 s^{-2}}{m} \cdot s = 1.95 \cdot 10^{-25} \cdot kg \cdot m \cdot s^{-2}$, which gives the energy of one structural wave $E = \frac{p^2}{2M_w}$, where $M_w = 1728 \cdot 1.660539 \cdot 10^{-27} kg = 2.869 \cdot 10^{-24} kg$ is the mass of carbn atoms which constitute the wave. Substitutting the values, the energy of one structural wave is $E = \frac{3.8025}{2 \cdot 2.869} \times \frac{10^{-50}}{10^{-24}} \cdot \frac{kg^2 \cdot m^2 \cdot s^{-2}}{kg} = 6.63 \cdot 10^{-27} kg \cdot m^2 \cdot s^{-2} = 4.14 \cdot 10^{-8} eV$ with the the energy per single carbon atom $E_a = 9.547 \cdot 10^{-7} \, kcal \cdot mol^{-1}/137 \, atoms = 6.97 \cdot 10^{-11} \, kcal \cdot mol^{-1} \cdot C \, atom^{-1}$. Negligibly small energy per carbon atom makes the structural wave of *phC(0,1)* $nA \times 1B$ dynamically stable.

*Estimation of internal pressures and temperatures of phC(0,1) structural waves*

For *phC(0,1)* $nA \times 1B$ lattice two significantly different structural waves were localized. The first one, of 5 unit cell wavelength and 1 unit cell width is characterized by 33.8Å wavelength, 6.9Å width and $9.2Å/2 = 4.6Å$ amplitude (see Figure S4). The volume of a box which contains one structure wave unit is:

$$V_{sw}^n = 33.8 \times 9.2 \times 6.9 \times 10^{-30} m^3 = 2.1 \times 10^{-27} m^3$$

Each single structure wave box contains 137 carbon atoms, so each single atom occupies $V_{at}^n = 2.1 \times 10^{-27} m^3 / 137 = 1.5 \times 10^{-29} m^3$. Averaged carbon-carbon bond length is 1.5Å, so a carbon atom occupies $1.5^3 Å^3 = 3.4 \times 10^{-30} m^3$ volume and the *phC(0,1) nA×1B* structure wave box can contain $2.1 \times 10^{-27} m^3 / 3.4 \times 10^{-30} m^3 = 618\ atoms$, which is 4.51 times greater than the number of atoms in one wavelength of the structure wave. Since structural positions of the atoms in infinite unconstrained and undisturbed low-dimensional lattice before the collapse of the wave function of any structure wave quasi-particle are completely undetermined, one can estimate the average carbon-carbon distance in the box as $1.5Å \times \sqrt[3]{4.51} = 2.5Å$.

Under normal conditions, each single atom of an ideal gas occupies $3.7 \times 10^{-26} m^3$. Since the atomic positions of an infinite structural wave are completely uncertain due to quantum effects, one can speculate that atoms are evenly distributed in the structural box determined by the effective wave width, wavelength and amplitude. The effective internal pressure of acting on each atom of the unconstrained and undisturbed infinite structural wave can be calculated as a quotient of the atomic volumes under normal conditions and in the structure wave: $3.7 \times 10^{-26} m^3 / 1.5 \times 10^{-29} m^3 = 2466.7\ atm$. Using the ideal gas laws the effective temperature inside the structural wave box can be estimated as

$$T = pV_{sw}^n / vR = \frac{249938378\ N \cdot m^{-2} \times 2.1 \times 10^{-27}\ m^3}{2.2 \times 10^{-22}\ moles \times 8.314\ N \cdot m \cdot mol^{-1} \cdot K^{-1}} = 2870\ K$$

where $T$ is temperature (*K*). For *phC(0,1) nA×1B* nanoribbon the pressure $p = 2466.7\ atm = 2466.7 \times 101325\ Pa = 249938378\ N \cdot m^{-2}$, amount of matter *v=137 atoms*$= 2.2 \times 10^{-22}\ moles$, the gas constant $R = 8.314\ J \cdot mol^{-1} \cdot K^{-1} = 8.314\ N \cdot m \cdot mol^{-1} \cdot K^{-1}$

The second structural wave of 18 unit cell wavelength is localized for *phC(0,1) nA×4B* nanoribbons and *nA×nB* (see above, Figure S5). It has 115.8 Å wavelength, the smallest width of 26.5 Å, $33.2Å/2 = 16.6Å$ amplitude and contains 1548 carbon atoms. Following the structural wave dimensions, the volume of structural wave box for *18A×4B* structural wave is:

$$V_{sw}^w = 115.8 \times 26.5 \times 33.2 \times 10^{-30} m^3 = 101880.810^{-30} m^3 = 1.0 \times 10^{-25} m^3$$

with averaged atomic volume of $6.5 \times 10^{-29} m^3$ which corresponds to 569.2 *atm* pressure and averaged carbon-carbon distance of 4.0Å.

Once again, using Ideal gas law for *18A×4B* structural wave the effective temperature is:

$$T = pV_{sw}^w / vR = \frac{57677307\ N \cdot m^{-2} \times 1.0 \times 10^{-25}\ m^3}{2.6 \times 10^{-21}\ moles \times 8.314\ N \cdot m \cdot mol^{-1} \cdot K^{-1}} = 2668\ K$$

with corresponding amount of matter *v=1548 atoms*$=2.6 \times 10^{-21}\ moles$.

# PENTA-GRAPHENE TOPOLOGICAL INSTABILITY

*Estimation of bending energy of penta-graphene clusters with taking into account of boundary effects*

For penta-graphene $3 \times 3$ $C_{72}H_{28}$ and $4 \times 4$ $C_{120}H_{36}$ clusters the boundary effects can be taken into account supposing that two distinctively different types of atoms at the edge of the clusters with reduced and preserved coordination numbers contribute differently to stabilization energy. Using linear approximation one can simply write two independent linear equations

$$\begin{cases} a_1 e_s + b_1 e_e = E_s \\ a_2 e_s + b_2 e_e = E_b \end{cases}$$

where $E_s$ and $E_b$ are stabilization energies for small $C_{72}H_{28}$ and extended $C_{120}H_{36}$ clusters, respectively, $e_s$ and $e_e$ are strain energies for perfectly coordinated and edge atoms, respectively, $b_1 = 26$ and $b_2 = 36$ are the numbers of edged carbon atoms for $C_{72}H_{28}$ and $C_{120}H_{36}$ clusters, respectively, and $a_1 = 46$ and $a_{21} = 84$, for $C_{72}H_{28}$ and $C_{120}H_{36}$ are the number of carbon atoms with preserved coordination numbers. Solution of the system returns B3LYP/6-31G$^*$ strain energy for non-edged atoms $e_s = 1.879\ kcal/mol/C\ atom$, which is way beyond vdW binding energy of $9.6 \times 10^{-1} - 9.6 \times 10^{-2}$ kcal/mol/atom, [3].

REFERENCES

[1] P.V. Avramov, D.G. Fedorov, P.B. Sorokin, S. Sakai, S. Entani, M Ohtomo, Y. Matsumoto, H. Naramoto, Intrinsic Edge Asymmetry in Narrow Zigzag Hexagonal Heteroatomic Nanoribbons Causes their Subtle Uniform Curvature, J. Phys. Chem. Lett. 3, pp 2003–2008 (2012).

[2] P. Avramov, V. Demin, M. Luo, C.H. Choi, P.B. Sorokin, B. Yakobson, L. Chernozatonskii, Translation Symmetry Breakdown in Low-Dimensional Lattices of Pentagonal Rings. J. Phys. Chem. Lett. 6, 4525–4531 (2015).

[3] P. Atkins, J. de Paula, Physical Chemistry for the Life Sciences; Oxford University Press: Oxford, U.K., 2006.